\documentclass[a4paper,11pt]{article}
\pdfoutput=1 

\usepackage{jcappub} 

\usepackage[T1]{fontenc} 

\usepackage{amsmath}
\DeclareMathOperator{\Tr}{Tr}
\usepackage{isomath}

\title{\boldmath Type Ia supernova Hubble diagrams with host galaxy photometric redshifts}

\author[a,1]{V. Ruhlmann-Kleider,\note{Corresponding author.}}
\author[b]{C. Lidman,}
\author[c]{A. Möller.}

\affiliation[a]{IRFU, CEA, Universit\'{e} Paris-Saclay, F-91191 Gif-sur-Yvette, France}
\affiliation[b]{The Research School of Astronomy and Astrophysics, Australian National University, Acton, ACT 2601, Australia}
\affiliation[c]{Centre for Astrophysics \& Supercomputing, Swinburne University of Technology, Victoria 3122, Australia}

\emailAdd{vanina.ruhlmann-kleider@cea.fr}
\emailAdd{christopher.lidman@anu.edu.au}
\emailAdd{amoller@swin.edu.au}

\abstract{
Systematic uncertainties associated to  type Ia supernova (SN Ia) Hubble diagrams from photometrically selected samples using photometric SN host galaxy redshifts are investigated. The host redshift 
uncertainties and the contamination by core-collapse SNe are both addressed. As a test case, we use the 3-year photometric SN Ia sample of the SuperNova Legacy Survey (SNLS), consisting of 437 objects between 0.1 and 1.05 in redshift with $4.7\%$ contamination. We combine this sample with non-SNLS objects of the spectroscopic sample from the joint analysis (JLA) of the SDSS-II and SNLS collaborations, consisting of 501 objects mostly below 0.4 in redshift. We study two options for the origin of the redshifts of the photometric sample, either provided entirely from the host photometric redshift catalogue used in the selection or a mixed origin where around 75$\%$ of the sample can be assigned spectroscopic redshifts from dedicated measurements or external catalogues. 
Using light curve simulations subject to the same photometric selection as data, we study the impact of photometric redshift uncertainties and contamination on flat $\Lambda_{CDM}$ fits to Hubble diagrams from such combined samples. Our primary finding is that photometric redshifts and contamination lead to biased cosmological parameters. The magnitude of the bias is found to be similar for both redshift options. This bias can be largely accounted for if photometric redshift uncertainties and contamination are taken into account  when computing the SN magnitude bias correction due to selection effects. To reduce the residual cosmological bias, we explore two methods to propagate redshift uncertainties into the cosmological likelihood computation, either by refitting photometric redshifts along with cosmology or by sampling the redshift resolution function. 
Redshift refitting fails at correcting the cosmological bias whatever the redshift origin, while sampling slightly reduces it in both cases.
Finally, for actual data, we find compatible results with those from the JLA diagram for mixed photometric and spectroscopic redshifts, while the full photometric option is biased upwards, but consistent with JLA when all statistical and systematic uncertainties are included.
}

\begin{document}
\maketitle
\flushbottom

\section{Introduction}
\label{sec:intro}
Distant type Ia supernovae, which gave evidence for the Universe expansion acceleration, are still an important probe to help characterising the equation of state of dark energy~\cite{Planck18,eBOSS20,Pantheon22}.
Standard SN~Ia selections rely on photometry for detection, supplemented by spectroscopic follow-up for typing and redshift assignment, which cannot be pursued in the future for the large SN~Ia samples expected from projects like the Vera C. Rubin Observatory Legacy Survey of Space and Time, the Euclid space mission or the Nancy Grace Roman Space Telescope. 
Purely photometrically selected SN~Ia samples were first used 
for rate measurements~\cite{Perrett12}, for cross-checking the Malmquist bias correction of a spectroscopic sample~\cite{Bazin11}, or to prove that spectroscopic observations of 
their hosts can result in large SN Ia samples with a good coverage of spectroscopic host redshifts, see e.g.~\cite{Lidman12}. The latter has since then become common practice in current SN~Ia projects~(\cite{Lidman20} for DES or ~\cite{Jones16} for Pan-STARRS) while photometric classification has become more complex, see e.g.~\cite{Moller22}.

Photometrically selected SN Ia samples have already been used for cosmology studies~(\cite{Sako11, Hlozek11, Campbell12} for SDSS and~\cite{Jones16, Jones17} for Pan-STARRS) but these used host galaxy spectroscopic redshifts. 
For these samples, the systematic uncertainty related to photometric classification is that of contamination, which can be overcome by both reducing the amount of contamination with more advanced classification and selection schemes~\cite{Vincenzi21} and estimating its impact more accurately with dedicated algorithms~\cite{Kunz07,Kessler17}. Recently, the use of host galaxy photometric redshifts for SN~Ia photometric samples of very low contamination has been explored~\cite{Chen22} and is expected to become of more importance for future large SN~Ia surveys~\cite{Mitra20,Linder19,Dai17} where contamination could also be an issue.

In this paper, we tackle the question of using photometric SN Ia samples with host photometric redshifts to derive cosmology constraints and study the impact of both photometric redshift resolution and core-collapse SN contamination on those constraints. For that purpose, we build mixed diagrams based on the 3-year photometric sample of SNLS~\cite{Bazin11} complemented by the non-SNLS part of the spectroscopic JLA sample~\cite{Betoule14}. 
We study the impact of using only host photometric redshifts or a mixture of photometric and spectroscopic redshifts for the SNSL photometric sample. This defines two possible mixed diagrams to be compared. 
We do this by means of Hubble diagrams derived from SN~Ia and core-collapse SN simulations and explore different approaches to propagate redshift resolution to the cosmological likelihood. We then fit the mixed diagrams from data. 

The outline of the paper is as follows. Section~\ref{sec:data} describes the SN~Ia photometric selection and the two mixed samples built from data, while section~\ref{sec:methodology} details the fitting methodology. Section~\ref{sec:bias} introduces the SN Ia light curve simulation and the photometric redshift model used to determine the SN magnitude bias of the photometric selection for pure SN Ia samples. Section~\ref{sec:redshift} describes two methods to propagate redshift resolution effects to cosmological fits and discusses the results on a set of simulated Hubble diagrams reproducing the redshift profile of the actual diagrams, assuming no contamination. Section~\ref{sec:conta} presents the core-collapse light curve simulation used to derive the expected contamination of the photometric selection and studies how the SN magnitude bias correction and the cosmological constraints evolve when that contamination is introduced in simulated diagrams. Finally, section~\ref{sec:resudata} discusses the results obtained on data and compare them to expectations from the simulation and to the JLA case.

\section{Data sample}
\label{sec:data}
We consider the JLA sample of high quality spectroscopically classified SN~Ia data assembled in~\cite{Betoule14} from the inter-calibrated SDSS-II and SNLS surveys, additional data from the HST and several nearby surveys from~\cite{Conley11}. This sample contains 740 events, among which 239 are from the SNLS survey. In this analysis, the latter subset is replaced by the 3-year photometric SNLS sample of~\cite{Bazin11}, hereafter briefly described.

\subsection{The SNLS 3-year photometric sample} 
\label{sec:cuts}

The SNLS photometric sample~\cite{Bazin11} contains 485 events, among which 150 are in common with the JLA spectroscopic SNLS subset  of 239 events. 
The first step of the photometric selection retains SN-like events, defined as 3-year light curves with a shape consistent with that expected from an SN event and sufficient temporal sampling in each of the four MegaCam bands. Each selected event was assigned a host photometric redshift from~\cite{Ilbert06} and its four light curves were simultaneously fit with the SALT2 light curve fitter~\cite{Guy10} at the assigned redshift.

In a second step, the SALT2 fit results were used to select SN~Ia compatible events. Restricting to well-sampled light curves, constraints on the SALT2 fit $\chi^2$, SN intrinsic colour and stretch were applied both to reject events with a poor fit and to reduce the contamination by core-collapse SNe or by SN~Ia events with incorrect redshift assignment. 

In the above procedure, light curves were calibrated using the SNLS tertiary standards of~\cite{Regnault09}. 
At the end of the selection procedure, the photometric redshift central resolution was $\sigma_{\Delta z/(1+z)} \sim 3\%$ and  the 5 (resp. 3) $\sigma$ outlier rates were 1.4\% (resp. 5.5\%).

\subsection{Mixed JLA samples}
\label{sec:mixed}
Prior to combining the above sample with non-SNLS events of the JLA sample, light curves of all 485 selected events were re-calibrated using the joint calibration of the SNLS and SDSS-II surveys~\cite{Betoule13} and re-fit with the same SALT2 version as in~\cite{Betoule14} at their assigned host photometric redshift. 
Combining the SNLS sample treated that way with the non-SNLS JLA events results in a first mixed JLA sample, referred as to the JLA-P sample in the following.

Part of the SNLS photometric events can also be assigned a spectroscopic redshift, either because they benefitted from a spectroscopic follow-up during the lifetime of SNLS, or because their host was observed in dedicated observations with the AAOmega spectrograph at the AAT~\cite{Lidman12}, or because they could be matched to an external catalogue of spectroscopic redshifts.  This is the case for 357 of the 485 SNLS photometric events. 
Replacing their photometric redshift by the spectroscopic one, and re-fitting those events with SALT2 at their spectroscopic redshift defines a second mixed JLA sample, hereafter called the JLA-B sample. 

In both samples, there are events at redshift above 1.05, a region where SALT2 results become unreliable for SNLS events due to vanishing signal in the $r_M$-band. We thus restrict SNLS SNe to have a redshift below that value in the two mixed samples, whose final composition is described in table~\ref{tab:samples}.
Note that compared to the JLA-P sample, the number of photometric redshifts in the JLA-B sample is divided by a factor of 4 and 90$\%$ of them are present at high redshift, $z>0.5$.

\begin{table}[tbp]
\centering
\begin{tabular}{|l|ccccc|}
\hline
sample & $N_{total}$ & $N_{spectro}$ & $N_{photo}$ & $N_{z_{pho}}$ & $N_{z_{spe}}$ \\
\hline
 JLA-P &  938 & 501 & 437 & 437 & 0 \\
JLA-B &  943 & 501 & 442 & 99 &  343 \\
\hline
\end{tabular}
\caption{\label{tab:samples} Composition of the two SN~Ia samples built from JLA non-SNLS spectroscopic events and SNLS photometric events. The photometric subsample uses host photometric redshifts for all events (line 1) or only for those events which do not have a spectroscopic redshift (line 2). While the initial SNLS photometric sample contains 485 events, the SNLS subsample in the diagram is restricted to redshifts below 1.05. Redshift migration above and below the redshift cut at 1.05 is responsible for the difference in the total number of SNe between the two diagrams.}
\end{table}

\section{Methodology for Hubble diagram systematic uncertainties}
\label{sec:methodology}
This paper relies on the method of~\cite{Betoule14} to propagate known uncertainties to SN Ia Hubble diagram fits. This method and the cosmological fitter we use are briefly described here. Changes  for the mixed JLA samples are detailed in the next three sections. 

\subsection{The JLA method}
\label{sec:method}
In the JLA method, the fit minimises  the following function:

\begin{equation}
\label{eq:chi2}
 \chi^2  = ( \vec{\mu}  -  \vec{\mu}_{model} )^\dag C_{\rm cov}^{-1} ( \vec{\mu}  -  \vec{\mu}_{model} )
 \end{equation} with $\vec{\mu}$ 
 the vector of distance estimates of SNe in the Hubble diagram, $\vec{\mu}_{model}$ 
the vector of their distance moduli predicted in the cosmology under test, and $C$ the covariance matrix of $\vec{\mu}$. 
For an SN with redshift $z$, the components of the $\vec{\mu}$  vector  are: 
\begin{equation}
 \label{eq:mu}
 \mu = m_B^{*} - (M_B - \alpha X_1 + \beta C) + \Delta m_B^{*}
 \end{equation}
where $m_B^*$, $X_1$ and $C$ are the SN parameters (apparent B-band rest-frame magnitude, stretch and colour) and $\Delta m_B^{*}$ is the SN magnitude bias correction due to selection effects. 
$M_B$ is the SN~Ia absolute magnitude and $\alpha$, $\beta$ are the nuisance parameters that describe the relation between the SN~Ia  peak magnitude and stretch and colour, respectively. The three SN parameters are provided by fitting the SN light curves by SALT2 at the SN redshift, but a peculiar velocity correction - important only at low redshift - is added to the SALT2 rest-frame magnitude to define $m_B^*$. As the absolute magnitude $M_B$ was found to depend on host galaxy properties~\cite{Sullivan11}, two possible values are usually considered for $M_B$, depending on the host galaxy stellar mass~\cite{Conley11,Betoule14}. Since our aim is to study systematic uncertainties in Hubble diagrams with host photometric redshifts and not to derive the most accurate cosmological results, we consider only one value for $M_B$  throughout this paper.

As for the vector of SN distance moduli, the components are:
\begin{equation}
\label{eq:model}
\mu_{model} = 5{\rm log_{10}}(d_L(z;\Omega_m)/10\rm{pc})
\end{equation}
where  $d_L$ is the luminosity distance at the SN redshift in the tested cosmology. The covariance matrix $C_{\rm cov}$ is expressed as:
\begin{equation}
\label{eq:covariance}
C_{\rm cov} = A C_{\eta} A^{\dag} + {\rm diag}\left(\frac{5\sigma_z}{z{\rm log10}}\right)^2 + {\rm diag} \left(\sigma^2_{lens}\right) + {\rm diag}  \left(\sigma^2_{int}\right)
\end{equation}
The last three terms, purely diagonal, account for redshift uncertainties due to peculiar velocities, magnitude uncertainties due to gravitational lensing and magnitude intrinsic dispersion not described by the other terms. Note that the peculiar velocity systematic term matters only at low redshift, in which case redshift uncertainties can be approximated by magnitude uncertainties. Values of $\sigma_z$, $\sigma_{lens}$ and $\sigma_{int}$ for the JLA sample are given in~\cite{Betoule14}. 
$C_{\eta}$ is the covariance matrix for the SN light curve parameters $\{m_B^{*}, X_1, C\}$ that includes statistical and systematic uncertainties:
\begin{equation}
\label{eq:eta}
C_{\eta} = C_{stat} + (C_{cal} + C_{LC mod} + C_{bias} + C_{dust} + C_{pecvel} + C_{nonIa})
\end{equation}
In the above, $C_{stat}$ propagates the light curve measurement uncertainties and the uncertainties due to the size of the SALT2 training sample, while the other matrices 
 deal with systematic uncertainties. Those are related to light curve calibration, $C_{cal}$, light curve model uncertainty, $C_{LC mod}$, SN magnitude bias correction estimate, $C_{bias}$, 
dust correction, $C_{dust}$,
peculiar velocity correction, $C_{pecvel}$ and contamination of the Hubble diagram by non Ia SNe, $C_{nonIa}$. The $A$ matrix in~\eqref{eq:covariance} obeys the following definition:
\begin{equation}
\label{eq:mata}
\vec{\mu} = A \vec{\eta} - \vec{u}(M_B)
\end{equation}
with $\vec{\eta}$ the vector of SN light curve parameters $\{...\, m_B^{*}, X_1, C \,...\}^\dag$ and $\vec{u}=\{...\,1, 0, 0 \,...\}^\dag$.

When only light curve measurement uncertainties are included, $C_{stat}$ generates the following purely diagonal term in 
$C_{\rm cov}$:
\begin{equation}
\label{eq:cstat}
\begin{split}
C_{LC} 
 &=  {\rm diag} \bigl( Var(m_B^{*}) + Var(\Delta m_B^{*}) + \alpha^2 Var( X_1) + \beta^2 Var( C) \\
& +2\alpha Cov_{mX}   -  2\beta Cov_{mC} - 2\alpha\beta Cov_{XC}  \bigr) 
\end{split}
\end{equation}
where variances of $m_B^{*}, X_1, C,\Delta m_B^{*}$ and covariances between $m_B^{*}, X_1, C$ contribute.
Adding to this the three diagonal terms of~\eqref{eq:covariance} generates what is referred to as 'diagonal' case in the following.

\subsection{Cosmology fitter}
\label{sec:fitter}
In this paper, the fixed-grid fitter of~\cite{Conley11} is used with the flat $\Lambda_{CDM}$ model as a benchmark. 
Following~\eqref{eq:chi2}, the fitter computes $\chi^2$ values over a grid in the $\{ \Omega_{m}, \alpha, \beta \}$ parameter space and convert them into likelihood probabilities via:
\begin{equation}
\label{eq:likel}
{\cal L} \propto \exp \left( -\frac{1}{2} \chi^2 +\frac{N_{SN}}{2} \right) 
\end{equation} 
with $N_{SN}$ the total number of SNe Ia in the Hubble diagram. The proportionality is obtained by normalisation over the grid. This neglects the fact that the correct normalisation depends on $\alpha, \beta$, in order to spare computation time, at the cost of producing biased uncertainties for these two nuisance parameters. The fitter reports the median value of each parameter, all other parameters being marginalised over (with flat priors). Confidence intervals for each parameter are derived from the cumulative probability distribution of that parameter, all other parameters being marginalised over. The interval limits are determined such that $68.3\%$ of the probability is enclosed. 

As $M_B$ is a simple additive parameter in the apparent magnitude predictions, it does not contribute to errors at each point of the grid (unlike $\alpha$ and $\beta$) and can be marginalised over analytically without generating a biased estimate. The fitter uses the $c/H_0$ reduced luminosity distances defined as:
\begin{equation}
\label{eq:reduced}
D_L= \frac{H_0}{c}d_L
\end{equation}
instead of $d_L$ in~\eqref{eq:model}. Consequently, the SN~Ia absolute magnitude which is marginalised over in the fitter is:
\begin{equation}
\label{eq:mb}
M_B'= M_B+ 5{\rm{log}_{10}} \frac{c}{H_0} + 25
\end{equation}
with $c$ expressed in km/s. 
This equation can be used to convert fitted values of $M_B'$ reported in this paper (see section~\ref{sec:fitdata}) into the usual definition of the SN~Ia absolute magnitude $M_B$.
Throughout this paper, a value of 70km/s/Mpc is assumed for $H_0$ (e.g. when quoting $M_B$ values from a cosmological fit or when computing luminosity distances).

Using the published JLA light-curves, we checked that 
 we could reproduce the results of~\cite{Betoule14} with our fitter, whether covariance matrices are the published ones or matrices that we reproduced following~\cite{Betoule14}. 
The next three sections describe how we treat the SN magnitude bias, photometric redshifts and core-collapse contamination for the mixed JLA samples.

\section{SN~Ia magnitude bias correction}
\label{sec:bias}
SN~Ia selection in a magnitude-limited survey generates incompleteness of the sample at high redshift which results in a bias of the reconstructed distances. For the SNLS component of the mixed JLA samples, this bias can be estimated from simulated SN light curves submitted to the same selection cuts as real SNLS light curves. 
In this section, we study the case of negligible contamination and consider SN~Ia simulation only.
The latter is described in the next section for the light curve part and in section~\ref{sec:zphomodel}  for the host photometric redshift modelling. The SN~Ia magnitude bias 
is presented in section~\ref{sec:biasres}.

\subsection{SN~Ia light curve simulation}
\label{sec:simul}
The light curve simulation is that of~\cite{Bazin11} with minor modifications. SN~Ia light curves were simulated from the SALT2 model of~\cite{Guy10} assuming a flat $\Lambda_{CDM}$ cosmology and a redshift volumetric distribution in the range 0.1 to 1.2. $X_1$ and $C$ parameter values were generated according to model distributions fitted to the set of spectroscopically identified SNe Ia present in our photometric sample, restricted to $z_{spe}<0.7$ to avoid spectroscopic selection biases. For each generated event, a magnitude $m_B^{*}$ was computed as a function of $X_1, C$ and the luminosity distance at the generated redshift in the assumed cosmology, $d_L(z, \Omega_M)$:
\begin{equation}
 \label{eq:mbstar}
 m_B^{*} = M_B - \alpha X_1 + \beta C +  5{\rm log_{10}}(d_L(z, \Omega_M)/10\rm{pc})
\end{equation}
A Gaussian spread  $\sigma_{int}$ was applied to $m_B^{*}$ in order to allow for SN~Ia intrinsic dispersion. Numerical values of $\Omega_M$, $M_B$, $\alpha$, $\beta$ and $\sigma_{int}$ were  updated w.r.t. those in~\cite{Bazin11}. The new values,  from the flat $\Lambda_{CDM}$ best fit of~\cite{Betoule14}, are reported in table~\ref{tab:simul}. A value of 70km/s/Mpc was assumed for $H_0$ when computing luminosity distances.

The magnitude of each SN after dispersion was converted into a global normalisation factor $X_0$ according to the relationship between $m_B^{*}$ and $X_0$  as measured from the spectroscopically identified SNe Ia in the photometric sample, with their light curves fitted with the same SALT2 version as in the simulation. We found:
\begin{equation}
 \label{eq:offset}
 -2.5 \rm{log}_{10}(X_0)  = m_B^{*} - 10.63935
\end{equation}
Contrary to what was done in~\cite{Bazin11},  no dependence in colour nor width was included in the above relationship as its effect (a few mmag at high colour values) was found to have a negligible impact on cosmological fits to our simulated samples given the intrinsic dispersion of 0.08.
 
For each event, the generated values of $X_0$, $X_1$, $C$ and redshift were then passed to SALT2 to produce light curves in the four MegaCam passbands.
All remaining steps of the light curve simulation, in particular the inclusion of detection and instrumental effects, is identical to what was done in~\cite{Bazin11}, to which we refer the reader for more details and for a comparison between data and simulation on variables used in the photometric selection of section~\ref{sec:cuts}. A total of 100,000 light curves were generated with the parameters values in the first line of table~\ref{tab:simul}. Lensing and peculiar velocity effects were not included in the simulation.

\begin{table}[tbp]
\centering
\begin{tabular}{|c|ccc|c|c|}
\hline
$\Omega_M$ & $M_B$ & $\alpha$ & $\beta$ & $\sigma_{int}$& $W_{out}$ \\
\hline
 0.295 &  -19.085 & 0.141 & 3.101 & 0.08 & 1 \\ \hline
 0.261 &  -19.085 & 0.135 & 3.026 & 0.06 & 0.94 \\
 0.319 &  -19.085 & 0.147 & 3.176 & 0.10 & 1.06 \\ 
\hline
\end{tabular}
\caption{\label{tab:simul} Parameter values used in our SN~Ia light curve simulations. The first row is for the main simulation, the second and third ones give extreme values in the ten simulations used to derive systematics. $W_{out}$ is introduced in section~\ref{sec:zphomodel}.}
\end{table}

\subsection{Host galaxy photometric redshift modelling}
\label{sec:zphomodel}

\begin{figure}[tbp]
\centering
\includegraphics[width=1.10\textwidth]{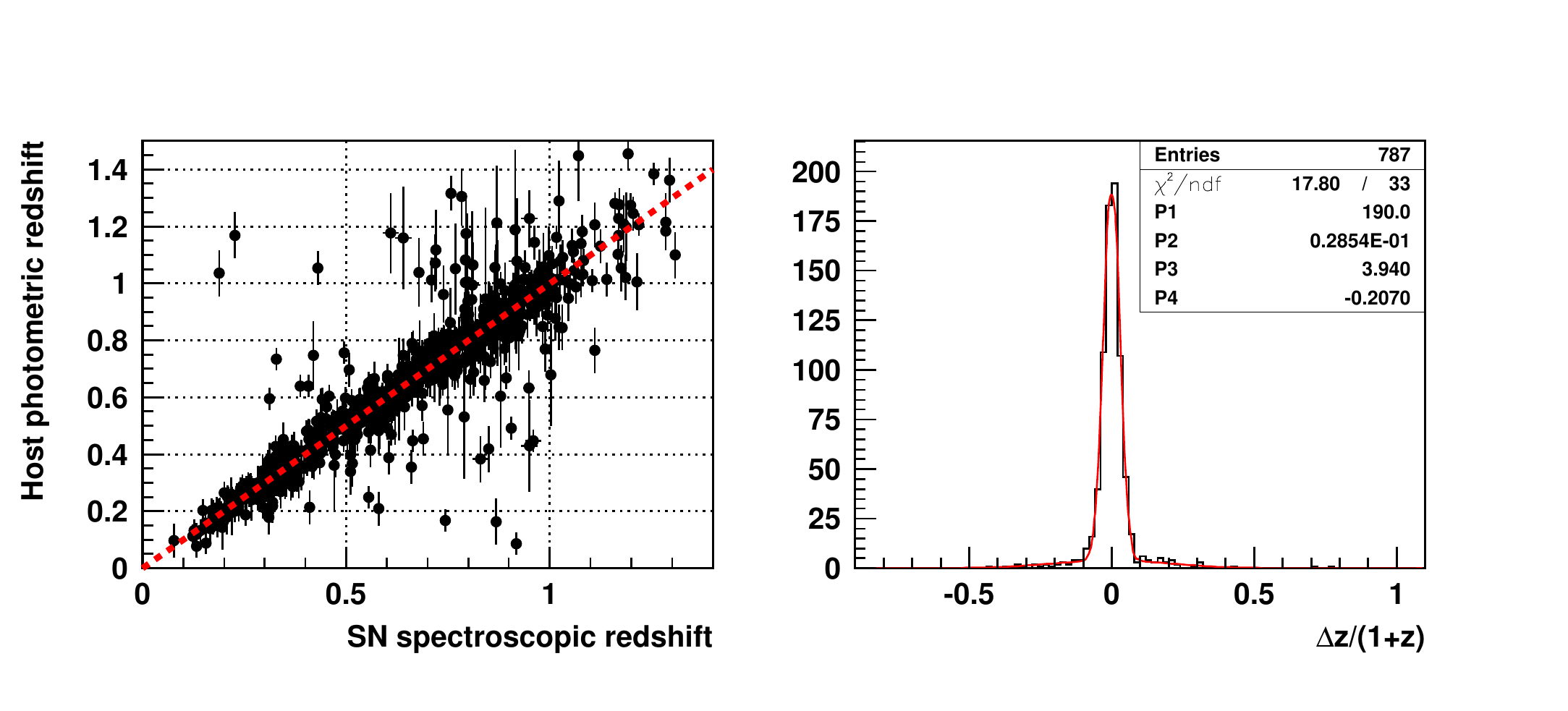}
\caption{\label{fig:zhostmodel} Host galaxy photometric redshifts compared with SN spectroscopic redshifts (either from the SN itself or its host) in the subsample of SNLS 3-year events with both measurements available. The comparison is done at the end of the SN selection first stage. The result of a double Gaussian fit to those data is presented in the righthand plot. Parameters P1 to P4 are the amplitudes and widths of the narrow and broad Gaussian components of the fitted model, respectively. The mean bias in redshift is $0.4\times10^{-3}$. }
\end{figure}

The model used to assign photometric redshifts to simulated events was changed with respect to that in~\cite{Bazin11} which was calibrated on a test-sample of 346 SN-like events having both host photometric redshifts and SN spectroscopic redshifts from the SNLS follow-up spectroscopy. Since that publication, the spectroscopic redshift coverage of 
SN-like events was enlarged, thanks to dedicated spectroscopic measurements of their hosts in several observation campaigns undertaken from 2012 to 2016 at the AAT~\cite{Lidman12} and to compilation of external catalogues of spectroscopic redshifts from surveys encompassing SNLS fields. As a result, the test-sample with redshifts of both origins now contains 787 events. 

From that sample, an updated host photometric model was derived, that describes the central resolution and the outlier distribution with two Gaussian distributions in $\Delta z/(1+z)$, since such a model provides an excellent fit to the $\Delta z/(1+z)$ distribution of the test-sample, as can be seen in figure~\ref{fig:zhostmodel}. The amplitudes of the two distributions are assumed to vary linearly with redshift in order to reproduce the observed increase in the number of outliers as redshift increases. Numerical values of the model parameters, derived from the test-sample, are summarised in table~\ref{tab:zoutliers}. The $3\sigma$ and $5\sigma$ outlier rates as measured in four redshift bins of the test-sample are found to be correctly reproduced by the simulation within uncertainties. Note that the mean bias in redshift, measured at $0.4\,10^{-3}$, is neglected in the model.

In the following, we consider the global normalisation of the outlier rate as the main source of systematic uncertainty in this modelling and account for it by scaling the $a_2$ and $b_2$ parameters w.r.t. their base values in table~\ref{tab:zoutliers}. The scaling factor is denoted $W_{out}$ in the following.

\begin{table}[tbp]
\centering
\begin{tabular}{|ccc|ccc|}
\hline
$\sigma_1$  & $a_1$ & $b_1$ & $\sigma_2$ &  $a_2$& $b_2$ \\
\hline
0.029 & -0.045 & 0.265 & 0.20 & 0.006 & 0.004 \\
\hline
\end{tabular}
\caption{\label{tab:zoutliers} Values of the parameters of the host photometric redshift model used in our light curve simulations. The model uses two Gaussian functions, one for the central resolution (index 1), the other for the outlier distribution (index 2). For each function, parameters are the Gaussian width ($\sigma$), and the slope ($a$) and offset ($b$) of the Gaussian normalisation ${\cal N}$, assumed to vary linearly with redshift, i.e. $\cal{N}$$=a \times z + b$. 
}
\end{table}

\subsection{SN~Ia magnitude bias computation}
\label{sec:biasres}
The SN~Ia magnitude bias of the SNLS photometric sample was estimated from simulated light curves subject exactly to the same selection algorithm as real data. 
We ran the SN~Ia selection algorithm of section~\ref{sec:cuts} on the reconstructed light curves (i.e. including instrumental noise) of our main SN~Ia simulation sample, which leaves a total of about 41,000 simulated events. Note that these selections make use of the SALT2 parameters derived from fits to the above light curves at their assigned host photometric redshift, as in real data for the two mixed samples. 

The selection bias $\Delta \mu(z)$ is computed in redshift bins, by taking, in each bin, the difference $\delta \mu (z)$ 
 between the event fitted distance modulus~\eqref{eq:mu} and the generated event distance modulus~\eqref{eq:model}:
\begin{equation}
 \label{eq:bias}
\delta \mu (z)  =  m_B^{*} - (M_B - \alpha X_1 + \beta C) -  5{\rm log_{10}}(d_L(z, \Omega_M)/10\rm{pc}) 
\end{equation}
and averaging it over all selected events in the bin. In the average, each difference is weighted by its statistical error, which encompasses light curve uncertainties, statistical uncertainties on the 
values of $\alpha, \beta$ and the SN~Ia intrinsic dispersion used in the simulation. These statistical uncertainties are also propagated in the averaging procedure to define the statistical uncertainty on $\Delta \mu(z)$. Note that the selection bias is the opposite to the magnitude correction~\eqref{eq:mu} to be used in the fit, $\Delta \mu(z)\equiv - \Delta m_B^{*}(z)$.

We first estimate what will be called hereafter the intrinsic bias of the selections. To do so, $m_B^{*}$, $X_1$ and $C$ in eq.\eqref{eq:bias} come from fitting the event generated light curves (i.e. without instrumental noise) with SALT2 at the event true redshift, which we consider as the event spectroscopic redshift and also used in the distance luminosity computation. $\Omega_M$ and $M_B$ are set to their input values in the simulation. For eq.~\eqref{eq:bias} to describe the effect of the selections on the apparent magnitude $m_B^*$ once corrected for variability through the $\alpha$ and $\beta$ terms, the values to be used for $\alpha$ and $\beta$ must reflect the variability of the initial, uncut SN~Ia sample. As eq.~\eqref{eq:bias} is intended to correct SN~Ia apparent magnitudes to be used in cosmological fits, it must also be coherent with the cosmological fitting procedure applied to real data.

\begin{figure}[tbp]
\centering
\includegraphics[width=.8\textwidth]{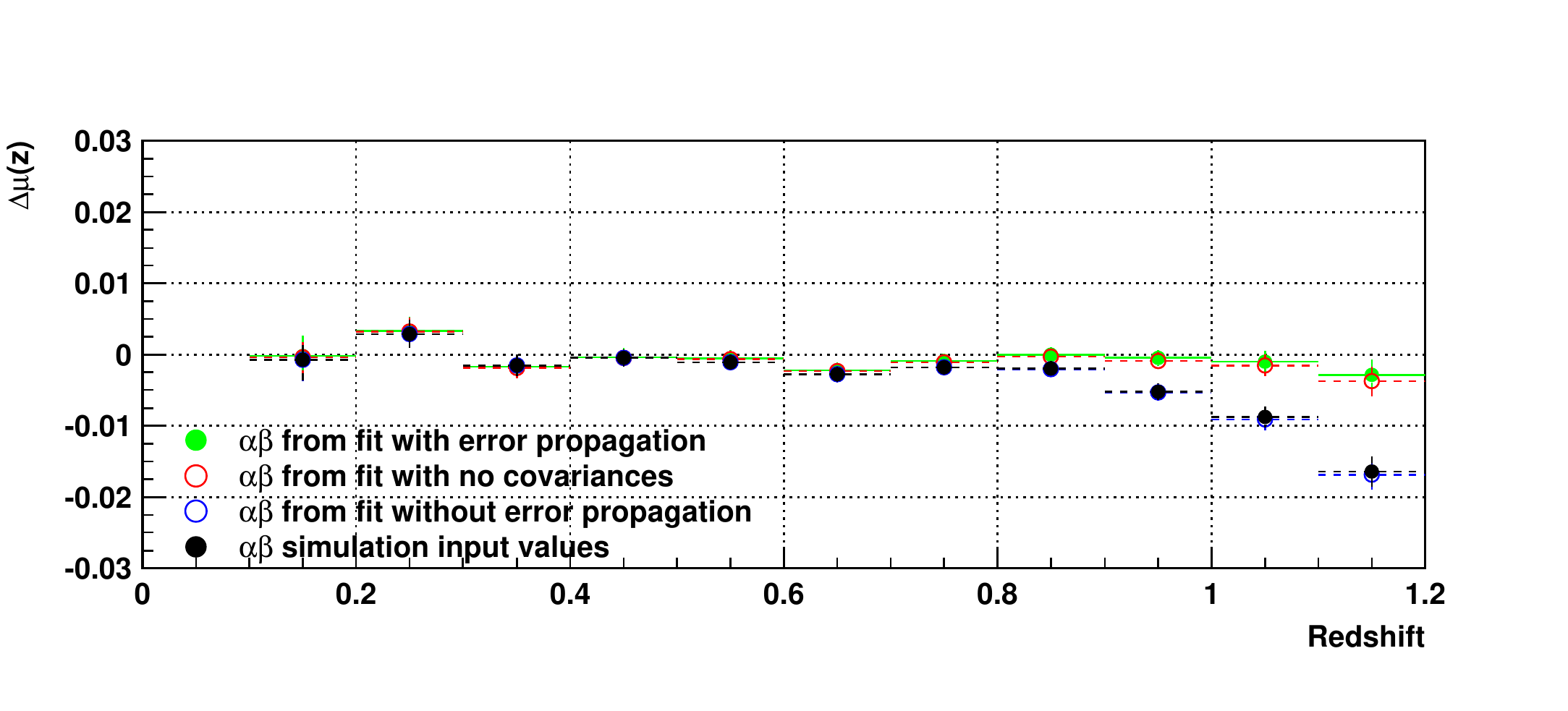}\\
\vspace{-1.5cm}
\includegraphics[width=.8\textwidth]{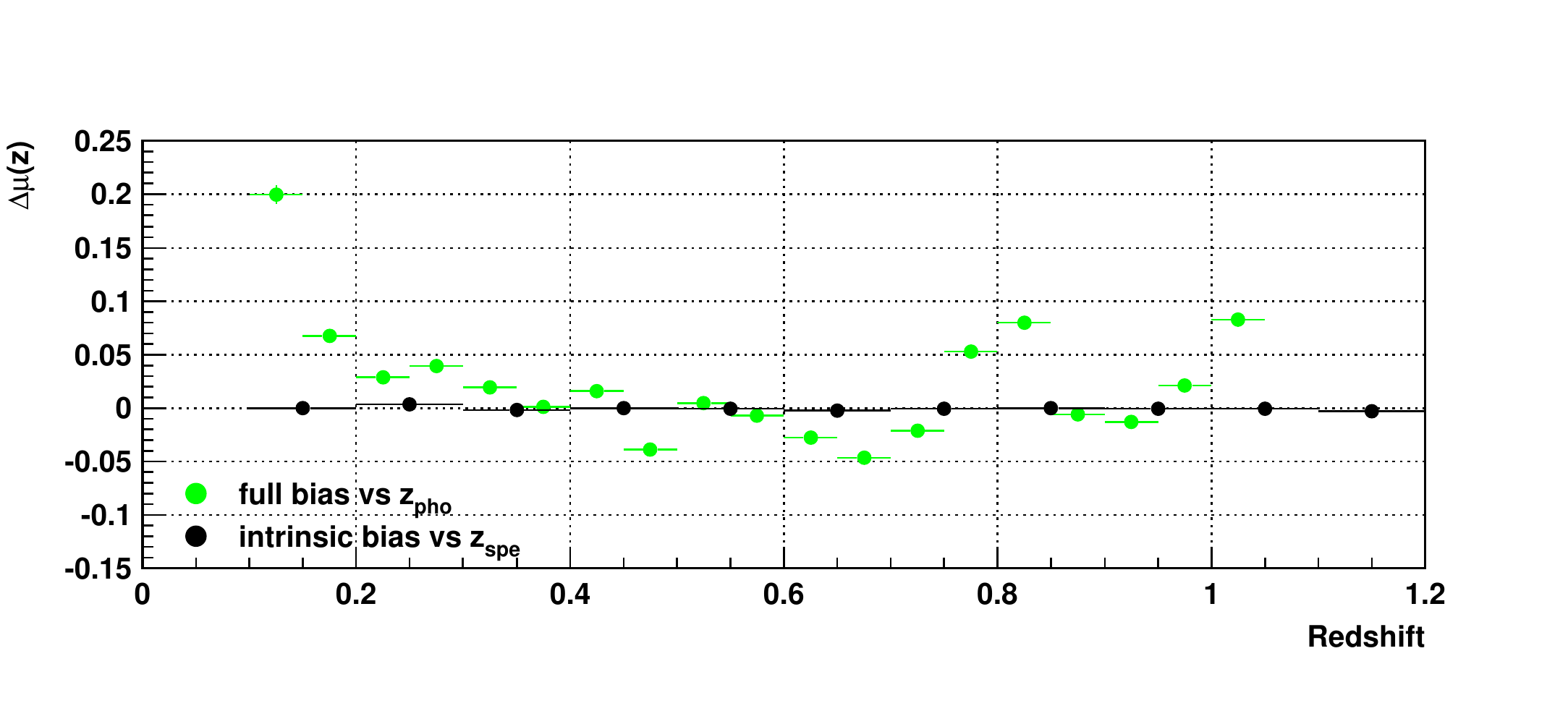}\\
\caption{\label{fig:bias} {\it Top.} SN~Ia photometric selection magnitude bias as a function of the spectroscopic redshift, from simulated SN~Ia light curves without instrumental noise, adjusted by SALT2 at their spectroscopic redshift (bias called intrinsic bias). Different choices of $\alpha$ and $\beta$ values are shown, either from cosmological fits with full light curve parameter error propagation (green, baseline), neglecting covariances in error propagation (red), with no error propagation at all (blue), or  taking simulation input values (black).
{\it Bottom.} SN~Ia photometric selection magnitude bias as a function of the host photometric redshift (green), from simulated SN~Ia light curves with instrumental noise, adjusted by SALT2 at their host photometric redshift. This bias, called full bias, is compared to the intrinsic bias as a function of the spectroscopic redshift (black), both biases being computed with the baseline values of $\alpha$ and $\beta$. 
All error bars are statistical.}
\end{figure}

 For the above reasons, the $\alpha$ and $\beta$ values in eq.~\eqref{eq:bias} result from a  flat $\Lambda_{CDM}$ fit to a large sample of simulated light curves without selections, fitting again generated light curves with SALT2 at their true, i.e. spectroscopic, redshifts. We used 25,000 events from the main simulation,  kept $\Omega_M$ fixed at the simulation input value (0.295) and propagated light curve parameter errors,  as in real data. 
 The $\sigma_{int}$ value in this fit was chosen so as to obtain a $\chi^2$ value close to the total number of SNe in the sample, as is common in SN~Ia cosmological analyses.
That fit gave:
\begin{equation}
 \label{eq:nuisanceforbias}
 \alpha = 0.14098 \pm 0.00045, \beta = 3.2452^{+0.0067}_{-0.0072}, M_B=-19.084
 \end{equation}
 We note that the simulation input value (see table~\ref{tab:simul}) is recovered for $\alpha$ and $M_B$ but not for $\beta$, an effect which is due to light curve parameter error propagation in the cosmological fit. Indeed, disabling error propagation in the fit  would give $\alpha = 0.14097\pm0.00044$ and $\beta = 3.0964^{+0.0056}_{-0.0079}$ in good agreement with simulation input values.

The bias computed with the values in~\eqref{eq:nuisanceforbias}
accounts for both the effect of the photometric selections and the fitting procedure, as advocated in~\cite{Mosher}. The top plot of figure~\ref{fig:bias} presents its evolution as a function of the SN~Ia spectroscopic redshift (green curve) and illustrates the effect of different choices of $\alpha$ and $\beta$ values in the bias computation. This plot shows that throughout the entire redshift interval used in the cosmological analysis (up to 1.05), our photometric selection generates a negligible bias, provided light curve error propagation in the cosmological fitting procedure is accounted for in the nuisance parameter values chosen to compute the bias. This is what we will do in the remaining of this paper. 

In actual data, only host galaxy photometric redshifts are available for all photometrically selected events. A magnitude bias including also that effect can be computed using assigned host redshifts in eq.~\eqref{eq:bias} for both luminosity distances and SALT2 parameter derivation. The latter must then use reconstructed light curves, that is, including instrumental noise, to allow SALT2 to fit light curves at redshifts far from the true ones. Nuisance parameters values remain the same as in the intrinsic bias computation, see eq.~\eqref{eq:nuisanceforbias}. The result, referred to as full SN~Ia magnitude bias in the following, is presented in figure~\ref{fig:bias} (bottom plot) as a function of the host photometric redshift and compared to the intrinsic bias as a function of the SN~Ia spectroscopic redshift. Once redshift migration due to resolution and outliers is accounted for, the magnitude bias becomes much larger, especially at low redshift due to strong variations of luminosity distances. As reconstructed light curves are used, the bias estimate becomes also noisier. To check the trend with redshift, the full bias was computed separately in the four SNLS fields. The redshift evolution is similar in the four fields up to $z=0.85$, indicating that redshift uncertainty is the dominant effect there, while light curve noise takes over at higher redshift. The noticeable bias increase  between redshifts 0.65 and 0.85 is likely related to the threshold at $z=0.68$, above which the $g_M$ band is no longer used in SALT2.
We discuss which magnitude bias correction to use in fits to the JLA-P and JLA-B diagrams in section~\ref{sec:toymcfits}. Systematic uncertainties in the bias estimates are described in Appendix~\ref{app:A}.


\section{Photometric redshift systematic uncertainties}
\label{sec:redshift}

 Accounting for photometric redshift uncertainty in Hubble diagram (HD) fits is difficult  since redshift uncertainties impact both the model (luminosity distance) and the data (corrected apparent magnitude) to be compared. 
 Two possible methods are proposed to properly propagate redshift uncertainties to cosmological fits. The first one refits individual photometric redshifts along with cosmology, while the second one samples the redshift resolution function to statistically propagate photometric redshift uncertainties to the cosmological fit $\chi^2$. In the following, the two methods are first described and then tested with a batch of simulated HDs representative of the sample size and redshift profile of the mixed samples from data. 

\subsection{Modelling SALT2 results}
\label{sec:shiftparam}
\begin{figure}[tbp]
\centering
\begin{tabular}{ccc}\includegraphics[width=.31\textwidth]{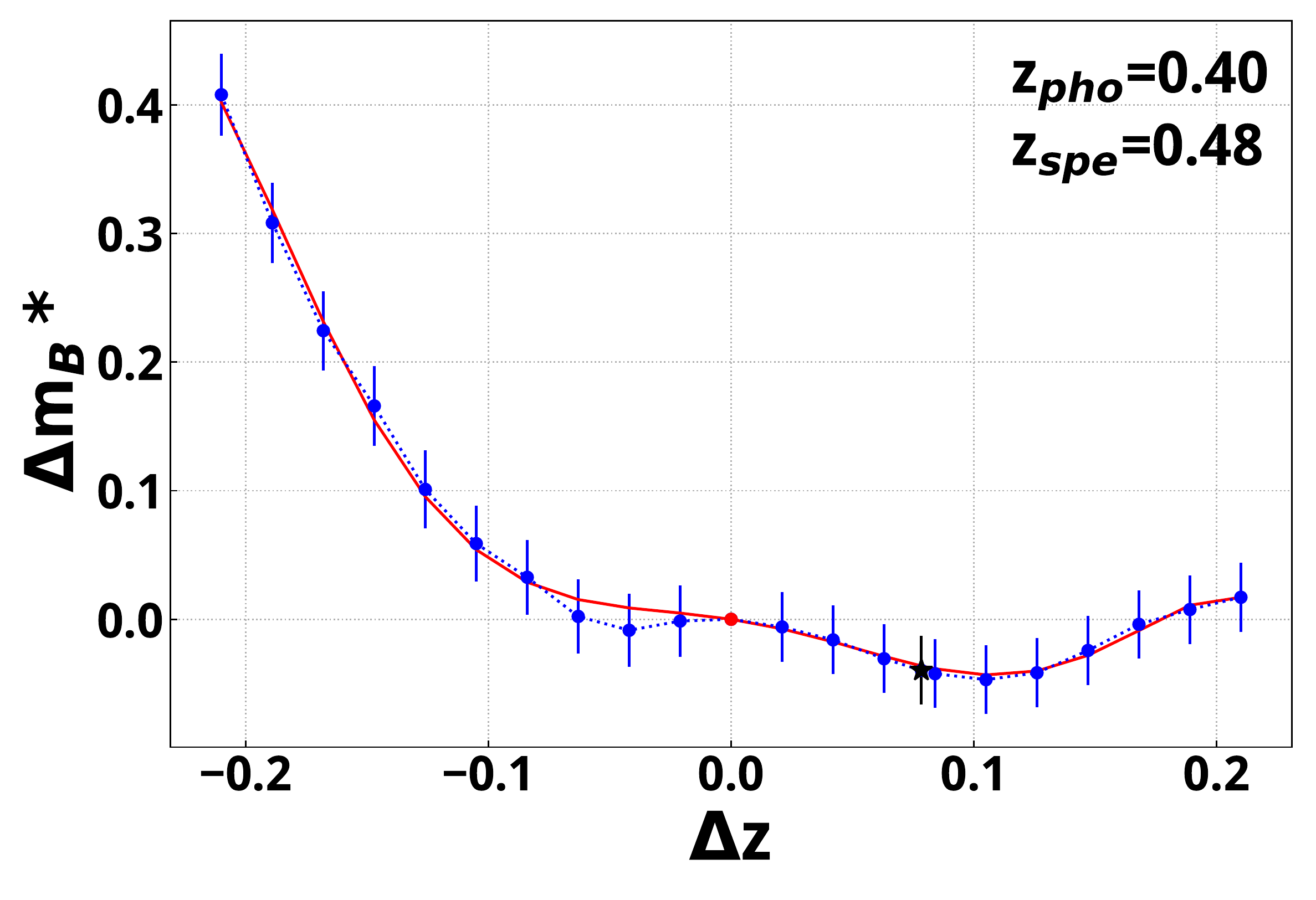} & \includegraphics[width=.31\textwidth]{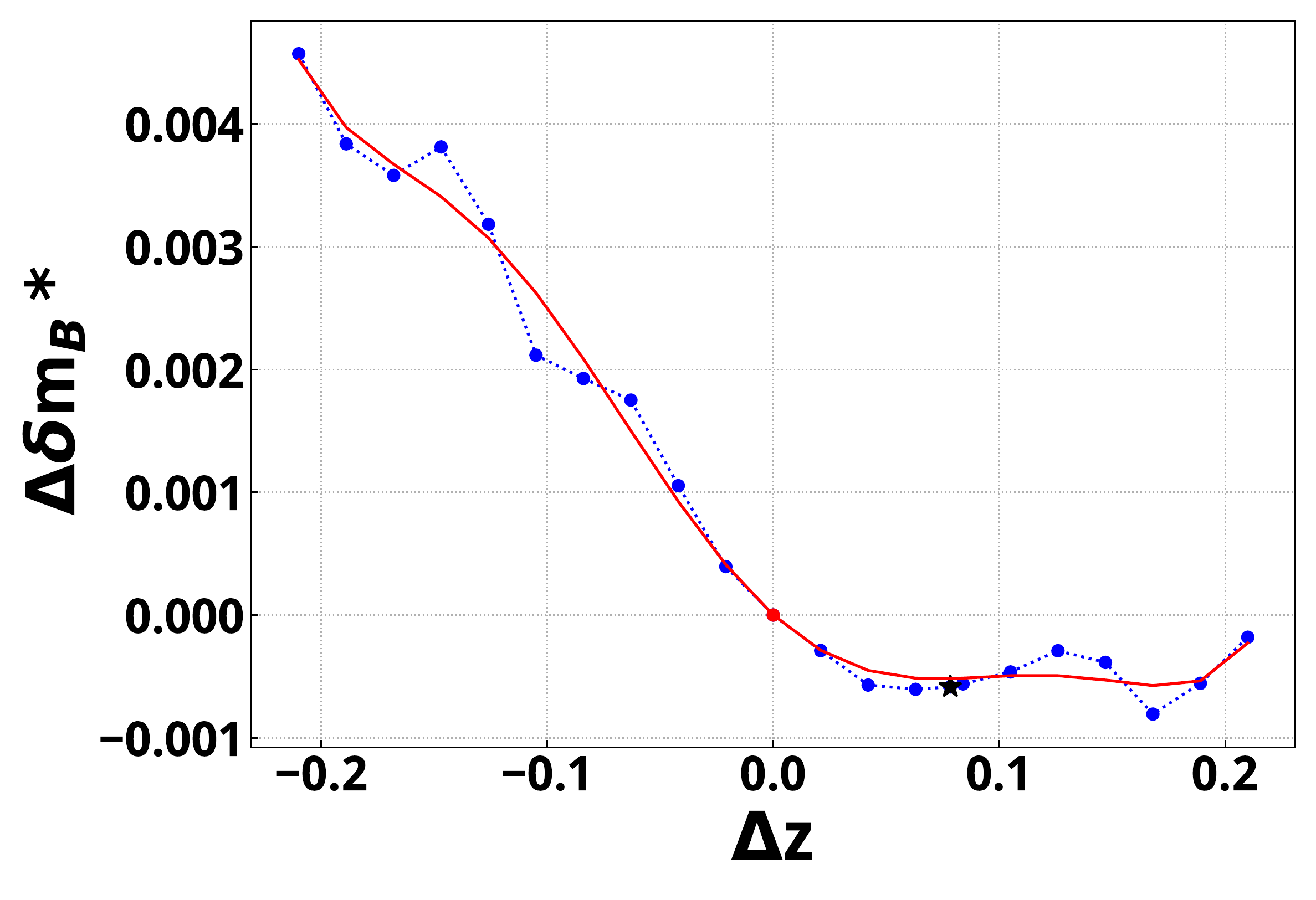}  & \includegraphics[width=.31\textwidth]{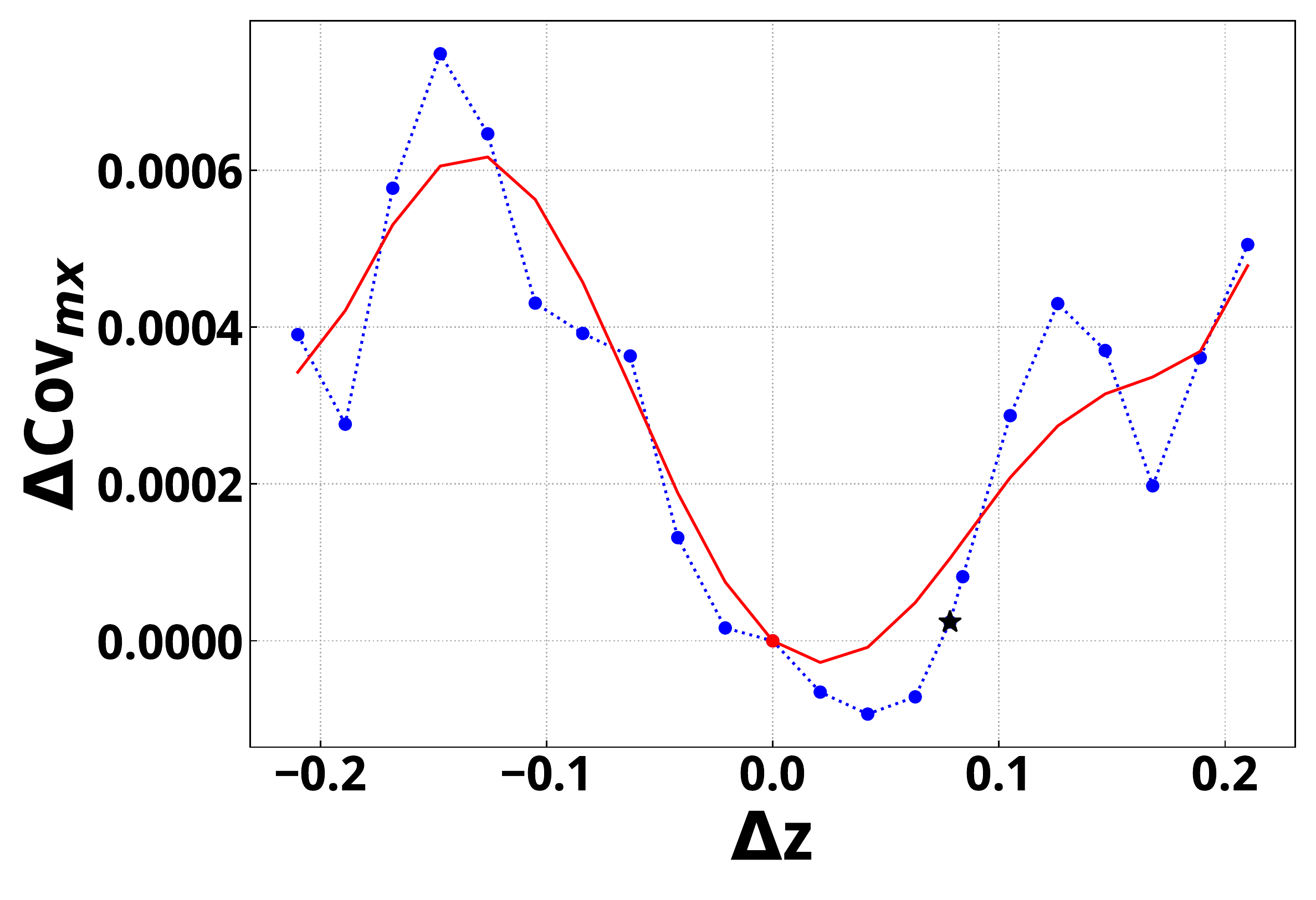}   \\
 \includegraphics[width=.31\textwidth]{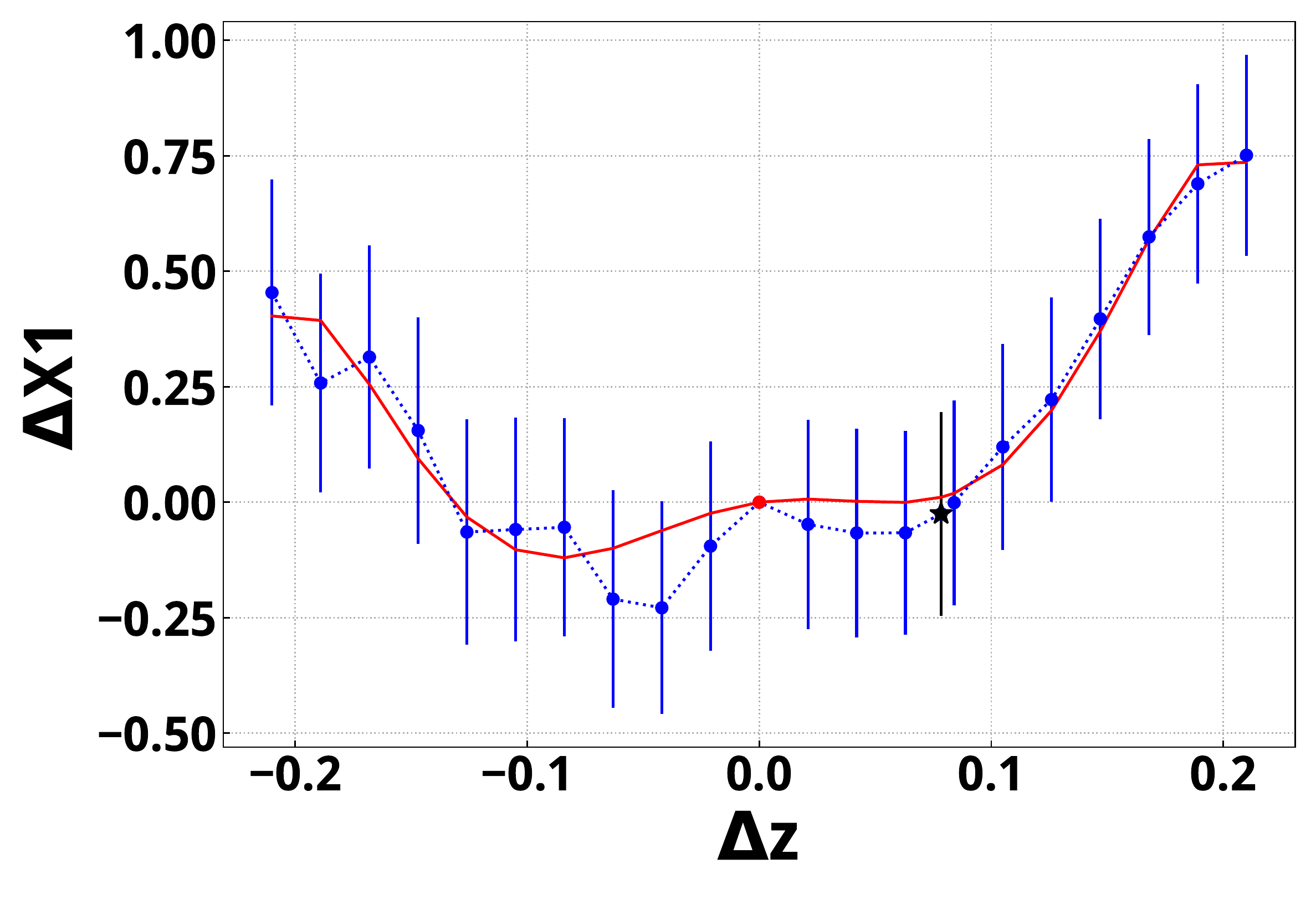} &  \includegraphics[width=.31\textwidth]{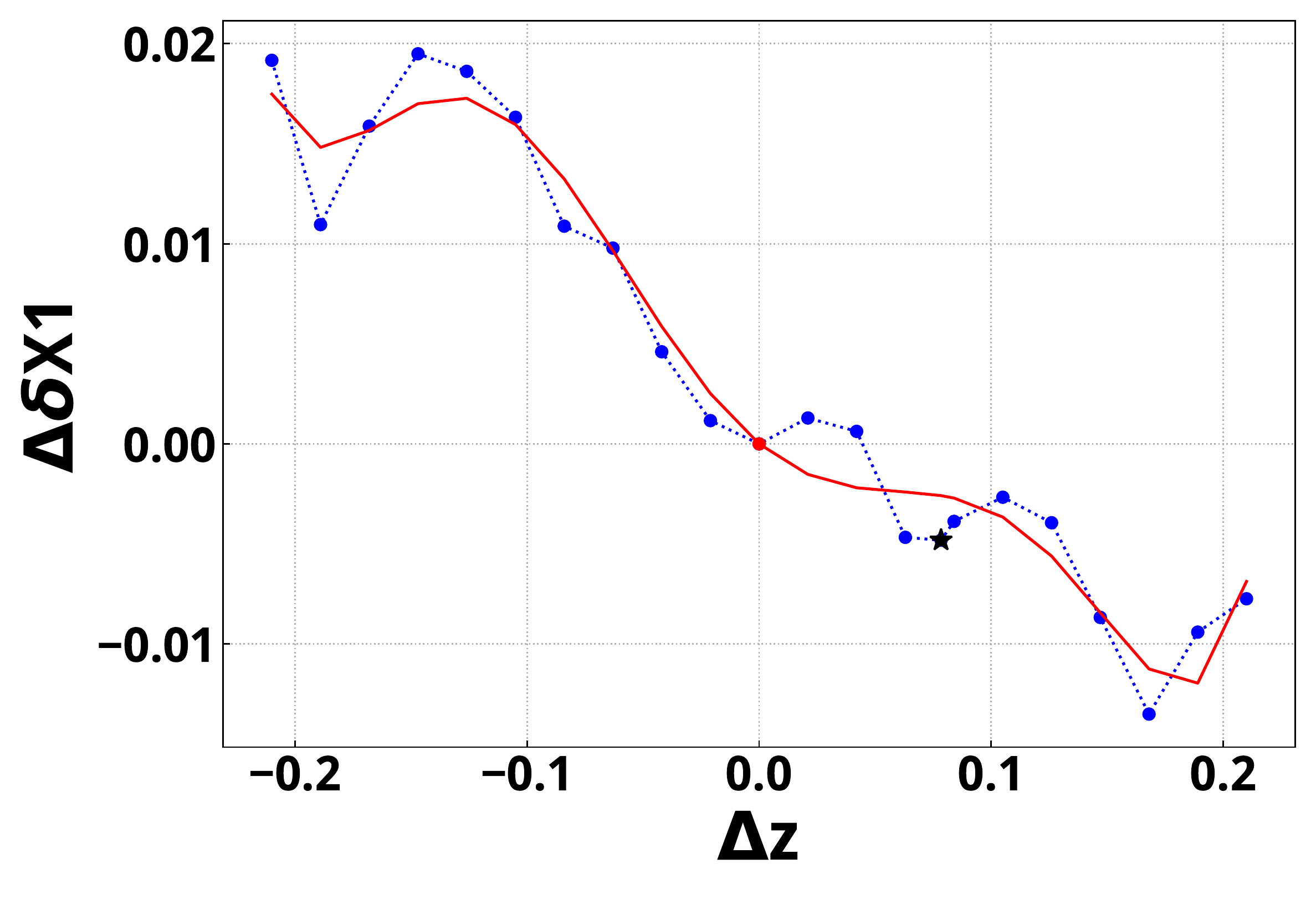} & \includegraphics[width=.31\textwidth]{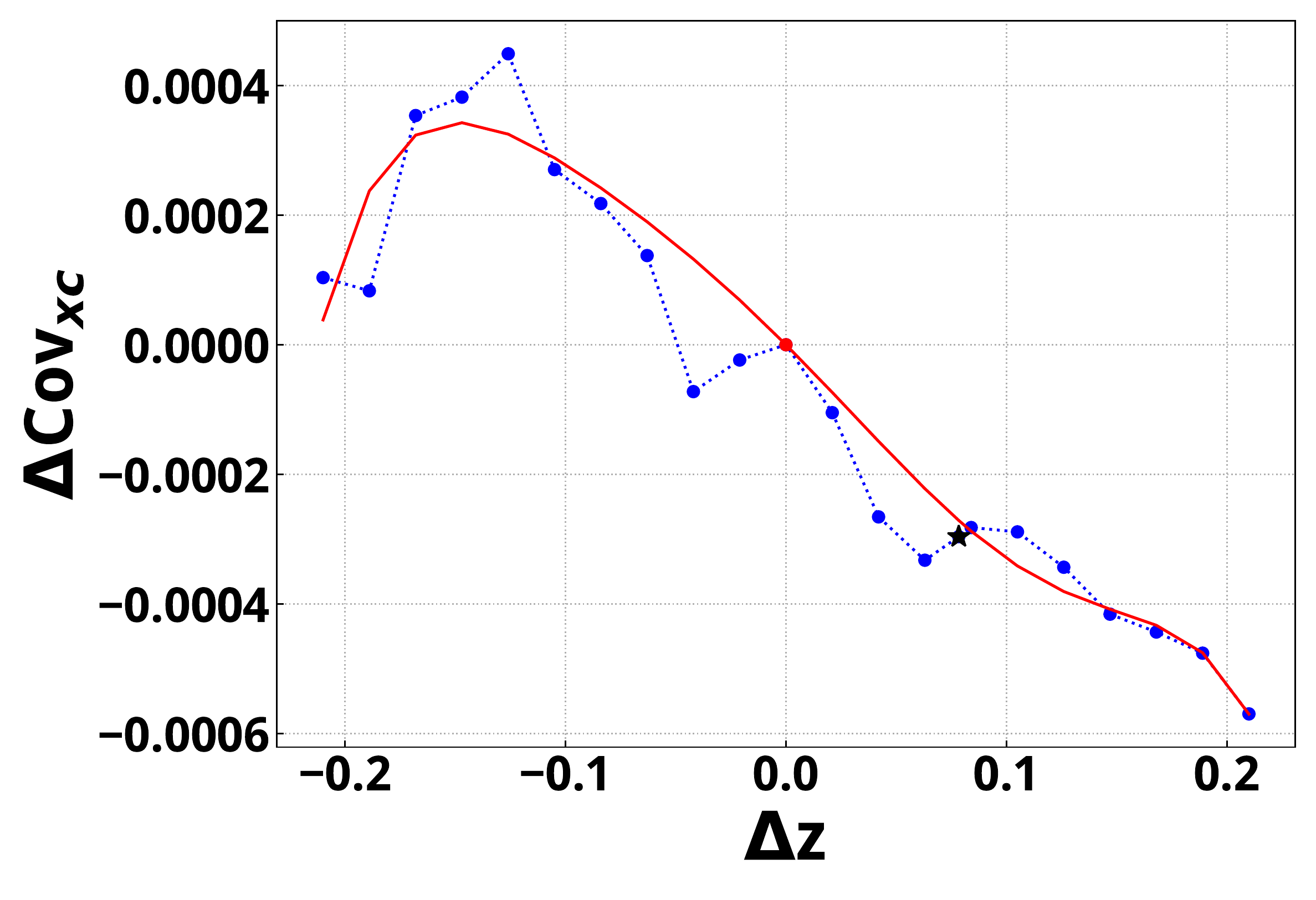}  \\
 \includegraphics[width=.31\textwidth]{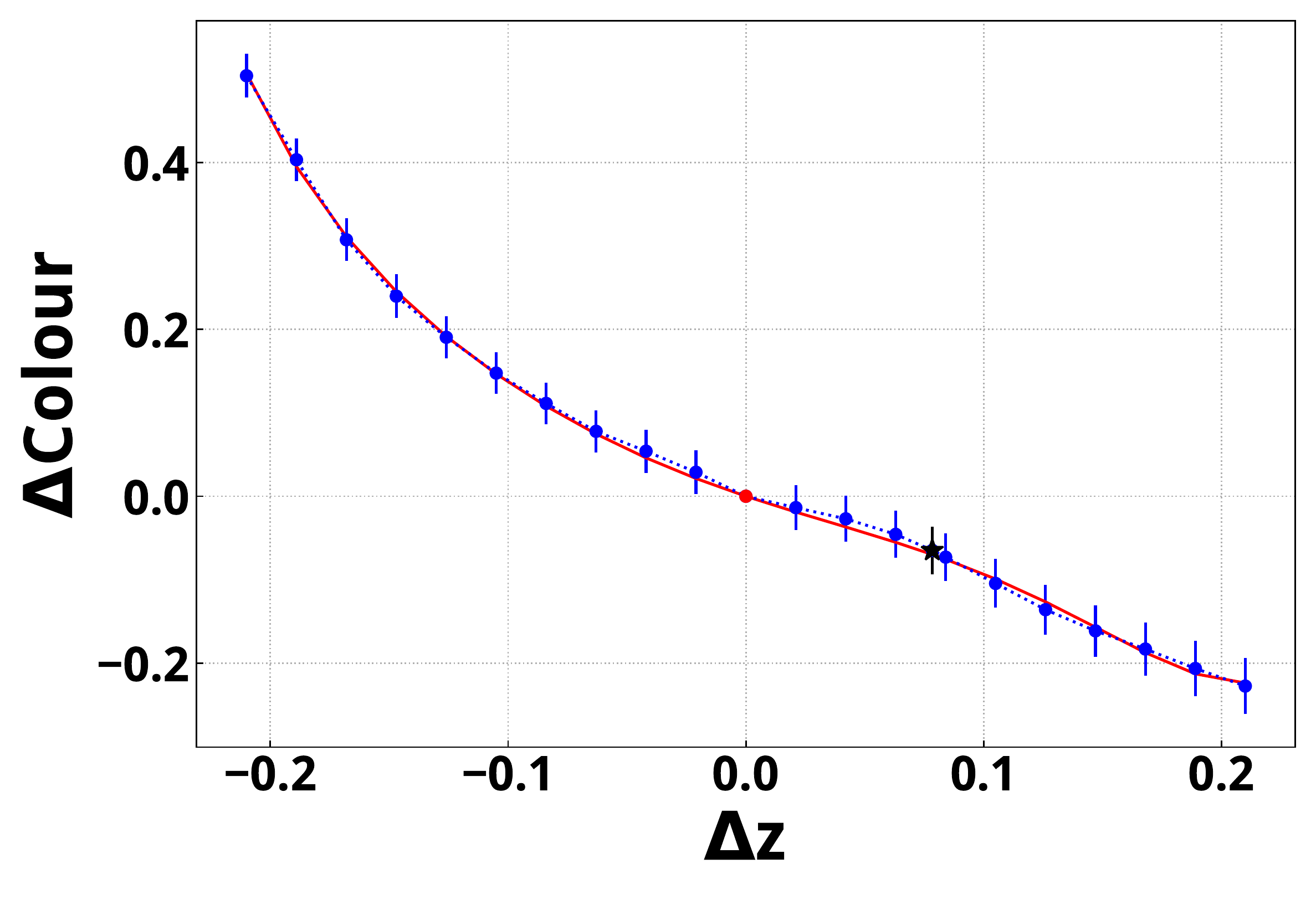} &  \includegraphics[width=.31\textwidth]{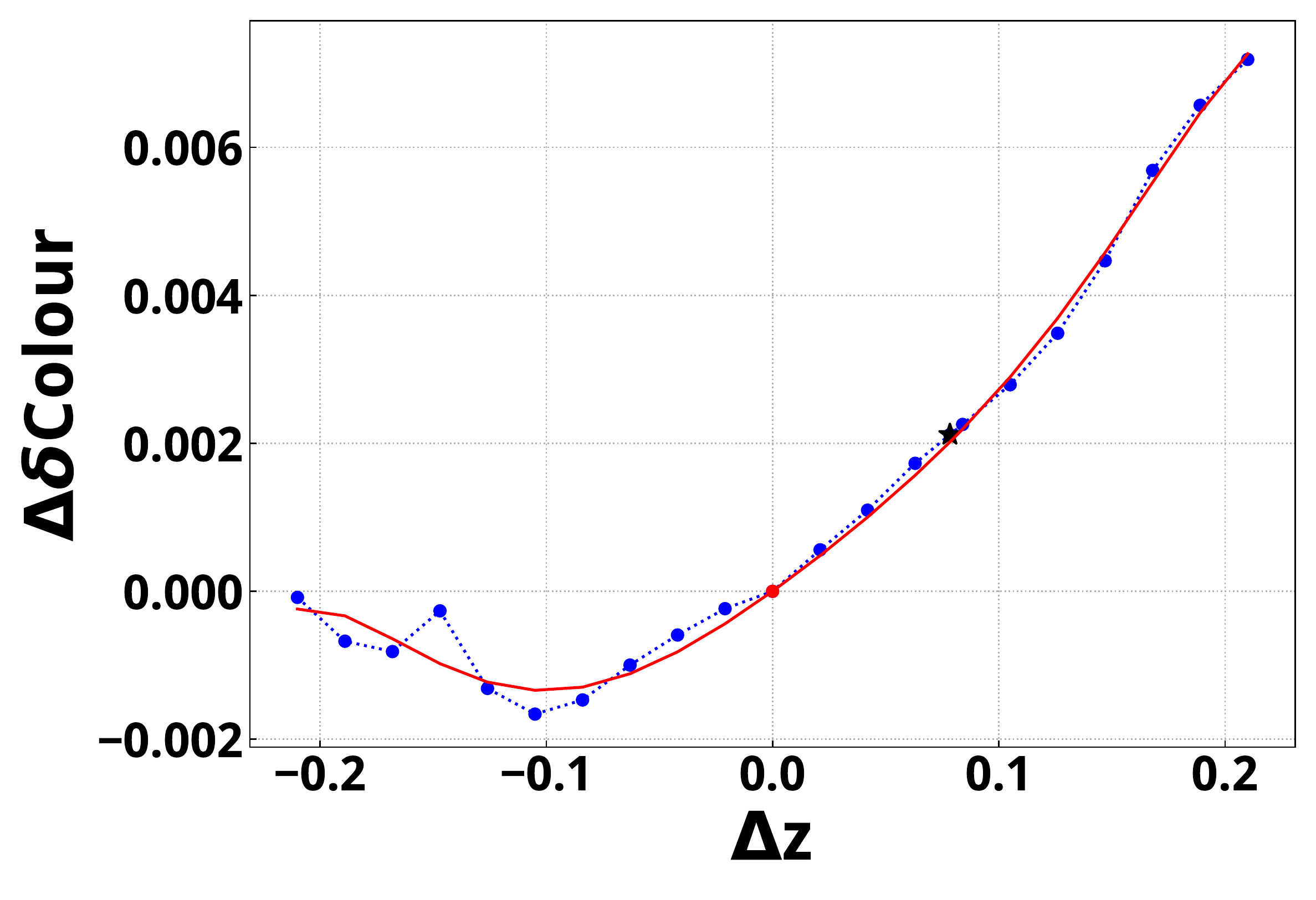} & \includegraphics[width=.31\textwidth]{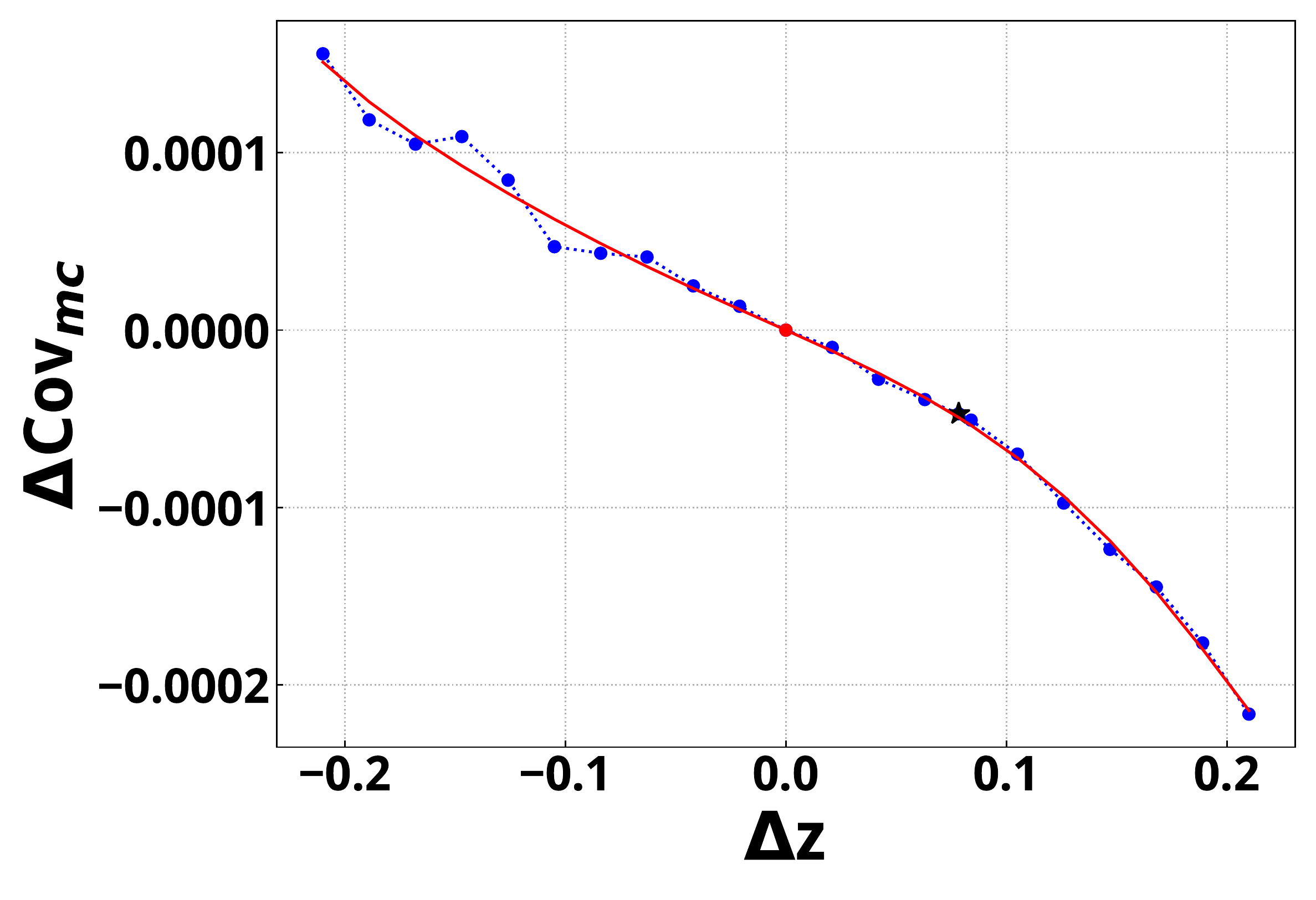}  \\
\end{tabular}
\caption{\label{fig:shifts} Variation of SALT2 parameters, their errors and covariances 
when changing the redshift hypothesis in SALT2 fits to light curves of one simulated SN~Ia event.  
The redshift change ($\Delta z)$ is defined with respect to the  event photometric redshift, $z_{pho}$. The star marks results obtained at the true redshift, $z_{spe}$. Red lines are degree 6 polynomial fits to the variations. Error bars on the left-hand plots are SALT2 errors from fits at the new redshift hypothesis. They are not used in the polynomial fits since they do not vary much from one point to the other but give an idea of the polynomial fit reliability.}
\end{figure}

Both methods require to know how the SALT2 parameters of the photometrically selected events would evolve when changing the redshift used in SALT2 fitting. An analytical modelling of those variations was adopted for this purpose. To do so, the light curves of all photometric events to be treated by either method are fit with different redshift hypotheses. In addition to the initial fit based on the assigned photometric redshift, $z_{pho}$, we run 20 other fits at redshifts defined in the following way:
\begin{equation}
 \label{eq:shifts}
z_{test} = z_{pho} + n  \times 0.03(1+z_{pho}) \quad \text{with} \quad n = \pm 0.5, \pm 1, \pm 1.5, \dots, \pm 5 
\end{equation}
These fits encompass redshift variations up to 5 times the central redshift resolution observed in data (see section~\ref{sec:zphomodel}). The shifts 
of $m_B^{*}$, $X_1$, $C$, their errors, covariances, and the shifts of the global SALT2 $\chi^2$ with respect to the initial fit results are then modelled with a degree 6 polynomial as a function of the change in redshift. In order to ensure reliable polynomial fits, points showing strong deviations w.r.t. the bulk of the results were automatically removed from the fit. This eliminates either single outliers or groups of points close to the minimal redshift accessible to SNLS, a region where SALT2 results become unreliable. Results of the polynomial fits are illustrated in figure~\ref{fig:shifts} for one simulated SN~Ia. Such a polynomial provides a very good fit to most variations, the most difficult case being shifts of $X_1$, its error and its covariances with the other two parameters. On the other hand, colour is observed to vary essentially linearly with redshift changes. Variations of the global SALT2 $\chi^2$, not represented in figure~\ref{fig:shifts}, are also smooth.

Note that we use the photometric redshift model of section~\ref{sec:zphomodel} to describe redshift uncertainties rather than the individual uncertainties provided in the catalogue (which are only crude estimates) but the method could be easily adapted to robust individual redshift uncertainties.

\subsection{Refitting redshifts with cosmology}
Redshift refitting was implemented in the cosmological fitter of section~\ref{sec:fitter}.
At each ($\Omega_M$, $\alpha$,$\beta$) point of the fitter grid, all photometric redshifts are varied, and these variations are propagated to distance luminosities, magnitude bias correction and, by use of the above shift model, to light curve parameters, their errors, covariances and the SALT2 $\chi^2$. The cosmological fit $\chi^2$ at the tested point is updated accordingly. Redshift variation is  pursued until minimisation of the $\chi^2$ is achieved and the redshifts at that minimum value are used to define the likelihood value in eq.~\eqref{eq:likel} and the analytically marginalized value of $M_B$. In addition, redshifts obtained after minimisation are marginalised over the grid to define new redshifts  of the photometric SNe. As redshift refitting aims at moving photometric SNe closer to their true redshift in the Hubble diagram, the magnitude bias correction should in principle be the correction to the intrinsic magnitude bias of sec~\ref{sec:bias}.

More precisely, the $\chi^2$ computed at each iteration of the minimisation is:
\begin{equation}
\label{eq:chi2refit}
\begin{split}
\chi^2(\vec{z}+\vec{\delta z}) &= \big[ \vec{\mu}(\vec{z}+\vec{\delta z})  -  \vec{\mu}_{model} (\vec{z}+\vec{\delta z})\big]^\dag C_{\rm cov}^{-1}(\vec{z}+\vec{\delta z}) \big[ \vec{\mu}(\vec{z}+\vec{\delta z})  -  \vec{\mu}_{model}(\vec{z}+\vec{\delta z}) \big] \\  
&+  \Tr \big[  {\rm diag} \left( \frac{\delta z} {0.03(1+z)}\right)^2  \big]  
+ \Tr \big[ {\rm diag} (\chi^2_{Salt}(\vec{z}+\vec{\delta z})-\chi^2_{Salt}(\vec{z}) )\big]
\end{split}
  \end{equation}
where $\vec{\delta z}$ is the vector of SN redshift changes, which are non zero only for photometric redshifts, where diag stands for a diagonal matrix and $\Tr$ for its trace.
In the above, components of $\vec{\mu}$ and $\vec{\mu}_{model}$ are still defined by equations~\eqref{eq:mu} and~\eqref{eq:model}, with $m_B^*, \Delta m_B^*, X_1, C$ and ${\rm d_L}$ computed at a redshift $z+\delta z$ instead of the initial photometric redshift $z$. 
In principle, the covariance matrix $C_{\rm cov}$, still given by~\eqref{eq:covariance} should have all its SN redshift dependent terms evaluated at redshifts $\vec{z}+\vec{\delta z}$, but this is impossible to put in practice. 
As the purpose of this paper is to test whether fitting photometric redshift offsets along with cosmology is feasible, we restrict to the purely diagonal case, for which $C_{\rm cov}=C_{LC}$ whose variation with redshift changes only requires to evaluate variances and covariances of $m_B^{*}, X_1, C$ and variances of $\Delta m_B^*$ at $\vec{z}+\vec{\delta z}$ in~\eqref{eq:cstat}. 

Eq.~\eqref{eq:chi2refit} contains two extra terms compared to~\eqref{eq:chi2}. The first one is a Gaussian prior to limit redshift excursion in agreement with the (central) resolution in the model of section~\ref{sec:zphomodel}. The second term is the SALT2 $\chi^2$ difference between fits run at $\vec{z}+\vec{\delta z}$ and at the initial redshift. This term intends to disfavour redshift variations that would improve the cosmological fit at the cost of a degradation of the light curve fit. As shown in figure~\ref{fig:chi2} in appendix~\ref{app:B}, globally one third of the events have a better SALT2 $\chi^2$ when light curves are fit at the event photometric redshift instead of the true one, but this fraction drops for HD outliers.
The SALT2 $\chi^2$ term in eq.~\eqref{eq:chi2refit} should thus help correcting redshifts more efficiently. Finally, to avoid testing redshift variations outside the applicability range of our SALT2 shift model, redshift variations are required to remain below $0.15(1+z)$, that is below $5\sigma$ and to give new redshifts between 0.1 (the minimal redshift in the SNLS sample) and 1.05.

Given the large number of photometric redshifs to fit (see $N_{z_{pho}}$ in table~\ref{tab:samples}), we used a multidimensional minimiser with derivatives to perform the minimisation of $\chi^2(\vec{z}+\vec{\delta z})$ over redshift changes $\vec{\delta z}$. We chose the vector Broyden-Fletcher-Goldfarb-Shanno (BFGS) algorithm in its second version implemented in the GSL library. The minimiser parameters were set to 0.03 for the initial trial step on $\vec{\delta z}$, 
$1.5\times N_{z_{pho}}$ for the iteration stopping criterion on the $\chi^2$ gradient norm, 5 for the maximum number of iterations and 0.1 (the default value) for the so-called tolerance on the line minimisation accuracy,  which defines the extent of the parameter space domain explored by the minimisation.

\subsection{Sampling the photometric redshift resolution function}
The second method propagates redshift uncertainties to the cosmology fit $\chi^2$ without attempting to refit individual photometric redshifts. At each ($\Omega_M$, $\alpha$,$\beta$) point of the fitter grid, all photometric redshifts are varied according to the central Gaussian resolution in the model of section~\ref{sec:zphomodel} and these variations are propagated to the $\chi^2$ as in the previous method. This is repeated a number of times and the $\chi^2$ values are averaged over all repetitions to define the $\chi^2$ at the current grid point in eq.~\eqref{eq:likel}. A similar procedure is used to propagate redshift uncertainties to the marginalised value of $M_B$. To ensure that $\chi^2$ and fitted parameter values are reproducible from one scan of the same grid point to the other, the number of repetitions, $N$, must be large. Tests on simulated events showed that $N=1000$ ensures stable enough results (absolute $\chi^2$ variations much below 1 from one scan to the other).

The $\chi^2$ computed at each point of the grid is thus:
\begin{equation}
\chi^2 = \frac{1}{N}\sum
\big[ \vec{\mu}(\vec{z}+\vec{\delta z})  -  \vec{\mu}_{model} (\vec{z}+\vec{\delta z})\big]^\dag C_{\rm cov}^{-1}(\vec{z}+\vec{\delta z}) \big[ \vec{\mu}(\vec{z}+\vec{\delta z})  -  \vec{\mu}_{model}(\vec{z}+\vec{\delta z}) \big]   
 \end{equation}
where the sum extends over redshift variations applied to photometric redshifts, which are denoted $\vec{\delta z}$. Requirements imposed on $\delta z$ values are the same as in the previous method. The last two terms in~\eqref{eq:chi2refit} are not added, since the prior on the  SALT2 $\chi^2$ difference term is not justified here, while the Gaussian prior in $\delta z$ is not required since redshift variations are directly drawn from a Gaussian distribution with $\sigma=0.03(1+z)$.

Contrary to the first method which aims at correcting HD outliers through redshift refitting, there is no natural protection in the sampling method against outliers generated by redshift variations.
For that purpose, in case a redshift variation produces an outlier (i.e. an individual SN $\chi^2$ above 3), up to 10 new variations in redshift are drawn for that event and, if none of them leads to a $\chi^2$ below 3, the null variation is kept.

\subsection{Toy Monte Carlo samples}
\label{sec:toymc}
The two methods were tested with a batch of simulated HDs representative of the sample size and redshift profile of the actual mixed samples, 
except for the HST subsample that contains very few events and was ignored.  All simulated HDs have a common spectroscopic subsample at low redshift that was built with our SN~Ia light curve simulation run in appropriate low redshift intervals\footnote{As the role of the low redshift spectroscopic events is only to provide anchorage of Hubble diagrams by tightly constraining $M_B'$, our low redshift simulations suffice for that purpose, there is no need to simulate the true low redshift and SDSS selection functions.}. We used the parameter values of table~\ref{tab:simul} (first row), except for $\sigma_{int}$  
which was set to values suitable for the low redshift and SDSS components of the JLA sample, as reported in~\cite{Betoule14}. Events from these specific simulations passing our photometric selections (section~\ref{sec:cuts}) were then used to build two sets of events representative of the low redshift and SDSS components. For these events, their true redshifts are considered as their spectroscopic redshifts and used to compute eq.~\eqref{eq:chi2}.

To obtain the high redshift part of the simulated HDs, photometrically selected events of our main SN~Ia simulation were used to generate 50 independent sets of events representative of the JLA-P SNLS subsample.
The reconstructed light curves of these events were fit with SALT2 at their photometric redshifts to define parameters entering in eq.~\eqref{eq:chi2}.
Using instead SALT2 fits at the true redshifts for part of the events, each of the 50 above sets was converted into a JLA-B SNLS subsample. These replacements were performed so as to reproduce the total sample size and the redshift profiles of the populations with photometric and spectroscopic redshifts in the real JLA-B SNLS subsample (see table~\ref{tab:samples}). 
Possible redshift migrations due to the upper cut at 1.05 in redshift were accounted for. 

Extending these replacements to all events defines a third type of SNLS subsample, hereafter referred as to JLA-S,
 where all photometric events have SALT2 parameters evaluated at their true redshifts. 
These diagrams have no counterpart in real data because of our incomplete spectroscopic redshift coverage but describe what could be optimal photometrically selected samples with spectroscopic redshifts for all events.
Finally, the simulated spectroscopic sample at low redshift was combined with each of the simulated SNLS subsamples to generate 50 independent HDs in each of the JLA-P, JLA-B and JLA-S categories. 

\subsection{Fits to simulated HDs}
\label{sec:toymcfits}
We performed flat $\Lambda_{CDM}$ fits to each simulated HD, using grids of 80x80x80 points in $\Omega_M, \alpha, \beta$ with steps of 0.0025, 0.001, 0.01, respectively.  We remind that $M_B$ is analytically marginalised over at each point of the grid and so does not add an extra dimension to the grid.
We tested both methods of redshift uncertainty propagation and ran also fits with no redshift error propagation, hereafter named standard fits, as a comparison. 
In all cases, we checked the impact of $3\sigma$ clipping aimed at eliminating outliers and varied the SN~Ia magnitude bias correction applied to SNLS SNe.
For the latter, we either use the correction to the intrinsic SN~Ia magnitude bias of section~\ref{sec:biasres} for all SNLS events in the diagrams, 
or use the correction to the full SN~Ia magnitude bias of section~\ref{sec:biasres} for events with SALT2 parameters evaluated at their photometric redshift and the correction to the intrinsic magnitude bias for all other SNLS events, if any.  These two types of corrections will be referred to as 'intrinsic bias correction' and 'full bias correction' in the following.
The first option is appropriate for the JLA-S diagrams but was also tried for the two other diagram types, while the second option is appropriate for JLA-P diagrams and approximate for JLA-B ones, a point which will be further discussed in section~\ref{sec:ccdiscussion}. In all cases, the intrinsic bias correction was applied to low redshift events.

Systematic uncertainties other than those related to photometric redshift uncertainties were not included in the fits and the only statistical uncertainties accounted for were those from the light curve parameters,  the SN~Ia intrinsic dispersion $\sigma_{int}$ and the statistical uncertainty on the bias correction (which has a negligible impact). 
While running on the 512,000 grid points with standard fits takes less than a minute, the sampling method takes on average 8 hours on JLA-B HDs, four times more on JLA-P HDs and is twice as long as redshift refitting.

In order to test for biases in the fit results, we resort again to the sample of 25,000 simulated SN~Ia events used in section~\ref{sec:bias}. This sample was submitted to our photometric selections, their generated light curves were fit with SALT2 at their true redshifts and their SALT2 apparent magnitudes were corrected for the intrinsic magnitude bias. A flat $\Lambda_{CDM}$ fit to this sample gives 
$\Omega_M=0.294\pm0.003$,  $M_B=-19.085$, $\alpha = 0.1408^{+0.0007}_{-0.0009}$ and $\beta = 3.229\pm0.009 $. The fit residual dispersion is 0.08.
The above values define what can be expected from a photometrically selected SN~Ia sample 10 times larger than observed, with noiseless light curves and spectroscopic redshifts. As such, it can be considered as a reference to compare the toy Monte Carlo fit results with. The $\sigma_{int}$ value in this fit was again chosen so as to obtain a $\chi^2$ value approximately equal to the total number of SNe in the sample. As this would be too cumbersome to do for all toy Monte Carlo fits, the latter were performed with $\sigma_{int}$ set to the value used in the simulation for each the three surveys entering the diagrams (low $z$, SDSS, SNLS). We checked that this has a marginal effect on the fitted parameters except for $\beta$, a point we discuss again in section~\ref{sec:toymcav}.

\subsection{Parameter posteriors}
Prior to defining average results from the fits to simulated diagrams, which will be the subject of the next section, it is important to check that posterior distributions are close to Gaussian, otherwise average results may be unreliable. This is addressed in the following.

\begin{figure}[tpp]
\centering
\begin{tabular}{c}
\includegraphics[width=1.04\textwidth]{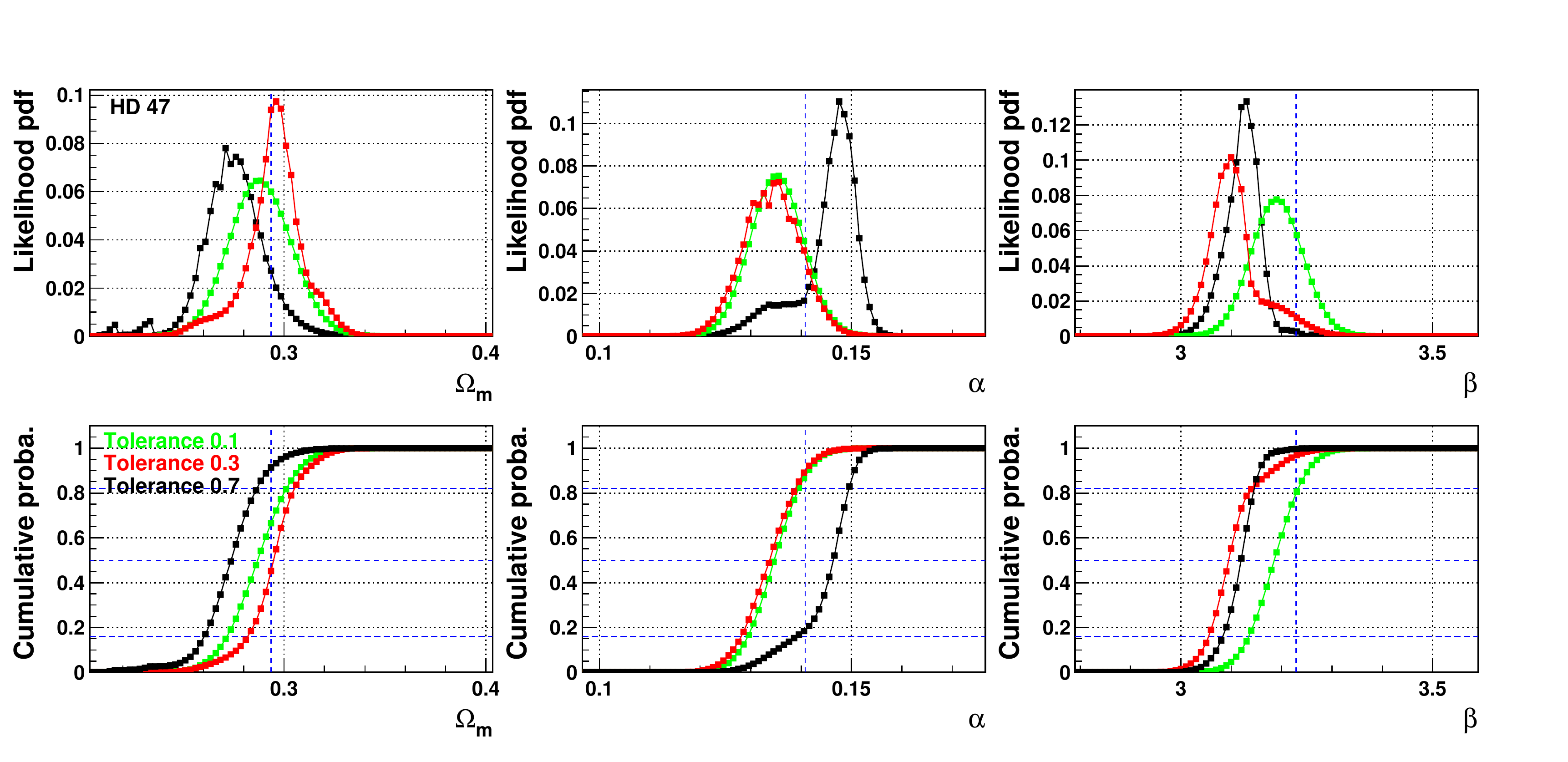} \\
\includegraphics[width=1.04\textwidth]{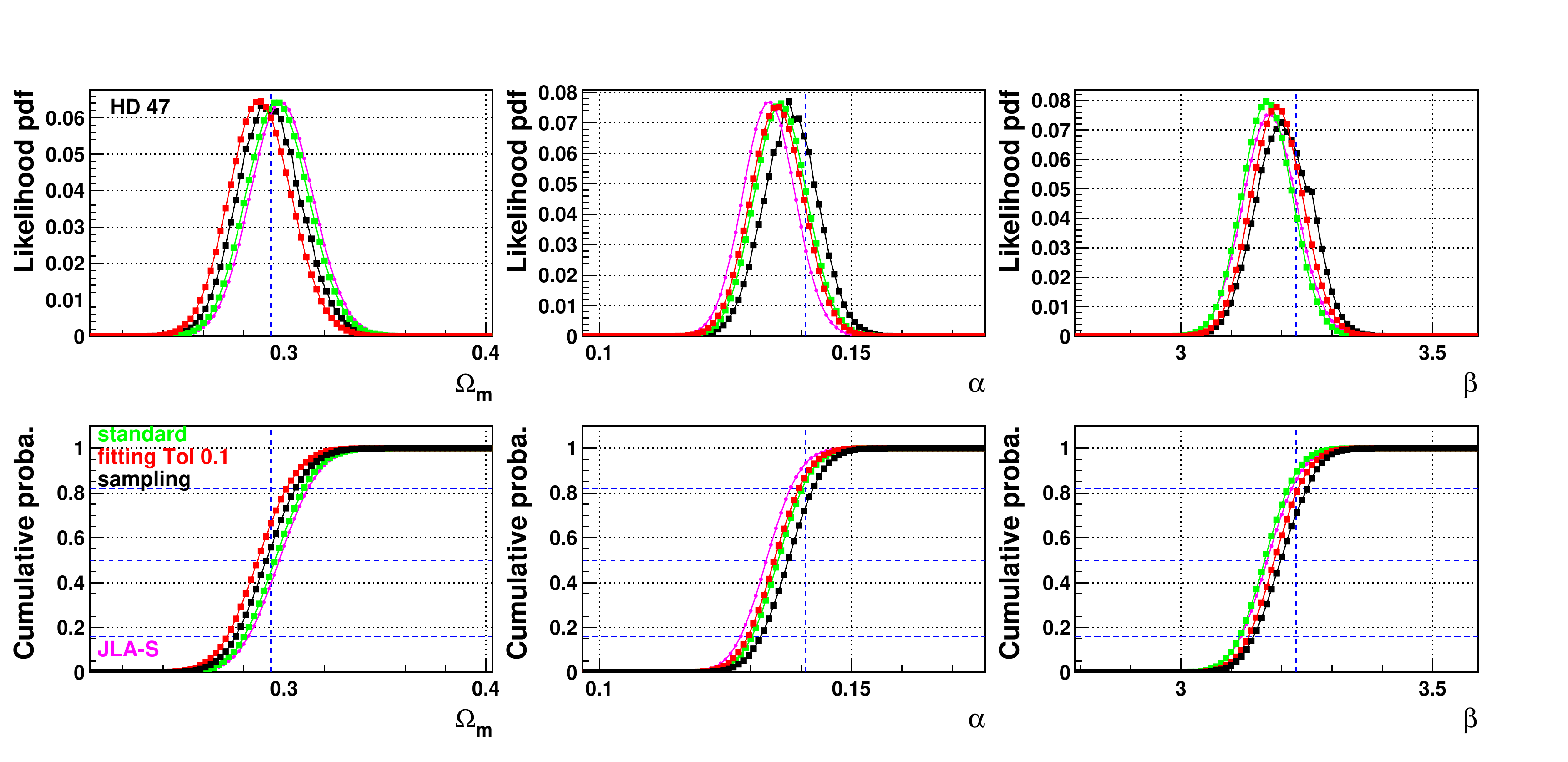} \\
\end{tabular}
\caption{\label{fig:postP} $\Omega_m, \alpha$ and $\beta$ posterior distributions and their cumulative distribution functions obtained for one JLA-P Hubble diagram in various cosmological fits. {\it Top:} fits with redshift refitting for three values of the tolerance 
that defines the extent of the domain to explore when minimising the $\chi^2$ over changes in the photometric redshifts at each point of the cosmological grid. Some tolerance values lead to non Gaussian posteriors, and thus unreliable fit results.
{\it Bottom:} standard fit (with clipping), redshift refitting (at the appropriate tolerance) and sampling fit. Results from a standard fit (with clipping) to the corresponding JLA-S HD are in purple. In all cases, the intrinsic SN~Ia magnitude bias correction is applied. Vertical blue dashed lines indicate the parameter expected value and horizontal ones define the parameter marginalised median value and error interval. }
\end{figure}

Posterior distributions for the cosmology and nuisance parameters were checked for all fits to simulated HDs.
We noticed that important distortions can affect parameter posteriors in the redshift refitting method (see top panel of figure~\ref{fig:postP}), leading to  incorrect marginalised fit results and underestimated error intervals. Such distortions are observed for 40\% (36\%) of the JLA-P (JLA-B) HDs in the case of $\Omega_M$ and are stronger in the JLA-P case.  In the sampling method, distortions can also be observed, but they are always moderate and do not degrade the cumulative probability function (and hence the fit results) significantly, unlike what happens with the refitting method.  We hereafter review possible causes for the observed distortions and explain how we overcame them.

Distortions in the $\Omega_M, \alpha$ and $\beta$ posteriors arise from discontinuities in the $\cal L$ map obtained on the cosmological grid, meaning that varying redshifts of the photometric SNe in each point of the grid, 
can produce strong  variations in $\cal L$ between two close points of the grid. 
One reason may be strong shifts of SALT2 parameters when the redshift variation crosses a threshold in the SALT2 procedure, such as that around redshift 0.68 due to the loss of the $g_M$  band. 
However, HDs with distorted posteriors do not exhibit an abnormally high number of photometric SNe with redshifts changes in the vicinity of that threshold, so this effect is likely to be sub-dominant. Other sub-dominant effects are unreliable polynomial fits of SALT2 parameter shifts, insufficient density of the cosmological grid and the SALT2 $\chi^2$ difference term in eq.~\eqref{eq:chi2refit}. All three effects were tested in the 
sampling method (and the first two in the refitting one). Relaxing cuts against unreliable polynomial fits, degrading the grid density in $\alpha, \beta$ by a factor of 2 or including the SALT2 $\chi^2$ difference term were each found to amplify the distortions, the largest impact being that of density. But none of these effects, even density, can be the main origin of the distortions. 

The most plausible scenario to explain the large discontinuities in the $\cal L$ map from the refitting method is the presence of local minima in the  $\chi^2(\vec{z}+\vec{\delta z})$ map in part of the cosmology grid points. From one grid point to the other, the minimiser can then end up on different minima, generating quite different $\chi^2$ values,
even for close points on the grid. To check this hypothesis, the refitting method was run on a few HDs for different values of the tolerance on the line minimisation accuracy, keeping all other conditions unchanged.
With looser tolerances, the $\vec{\delta z}$ parameter space enlarges and the minimisation is observed to lead to quite different best fits on the grid, a result favouring the hypothesis of the existence of multiple local minima in the $\chi^2(\vec{z}+\vec{\delta z})$ map. This hypothesis also explains why discontinuities are larger and more frequent in JLA-P HDs since those have a large number of photometric redshifts to refit, with many of them below $z\sim0.7$, a region where distance luminosities are prompt to strong variations with redshift.

In order to obtain reliable results for a maximum of HDs, the refitting method was run for each diagram with three tolerance values, 0.1, 0.3 and 0.7. For each diagram, we retain the results obtained for the tolerance that maximises the number of parameters with closest to Gaussian posteriors, and  in the case of posteriors of equivalent quality, we keep the results from the tolerance giving the highest best fit $\cal L$ value (see bottom plots of figure~\ref{fig:postP}).
HDs giving distorted parameter posteriors for the three tolerance values are discarded from the averaging procedure for the corresponding parameter. This removes 6\%, 8\% and 22\% of the JLA-P HDs for  $\Omega_M, \alpha$ and $\beta$, respectively, while only 2\% of the JLA-B HDs are removed for $\Omega_M$ and none for $\alpha$ and $\beta$. As discontinuities in the $\cal L$ map for the  sampling method are always minor and with negligible impact on the cumulative distributions, all HDs are considered for that method.

\subsection{Average results}
\label{sec:toymcav}
Each fit to a simulated HD retained by the above procedure gives marginalised median values of  $\Omega_M, \alpha, \beta, M_B$ and their errors (except for $M_B$), all derived from the parameter posterior cumulative distribution function (see section~\ref{sec:fitter} and figure~\ref{fig:postP}). Depending on the parameter, distributions of their marginalised
values and error intervals over all retained fits can be symmetric or skewed. We thus rely on the median values of these distributions as a reliable indicator of averaged results and characterise the dispersion of the results by the 15.9-84.1 percentile range of the distributions, which provides a good indicator of the error intervals 
expected from a single HD fit. Although median and percentile ranges are robust against outliers, values outside of the 15.9-84.1 percentile range expanded on either side by 1.5 times that range (equivalent to a 3.5$\sigma$ cut for a Gaussian distribution) were rejected from the average, and the 15.9-84.1 percentile range recomputed. 

\begin{table}[tbp]
\centering
{\scriptsize
\begin{tabular}{|l|c|c|c|c|c|c|}
\hline
 Fit & $\Delta\Omega_M$ & $\Delta M_B$ & $\Delta \alpha$ & $\Delta \beta$ & $n_{out}$ & rms\\ \hline
 
\multicolumn{7}{|c|} {{\bfseries JLA-S samples}}   \\ \hline
standard        & $-0.001\pm0.004$ & $-0.009\pm0.001$ & $-0.005\pm0.001$ & $-0.054\pm0.010$ & $2.00\pm0.23$ & 0.181\\\hline
standard, clip & $-0.001\pm0.004$ & $-0.008\pm0.001$ & $-0.005\pm0.001$ & $-0.053\pm0.010$ & $0.00\pm0.03$ & 0.178 \\\hline

 \multicolumn{7}{|c|} {{\bfseries JLA-B samples}}   \\ \hline
standard        & $-0.007\pm0.004$ & $-0.010\pm0.001$ & $-0.004\pm0.001$ & $-0.047\pm0.010$ & $3.00\pm0.24$  & 0.183 \\\hline 
standard, clip & $-0.006\pm0.004$ & $-0.011\pm0.001$ & $-0.005\pm0.001$ & $-0.052\pm0.010$ & $0.00\pm0.03$  & 0.178 \\\hline 
standard, full        & $-0.003\pm0.004$ & $-0.009\pm0.001$ & $-0.005\pm0.001$ & $-0.045\pm0.010$ & $3.00\pm0.27$  & 0.184 \\\hline 
standard, full, clip & $-0.001\pm0.004$ & $-0.009\pm0.001$ & $-0.005\pm0.001$ & $-0.052\pm0.010$ & $0.00\pm0.03$  & 0.178 \\\hline 

sampling         & $-0.003\pm0.004$ & $-0.009\pm0.001$ & $-0.005\pm0.001$ & $-0.053\pm0.010$  & $2.00\pm0.26$  & 0.187 \\\hline 
sampling, clip & $-0.002\pm0.004$ & $-0.009\pm0.001$ & $-0.005\pm0.001$ & $-0.059\pm0.010$  & $0.00\pm0.02$  & 0.182 \\\hline 
sampling, full         & $0.000\pm0.004$ & $-0.008\pm0.001$ & $-0.005\pm0.001$ & $-0.052\pm0.010$  & $2.00\pm0.26$  & 0.186 \\\hline 
sampling, full, clip & $0.001\pm0.004$ & $-0.008\pm0.001$ & $-0.005\pm0.001$ & $-0.056\pm0.010$  & $0.00\pm0.02$  & 0.182 \\\hline 

refitting & $-0.006\pm0.004$ & $-0.010\pm0.001$ & $-0.005\pm0.001$ & $-0.053\pm0.010$ & $2.00\pm0.24$  & 0.179 \\\hline

 \multicolumn{7}{|c|} {{\bfseries JLA-P samples }}   \\ \hline
standard        & $-0.008\pm0.004$ & $-0.009\pm0.001$ & $-0.001\pm0.001$ & $0.057\pm0.013$  & $9.00\pm0.46$  & 0.203 \\\hline 
standard, clip & $-0.008\pm0.004$ & $-0.008\pm0.001$ & $-0.003\pm0.001$ & $0.014\pm0.013$  & $0.00\pm0.07$  & 0.192 \\\hline 
standard, full        & $-0.002\pm0.004$ & $-0.009\pm0.001$ & $-0.001\pm0.001$ & $0.059\pm0.012$  & $8.00\pm0.44$  & 0.201 \\\hline 
standard, full, clip & $0.000\pm0.004$ & $-0.008\pm0.001$ & $-0.002\pm0.001$ & $0.019\pm0.012$  & $0.00\pm0.05$ & 0.191 \\\hline 

sampling        & $-0.002\pm0.004$ & $-0.006\pm0.001$ & $-0.001\pm0.001$ & $0.031\pm0.012$ & $5.00\pm0.42$  & 0.211 \\\hline 
sampling, clip & $-0.001\pm0.004$ & $-0.006\pm0.001$ & $-0.002\pm0.001$ & $-0.009\pm0.011$  & $0.00\pm0.02$  & 0.202  \\\hline 
sampling, full        & $0.004\pm0.004$ & $-0.005\pm0.001$ & $-0.001\pm0.001$ & $0.025\pm0.011$ & $5.50\pm0.39$  & 0.209 \\\hline 
sampling, full, clip & $0.005\pm0.004$ & $-0.004\pm0.001$ & $-0.002\pm0.001$ & $-0.010\pm0.011$  & $0.00\pm0.00$  & 0.201 \\\hline 

refitting & $-0.008\pm0.005$ & $-0.011\pm0.002$ & $-0.002\pm0.001$ & $-0.007\pm0.016$  & $4.00\pm0.49$  & 0.186\\\hline

\end{tabular}
}
\caption{\label{tab:toy} Cosmological fit results averaged over independent simulated photometric SN~Ia Hubble diagrams with no contamination, in each diagram category. 
Results are the median biases in $\Omega_M, M_B, \alpha$ and $\beta$
w.r.t. their expected values, the median number of  $3\sigma$ HD outliers, $n_{out}$, and the median HD dispersion, rms.
Fits with $3\sigma$ clipping are marked clip, those using the full SN~Ia magnitude bias correction instead of the intrinsic one are labelled full (see text).
For the refitting method, only diagrams leading to reasonably Gaussian parameter posteriors enter the average. 
The statistical uncertainty on the median is computed from that on the mean value assuming Gaussian distributions, and adding the statistical uncertainty on the reference value in quadrature for the median biases.}
\end{table}

\begin{table}[tbp]
\centering
{\footnotesize
\begin{tabular}{|l|c|c|c|c||c|c|c|c|}
\hline
Fit & $\Omega_M$  & $M_B$ & $\alpha$ & $\beta$  & $\Omega_M$  & $M_B$ & $\alpha$ & $\beta$  \\ \hline
 
& \multicolumn{8}{c|} {{\bfseries JLA-S samples}}   \\ \hline
standard        & $0.026$  & 0.008 & $0.005$ & $0.053$  & $0.026$  & 0.008 & $0.005$ & $0.053$ \\ \hline
standard, clip & $0.027$  &0.008 & $0.006$ & $0.055$  & $0.027$  &0.008 & $0.006$ & $0.055$ \\ \hline

& \multicolumn{4}{c||} {{\bfseries JLA-B samples}}  & \multicolumn{4}{|c|} {{\bfseries JLA-P samples }}   \\ \hline
standard        & $0.028$  & 0.007 & $0.007$ & $0.051$ &  $0.031$  & 0.012 & $0.010$ & $0.092$   \\ \hline 
standard, clip & $0.026$  & 0.007 & $0.006$ & $0.055$  & $0.028$  & 0.012 & $0.009$ & $0.097$   \\ \hline 
standard, full.       & $0.025$  & 0.006 & $0.007$ & $0.050$  & $0.033$  & 0.011 & $0.009$ & $0.085$  \\ \hline 
standard, full, clip & $0.028$  & 0.006 & $0.006$ & $0.060$ & $0.031$  & 0.011 & $0.008$ & $0.091$  \\ \hline 

sampling        & $0.027$  & 0.007 & $0.007$ & $0.055$ & $0.029$  & 0.011 & $0.008$ & $0.090$  \\ \hline 
sampling, clip & $0.026$  & 0.007 & $0.006$ & $0.051$ & $0.026$  & 0.011 & $0.007$ & $0.076$  \\ \hline 
sampling, full        & $0.027$  & 0.007 & $0.007$ & $0.054$ & $0.032$  & 0.011 & $0.008$ & $0.077$  \\ \hline 
sampling, full, clip & $0.026$  & 0.007 & $0.006$ & $0.060$  & $0.031$  & 0.011 & $0.007$ & $0.076$  \\ \hline 

refitting &  $0.028$  & 0.008 & $0.007$ & $0.053$  & $0.045$  & 0.021 & $0.011$ & $0.116$  \\ \hline
\end{tabular}
}
\caption{\label{tab:toy3} Same as table~\ref{tab:toy} 
for the 16-84 percentile ranges in the $\Omega_M, M_B, \alpha$ and $\beta$ marginalised values. }
\end{table}

Average results, as defined above, obtained on the simulated JLA-B and JLA-P HDs with the refitting and sampling methods are reported in tables~\ref{tab:toy} and~\ref{tab:toy3} and compared to standard fit results on the same samples and on the JLA-S ones. For sake of simplicity, the clipping in sampling fits is that given by the standard fits with the same magnitude bias correction. In the following, we first describe the results of the various fitting methods on the different types of diagrams and then summarise our findings at the end of the section.

For the JLA-S samples, we note important biases in $M_B$, $\alpha$ and $\beta$ with respect to the reference sample. For $\beta$, part of the bias comes from the $\sigma_{int}$ settings in the simulated diagram fits (see section~\ref{sec:toymcfits}) which do not lead to reduced $\chi^2$ values around 1 for the JLA-S diagrams as for the reference sample (the same is true for JLA-B diagrams, but not for JLA-P ones). This explains $40\%$ of the bias in beta for the JLA-S samples, the remaining part being due to instrumental noise included in the simulated diagrams but not in the reference sample. The latter effect completely explains the bias in $\alpha$ for the JLA-S diagrams, and part of the bias in $M_B$, the remaining effect being the different anchorage of the simulated diagrams at low redshift w.r.t. that of the reference sample. 

In table~\ref{tab:toy}, we first observe that without clipping the number of $3\sigma$ HD outliers and the HD dispersion are high for JLA-P HDs, as expected for HDs with only photometric redshifts at $z$$>$0.2, while they are similar for JLA-B and JLA-S samples. The refitting and sampling methods reduce the number of outliers but not as efficiently as clipping, which is also the most efficient way to reduce the HD dispersion.
Comparing the refitting and sampling methods with intrinsic bias correction and no clipping - a priori the most appropriate conditions for these methods - the refitting method results in a bias in $\Omega_M$ similar to that observed with standard fits, for both JLA-B and JLA-P samples, while the sampling method corrects that bias efficiently. As the refitting method fails at producing unbiased cosmological results on top of having non-Gaussian posteriors, we did not study this method any further.

Adding clipping and using the full bias correction reduces the $\Omega_M$ bias to a negligible level in all cases, except in sampling fits to the JLA-P HDs, for which clipping has essentially no effect and the full bias correction slightly increases the $\Omega_M$ bias.
The full bias correction accounts on average for redshift migration effects due to redshift resolution, while the sampling method reshuffles photometric redshifts according to resolution and propagates these individual changes to all photometric SN properties, including the magnitude bias correction which is taken at the reshuffled redshifts. This assumes that the full bias correction computed initially remains appropriate after redshift reshuffling, which is probably not the case for JLA-P HDs due to their large number of photometric redshifts. However, the  $\Omega_M$ bias for sampling fits with the full bias correction remains at an acceptable level.

As of nuisance parameters, the bias in $M_B$ is similar to that of the JLA-S samples in most cases, and lower for JLA-P samples when clipping and the full bias correction are used. Biases in $\alpha, \beta$ of JLA-B samples are similar to those of JLA-S HDs and do not vary significantly with the fitting method or conditions. We relate this stability to the small number of outliers (affected by clipping) and the small number of SNe with photometric redshifts (subject to the full bias correction and redshift reshuffling in the sampling method) in the JLA-B samples.  
JLA-P diagrams have biases in $\alpha$ and $\beta$ different from those of the two other diagram types, which we relate to their higher dispersion and higher number of outliers. To reduce these requires higher values of $\alpha$ and $\beta$ than those for JLA-S and JLA-B HDs. This translates into lower biases in $\alpha$, biases in $\beta$ becoming positive when no clipping is applied (with lower biases for sampling fits since redshift reshuffling helps reducing the dispersion) and reduced biases in $\beta$, either positive or negative depending on the fit method, when clipping is applied.

Finally, table~\ref{tab:toy3} shows that percentile ranges in the four parameters are similar for JLA-B and JLA-S HDs and do not evolve significantly with fitting methods or conditions. 
On the other hand, percentile ranges for JLA-P HDs are higher than those of JLA-S samples, especially for $M_B$ and $\beta$ where an increase of 70-80$\%$ is observed. 
Taking half of the percentile range as an estimate of the statistical uncertainty expected for one single fit, the biases in $\Omega_M$ commented above can be better quantified. For standard fits, the bias amounts to $0.5\,\sigma$ for both JLA-B and JLA-P diagrams. Sampling reduces this bias by a factor better than 2. Adding clipping and the full magnitude bias correction, the bias in $\Omega_M$ drops below $0.1\,\sigma$ for all fits and diagrams, while it remains at the level of $0.3\,\sigma$ for sampling fits to JLA-P diagrams. For these, the bias drops below $0.1\,\sigma$  if clipping is used together with the intrinsic bias correction.

In summary, fitting pure SN~Ia Hubble diagrams containing photometric redshifts leads to biased cosmological results if redshift uncertainties are not accounted for. This is true for both JLA-B diagrams, where photometric redshifts are located mostly at $z$$>$0.5 and make 10$\%$ of the total sample, and JLA-P diagrams, where photometric redshifts  start at $z$$\sim$0.2 and make about half of the total sample.
This effect cannot be corrected by refitting photometric redshifts with cosmology, but it 
can be corrected by sampling the redshift resolution function for SNe with photometric redshifts, even if the magnitude bias correction is computed as if all SNe had spectroscopic redshifts. Uncertainties in the cosmological parameters remain similar to those of spectroscopic HDs for JLA-B diagrams and are degraded for JLA-P ones, despite some improvement brought by clipping.
Standard fits offer similar performance only if the magnitude bias correction also includes the effect of redshift resolution. 

In the following, we discuss possible reasons for the failure of the redshift refitting method and then study how the previous results evolve when HDs are contaminated by core-collapse SNe, restricting to standard and sampling fits only.

\subsection{Discussion}

\begin{table}[tbp]
\centering
{\footnotesize
\begin{tabular}{|c|c|c|c|c|}
\hline
Case & $\sigma_{c} (\%)$  & $n_{out}^{10}$ & $n_{out}^{15}$ & $n_{out}^{20}$   \\ \hline
 \multicolumn{5}{|c|} {{\bfseries JLA-B samples $N_{z_{pho}}=99$}}   \\ \hline
initial & $ 3.3\pm0.1$ & $2.0\pm0.2$ & $1.0\pm0.2$ & $0.0\pm0.0$  \\ \hline
final  & $3.0\pm0.1$ & $2.0\pm0.2$ & $1.0\pm0.2$ & $0.0\pm0.0$   \\ \hline \hline

\multicolumn{5}{|c|} {{\bfseries JLA-P samples $N_{z_{pho}}=437$}}  \\ \hline
initial &  $2.83\pm0.02$ & $9.0\pm0.5$ & $3.0\pm0.2$ & $1.0\pm0.2$  \\ \hline
final  &  $2.58\pm0.04$ & $8.0\pm0.5$ & $3.0\pm0.2$ & $1.0\pm0.2$  \\ \hline

\end{tabular}
}

\caption{\label{tab:toy4}  Results of photometric redshift refitting averaged over nearly 50 independent simulated JLA-B and JLA-P Hubble diagrams with no contamination. 
The number of photometric redshifts to refit is indicated in lines 2 and 5.
We show the median redshift resolution in $\Delta z/(1+z)$, labelled $\sigma_c$, and the median number of redshift outliers with $|\Delta z/(1+z)|$ above 0.10, 0.15 and 0.20, in the initial samples and after redshift refitting (using the intrinsic SN~Ia magnitude bias correction). The redshift resolution is only slightly improved by redshift refitting with cosmology.}
\end{table}

\begin{figure}[tbp]
\centering
\includegraphics[width=.8\textwidth]{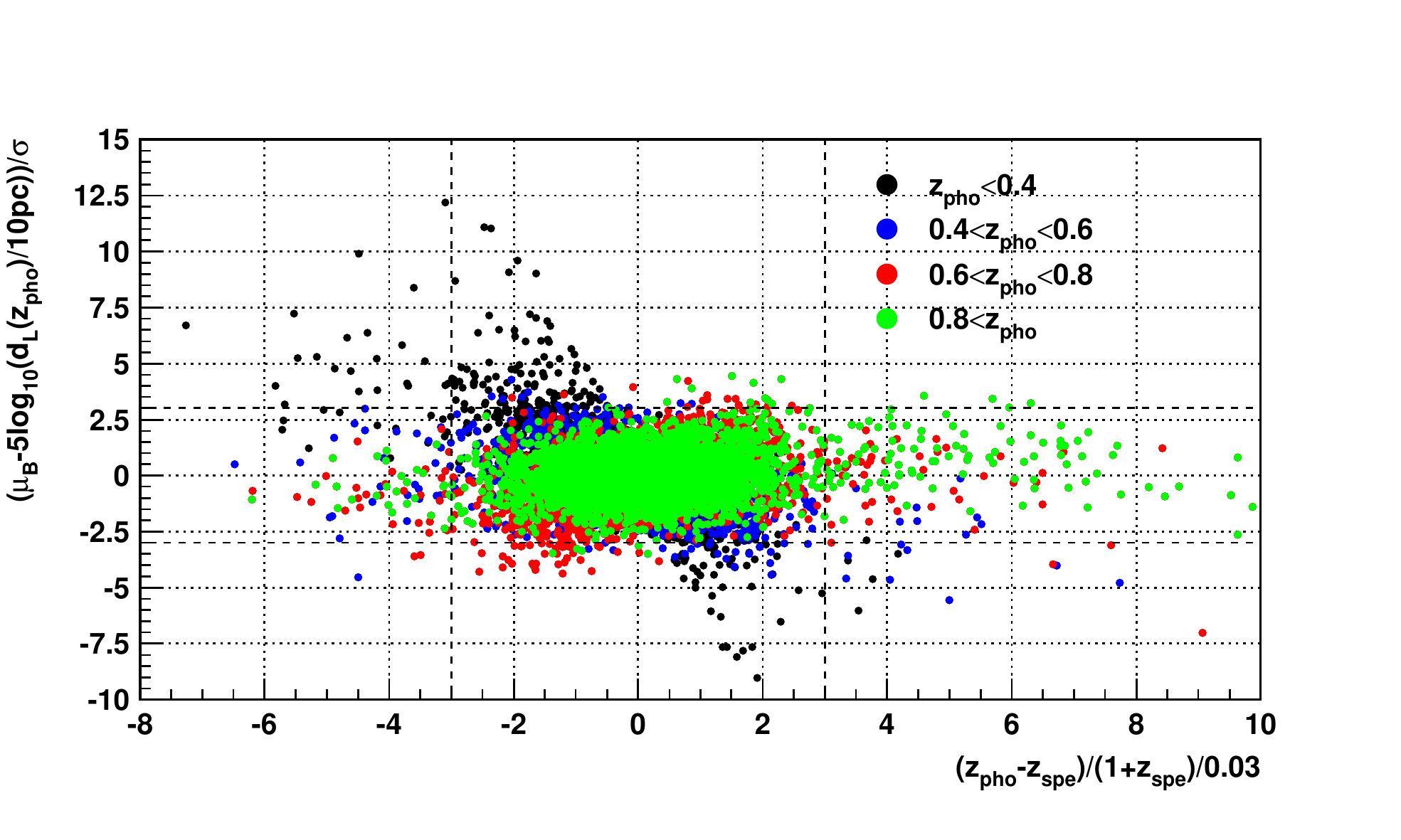}
\caption{\label{fig:outliers} Normalised Hubble diagram residual  (using the intrinsic SN~Ia magnitude bias correction) as a function of host photometric redshift bias normalised by central redshift resolution. We use a sample of around 12,000 simulated SN Ia light curves passing the SN Ia photometric selections and adjusted by SALT2 at their host photometric redshift. Different colours refer to different host photometric redshift ranges. 
Vertical and horizontal dashed lines define $3\sigma$ outliers. Redshift outliers are not necessarily Hubble diagram outliers, especially at high redshift, while most Hubble diagram outliers at low redshift are due to small redshift uncertainties generating large changes in luminosity distance.}
\end{figure}

\begin{table}[hbtp]
\centering
{\footnotesize
\begin{tabular}{|c|c|c||c|c|c|}
\hline
$r_{better}$ ($\%$) & old $\sigma_c$  $(\%)$ & new $\sigma_c$  ($\%$)  & $r_{worse}$ ($\%$) & old $\sigma_c$ ($\%$) & new $\sigma_c$ ($\%$) \\ \hline
 \multicolumn{6}{|c|} {{\bfseries JLA-B samples }}   \\ \hline
 $57.6\pm0.9$  & $3.8\pm0.1$  & $3.3\pm0.1$ &$42.4\pm0.9$  & $3.2\pm0.2$ & $3.5\pm0.2$ \\ \hline

\multicolumn{6}{|c|} {{\bfseries JLA-P samples  }}   \\ \hline
 $63.4\pm0.3$ & $3.2\pm0.1$   & $2.6\pm0.1$ & $36.6\pm0.3$   & $2.2\pm0.1$ &$2.5\pm0.1$ \\ \hline
\end{tabular}
}
\caption{\label{tab:toy5}  Results of photometric redshift refitting (using the intrinsic SN~Ia magnitude bias correction) averaged over nearly 50 independent simulated Hubble diagrams in each of the JLA-B and JLA-P categories. We show the median rate of better and worse redshifts, and for each of those categories, the evolution of the median central resolution with refitting. Redshift refitting produces slightly more better redshifts. }
\end{table}

Refitting photometric redshifts produces biased cosmological results. Moreover, as shown in table~\ref{tab:toy4}, 
 the redshift resolution obtained after redshift refitting is hardly changed (a few per mill improvement) and there is no visible reduction of the number of redshift outliers.

The basic reason is that fitting redshifts together with cosmology acts primarily on HD outliers. But our selection retains light curves which, once adjusted at their assigned photometric redshift, are compatible with being  SN Ia at this redshift. So, incorrect or even outlier photometric redshifts do not necessarily lead to HD outliers. This is illustrated in figure~\ref{fig:outliers} which shows that $3\sigma$ HD outliers represent only a small fraction of SNe in the photometric sample (around $3\%$) and only  part of them (14\%) are due to $3\sigma$ redshift outliers.

Also highlighted in this figure is a second, perhaps more fundamental reason for the modest change in redshift resolution. The strongest HD outliers come mostly from low redshift SNe, at $z$$<$0.4, a region where distance luminosities vary rapidly with redshift, which explains why such strong outliers can be generated even with a redshift uncertainty around 1\%. A small variation of their redshift is likely to move these HD outliers back to the Hubble flow with no significant change of the central redshift resolution. On the other hand, at higher redshift, 
strong redshift outliers do not lead to very strong HD outliers since they are sitting in a region where distance luminosities vary slowly with redshift. This, together with the usual sources of light curve parameter uncertainties (sampling, photometric accuracy, SN Ia intrinsic dispersion) gives a certain flexibility for a high redshift SN to be made compatible with the same cosmology for different redshift values, including values far from the true one. 

Altogether, above $z_{pho}\sim0.8$, redshift refitting likely produces redshift changes in either direction, 
 so that globally the central resolution is minimally affected. This is confirmed by table~\ref{tab:toy5} which breaks down the results of redshift refitting according to whether a better or worse redshift is obtained. The performance is only slightly better for redshifts which get improved, resulting in a very modest improvement of the global redshift resolution.

\section{Core-collapse contamination}
\label{sec:conta}
The previous sections dealt with purely SN~Ia mixed Hubble diagrams, while our photometric sample is contaminated by core-collapse (CC) events, as reported in~\cite{Bazin11}. In this section, we use the CC  simulation developed in~\cite{Bazin11} together with our main SN~Ia simulation to estimate more precisely this contamination and its redshift profile. Then we review the changes induced by this contamination on the SN magnitude bias estimate of section~\ref{sec:bias}, and on the  results of section~\ref{sec:toymc}, restricting to standard fits and fits that include sampling of the redshift resolution.

\subsection{CC simulation and magnitude bias correction}
\label{sec:biascc}
The effect of contamination by core-collapse SNe was studied with the CC light curve simulation described in~\cite{Bazin11}. 
The analytical light curve model used in this simulation, which is also that used in the SN Ia photometric selection of section~\ref{sec:cuts}, is general enough to describe the light curve shape of all types of supernovae.
Values of the model parameters and their evolution with redshift were set up separately for fast and slow declining CC SNe, respectively dubbed as SNIbcII and SNIIp events in the following. In the present work, this simulation was updated  with the value of $\Omega_M$ in table~\ref{tab:simul} and the model of photometric redshifts of section~\ref{sec:zphomodel}. 
A total of 100,000 light curves was generated in each of the SNIbcII and SNIIp classes (5 times more than in~\cite{Bazin11}), of which 486 and 8 remained after the SN Ia selection of section~\ref{sec:cuts}, respectively. The contamination of our SN Ia photometric sample is thus essentially due to fast-declining or non-plateau events.

\begin{figure}[tbp]
\centering
\begin{tabular}{cc}
\hspace{-0.46cm}\includegraphics[width=.55\textwidth]{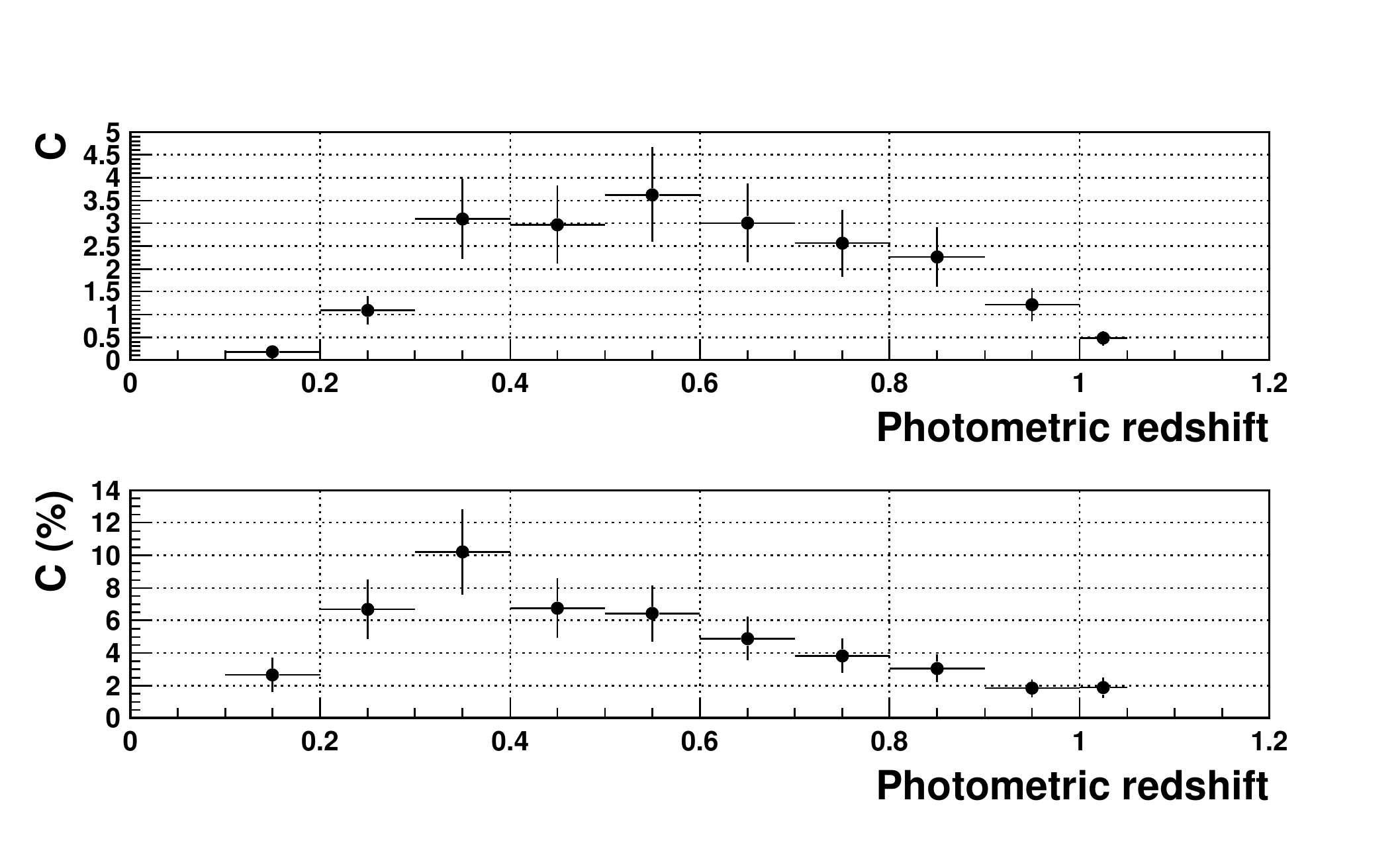}  &
\hspace{-1.1cm}
\includegraphics[width=.57\textwidth]{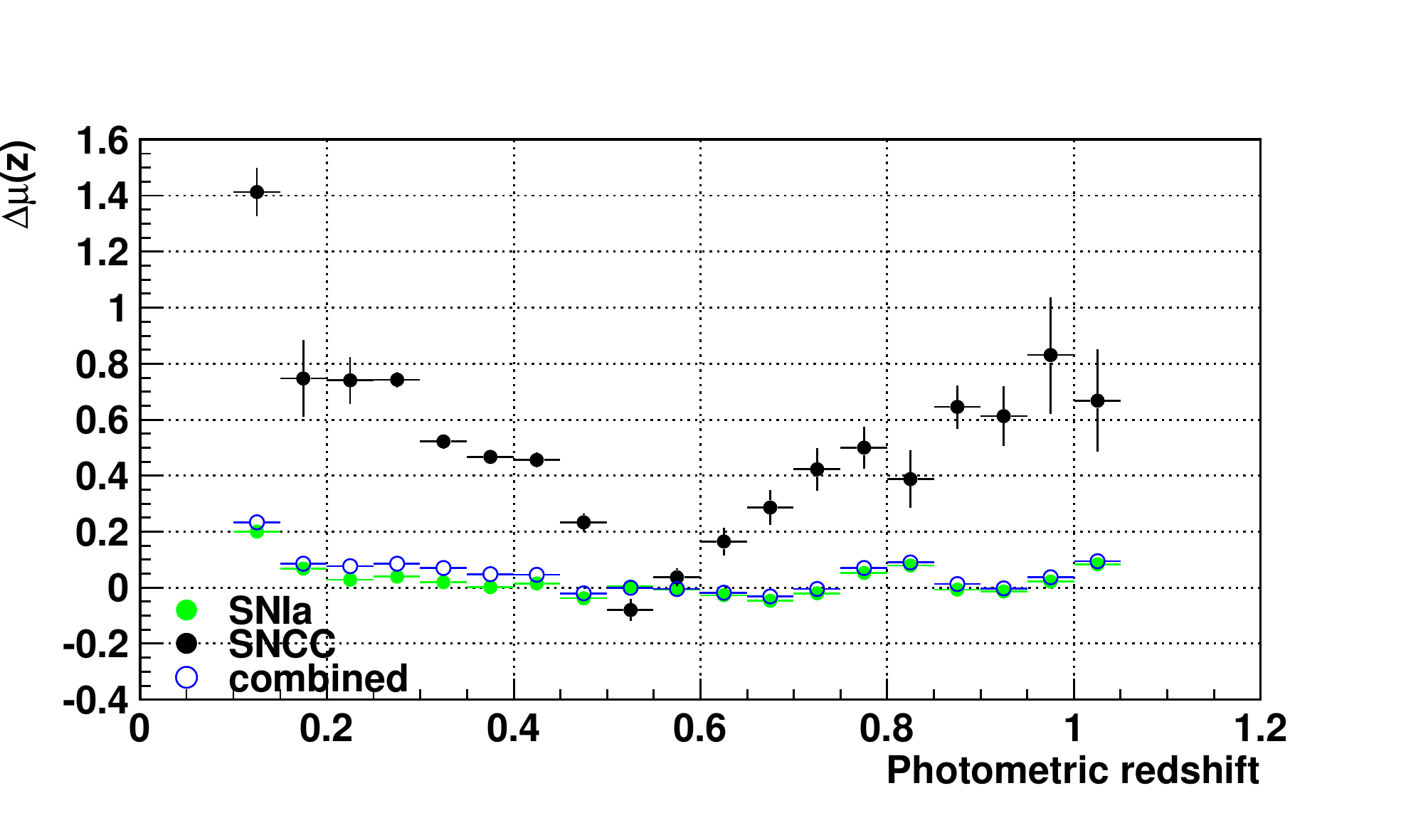}  \\
\end{tabular}

\caption{\label{fig:contacc} Estimated contamination of the SNLS 3-year photometric sample by core-collapse SNe as a function of the SN host photometric redshift (up to 1.05). {\it Top-left:} contamination as absolute number of events.  {\it Bottom-left:} contamination as a fraction of the 
data sample. {\it Right:} full magnitude bias for type Ia SNe (from section~\ref{sec:biasres}), for core-collapse SNe and their combination in the JLA-P sample, as a function of the SN photometric redshift. Errors combine statistical uncertainties due to the simulated sample sizes, as well as statistical and systematic uncertainties on the CC to Ia volumetric rate ratio used to normalise SN Ia and CC simulations.}
\end{figure}

The contamination of the SNLS 3-year photometric sample  by CC SNe was estimated from these simulated samples following the method described in~\cite{Bazin11}. CC and Ia simulations were normalised to each other using the CC to Ia volumetric rate ratio published in~\cite{Bazin09} 
and the relative fraction between fast and slow declining events reported in~\cite{Bazin11}. As a result, the total contamination of the SNLS photometric sample of 485 events is estimated to be $22.1\pm6.1$ events.
The quoted uncertainty comes from the statistical error due to the limited size of the simulated SN Ia and CC samples and from the statistical and systematic errors on the measured volumetric rate ratio of~\cite{Bazin09}. 
Figure~\ref{fig:contacc} presents the CC contamination of the SNLS photometric sample as a function the SN host photometric redshift (top-left panel). The right-hand plot shows the magnitude biases 
for CC and SN Ia events separately and once they are combined according to the contamination rate in the bottom-left panel. The SN Ia bias here is the full magnitude bias defined in section~\ref{sec:biasres} (i.e. SN Ia contribution only, using reconstructed light curves at the host photometric redshift). Altogether, the CC contamination induces a slight increase of the bias of the total sample, mostly at low redshift, $z_{pho}$$<$0.5. In the following, the correction to the above combined SNIa-CC magnitude bias will be referred to as 'full combined bias correction' and that deduced from the pure CC magnitude bias as 'full CC bias correction'.

\subsection{Fit results for contaminated toy Monte Carlo samples}
\label{sec:fitcc}
The CC simulations were used to include contamination in the simulated JLA-P and JLA-B samples of section~\ref{sec:toymc}. In each initial JLA-P HD, we removed a number of SNe Ia in each photometric redshift bin according to the left-hand top plot of figure~\ref{fig:contacc} and replaced them by an equal number of photometrically selected CC events. Each of these contaminated JLA-P samples was then converted into a contaminated JLA-B sample as explained in section~\ref{sec:toymc}. Both sets of contaminated HDs were then fit with the standard and sampling fitting methods. For the latter, the simulated CC light curves were fit as described in section~\ref{sec:shiftparam} to obtain polynomial models of the evolution of SALT2 parameter shifts with redshift. The SALT2 parameter shifts for CC SNe vary similarly as those of SNe Ia and are as correctly described by the same polynomial models.

\begin{table}[tbp]
\centering
{\scriptsize
\begin{tabular}{|l|c|c|c|c|c|c|}
\hline
 Fit & $\Delta\Omega_M$ & $\Delta M_B$ & $\Delta \alpha$ & $\Delta \beta$ & $n_{out}$ & rms \\ \hline
 
\multicolumn{7}{|c|} {{\bfseries JLA-S samples (without  CC contamination))}}   \\ \hline
standard        & $-0.001\pm0.004$ & $-0.009\pm0.001$ & $-0.005\pm0.001$ & $-0.054\pm0.010$ & $2.00\pm0.23$  & 0.181 \\\hline
standard, clip & $-0.001\pm0.004$ & $-0.008\pm0.001$ & $-0.005\pm0.001$ & $-0.053\pm0.010$ & $0.00\pm0.03$  & 0.178 \\\hline

 \multicolumn{7}{|c|} {{\bfseries JLA-B samples }}   \\ \hline
standard              & $-0.024\pm0.004$ & $-0.011\pm0.001$ & $-0.001\pm0.001$ & $-0.131\pm0.016$ & $8.50\pm0.42$  & 0.206 \\\hline  
standard, clip       & $-0.015\pm0.004$ & $-0.013\pm0.001$ & $-0.004\pm0.001$ & $-0.012\pm0.011$ & $0.00\pm0.11$  & 0.186 \\\hline 
standard, full        & $-0.016\pm0.004$ & $-0.009\pm0.001$ & $-0.002\pm0.001$ & $0.113\pm0.016$ & $8.00\pm0.40$  & 0.206 \\\hline 
standard, full, clip & $-0.008\pm0.004$ & $-0.010\pm0.001$ & $-0.004\pm0.001$ & $-0.004\pm0.010$ & $0.00\pm0.13$  & 0.187 \\\hline 

sampling                & $-0.020\pm0.004$ & $-0.010\pm0.001$ & $-0.002\pm0.001$ & $0.099\pm0.015$  & $8.00\pm0.38$  & 0.208 \\\hline 
sampling, clip        & $-0.011\pm0.004$ & $-0.011\pm0.001$  & $-0.004\pm0.001$ & $-0.029\pm0.011$  & $0.00\pm0.08$ &  0.189 \\\hline 
sampling, full         & $-0.012\pm0.004$ & $-0.008\pm0.001$ & $-0.002\pm0.001$ & $0.088\pm0.015$  & $7.50\pm0.40$  & 0.208 \\\hline 
sampling, full, clip  & $-0.005\pm0.004$ & $-0.009\pm0.001$ & $-0.004\pm0.001$ & $-0.017\pm0.011$  & $0.00\pm0.09$ & 0.190 \\\hline

 \multicolumn{7}{|c|} {{\bfseries JLA-P samples }}   \\ \hline
standard                & $-0.028\pm0.004$ & $-0.010\pm0.001$ & $-0.002\pm0.001$ & $0.206\pm0.016$  & $13.00\pm0.53$  & 0.224 \\\hline 
standard, clip        & $-0.018\pm0.004$ & $-0.011\pm0.001$ & $-0.002\pm0.001$ & $0.053\pm0.013$  & $1.00\pm0.18$  & 0.199 \\\hline 
standard, full         & $-0.001\pm0.004$ & $-0.010\pm0.001$ & $0.001\pm0.001$ & $0.184\pm0.015$  & $13.00\pm0.54$ & 0.220 \\\hline 
standard, full, clip & $0.006\pm0.004$ & $-0.008\pm0.001$ & $-0.002\pm0.001$ & $0.049\pm0.013$  & $1.00\pm0.18$  & 0.198 \\\hline

sampling               & $-0.022\pm0.004$ & $-0.008\pm0.001$ & $0.002\pm0.001$ & $0.176\pm0.015$  & $10.00\pm0.44$ & 0.230 \\\hline 
sampling, clip        & $-0.012\pm0.004$ & $-0.008\pm0.001$ & $0.001\pm0.001$ & $0.035\pm0.012$  & $0.00\pm0.08$  & 0.208 \\\hline 
sampling, full           & $0.004\pm0.004$ & $-0.006\pm0.001$ & $0.001\pm0.001$ & $0.146\pm0.015$  & $9.50\pm0.46$  & 0.228 \\\hline 
sampling, full, clip   & $0.013\pm0.004$ & $-0.004\pm0.001$ & $-0.001\pm0.001$ & $0.020\pm0.012$  & $0.00\pm0.09$  & 0.207 \\\hline 
\end{tabular}
}
\caption{\label{tab:toyCC}  
Same as  table~\ref{tab:toy} for simulated JLA-B and JLA-P Hubble diagrams with contamination by core-collapse SNe.
Fits labelled full use the magnitude bias correction including contamination (see text), all others use the intrinsic SN~Ia magnitude bias correction. 
}
\end{table}

\begin{table}[tbp]
\centering
{\footnotesize
\begin{tabular}{|l|c|c|c|c||c|c|c|c|}
\hline
Fit & $\Omega_M$  & $\Delta M_B$ & $\alpha$ & $\beta$  & $\Omega_M$  & $\Delta M_B$ & $\alpha$ & $\beta$  \\ \hline
 
& \multicolumn{8}{c|} {{\bfseries JLA-S samples (without CC contamination)}}   \\ \hline
standard       & $0.026$  & 0.008 & $0.005$ & $0.053$  & $0.026$  & 0.008 & $0.005$ & $0.053$ \\ \hline
standard, clip & $0.027$  & 0.008 & $0.006$ & $0.055$  & $0.027$  & 0.008 & $0.006$ & $0.055$ \\ \hline

& \multicolumn{4}{c||} {{\bfseries JLA-B samples }}  & \multicolumn{4}{|c|} {{\bfseries JLA-P samples }}   \\ \hline
standard        & $0.030$  & 0.009  & $0.007$ & $0.150$  &  $0.034$  &  0.015 & $0.011$ & $0.159$ \\ \hline 
standard, clip & $0.030$  & 0.009  & $0.007$ & $0.065$  &  $0.030$  &  0.015 & $0.011$ & $0.108$ \\ \hline 
standard, full        & $0.030$  & 0.008 & $0.007$ & $0.144$   & $0.035$  & 0.015& $0.011$ & $0.156$  \\ \hline 
standard, full, clip & $0.029$  & 0.008 & $0.007$ & $0.065$  & $0.030$  & 0.015 & $0.009$ & $0.100$ \\ \hline 

sampling         & $0.029$  & 0.009 & $0.008$ & $0.142$  & $0.032$  & 0.012& $0.010$ & $0.146$  \\ \hline 
sampling, clip  & $0.029$  & 0.009 & $0.007$ & $0.059$ & $0.028$  & 0.012 & $0.008$ & $0.089$  \\ \hline 
sampling, full        & $0.030$  & 0.009 & $0.007$ & $0.139$  & $0.031$  & 0.012 & $0.009$ & $0.138$  \\ \hline 
sampling, full, clip & $0.030$  & 0.009 & $0.007$ & $0.061$  & $0.029$  & 0.012 & $0.008$ & $0.090$  \\ \hline 

\end{tabular}
}
\caption{\label{tab:toyCC2} Same as table~\ref{tab:toyCC} 
for the 16-84 percentile ranges in the $\Omega_M, \Delta M_B, \alpha$ and $\beta$ marginalised values. }
\end{table}

Average results of the standard and sampling cosmological fits are given in tables~\ref{tab:toyCC} and~\ref{tab:toyCC2}, where they are compared to results for uncontaminated JLA-S HDs. 
Similar to what was done in section~\ref{sec:toymcfits}, we either use the intrinsic SN~Ia bias correction for all SNLS events in the diagrams, 
or use the full combined correction previously defined for events with SALT2 parameters evaluated at their photometric redshift, all other SNLS events, if any, being corrected for the intrinsic SN~Ia magnitude bias.
We first describe the fit results compared to the uncontaminated case and then summarise our findings at the end of the section.

CC contamination results in an increased number of outliers, by a factor of nearly 3 in JLA-B HDs and by $40\%$ in JLA-P ones, while the HD rms increases by 13 and 10 $\%$, respectively (see table~\ref{tab:toy}). For both types of fits and HDs, biases in $\Omega_M$ are significantly larger than in uncontaminated HDs, reaching $1.6\,\sigma$ for standard fits, where $\sigma$ is the expected statistical fit error, unless the sampling method with the full combined bias correction is applied, with (resp. without) clipping in JLA-B (resp. JLA-P) HDs.  
In that case, the $\Omega_M$ bias is reduced to an acceptable level (at most 0.005 or $0.3\,\sigma$) for both types of diagrams.
We note that clipping has a more significant effect than for uncontaminated HDs (see table~\ref{tab:toy}). It helps to reduce the $\Omega_M$ bias in all cases, whatever the HD type, fit method or magnitude bias correction, except for JLA-P HDs when the full combined bias correction is applied, whatever the fit method. 
This may reveal a negative interplay between clipping and the full combined bias correction for both standard and sampling fits for these diagrams.
The large number of HD outliers may make the full combined bias correction inappropriate when clipping is applied, especially for the sampling method, despite the fact that clipping w.r.t. the simulation cosmology was applied in the magnitude bias computation. JLA-B diagrams do not suffer from such an issue since the number of outliers is low, as well as the number of photometric SNe, to which the full combined correction is applied.

With the sampling method and full combined bias correction, biases in the nuisance parameters $M_B$ and $\alpha$ are similar to or better than those of JLA-S diagrams for both types of diagrams, with or without clipping, a trend already observed in the uncontaminated case for JLA-P diagrams. On the other hand, without clipping, biases in $\beta$ are significantly larger than without contamination, reaching 0.09 and 0.15 in JLA-B and JLA-P HDs respectively, to be compared to 0.06 at most without contamination (see table~\ref{tab:toy}).  With clipping this bias is reduced to 0.02 for both types of HDs. As JLA-P HDs are better treated without clipping to avoid biasing cosmological parameters, a bias in $\beta$ of order 0.15 should be expected. Similar conclusions are reached with standard fits, but with biases in all nuisance parameters slightly higher. 

Finally, table~\ref{tab:toyCC2} presents the 16-84 percentile ranges in the marginalised values of the four fit parameters. Compared to the uncontaminated case (see table~\ref{tab:toy3}), the percentile ranges in $\Omega_M$ and $M_B$ increase  moderately, by at most $15\%$ and 4$\,$mmag, respectively. Percentile ranges in $\alpha$ increase by at most $20\%$, while those in $\beta$ increase by a factor of 1.5 to 3, unless clipping is applied. While percentile ranges are similar (within a few $\%$ ) with both types of fits in JLA-B HDs, sampling gives reduced ranges (by 10 to 20$\%$) for JLA-P HDs, especially in $M_B$. 
Compared to JLA-S HDs and retaining only sampling fits with the best option for each diagram type ('full, clip' for JLA-B, 'full' for JLA-P), the percentile ranges of JLA-B HDs are larger by $15\%$ for $\Omega_M$ and 
12 to 16$\%$ for the nuisance parameters, while those of JLA-P HDs are higher by $19\%$ for $\Omega_M$, and by a factor 1.5 to 2.6 for the nuisance parameters.

In summary, CC contamination increases the bias in $\Omega_M$ by a factor 2 to 3 w.r.t. that of purely SN Ia JLA-B and JLA-P HDs treated with the intrinsic SN~Ia magnitude bias correction, depending whether clipping is applied or not. To reduce the cosmological bias down to a fraction (one-third) of the statistical error, both types of diagrams should be treated with the full combined correction, with (resp. without) clipping in the JLA-B (resp. JLA-P) case. The price to pay for JLA-P HDs is a significant bias in $\beta$ of the order of $0.15$ (5$\%$ in relative value) and an increase of its interval error by 50$\%$, making the $\beta$ bias a 2$\sigma$ effect for the sampling method ($2.4\sigma$ for standard fits).
 Biases in $M_B$ and $\alpha$ are expected to remain at the same level as in fully spectroscopic SN Ia HDs, despite the contamination. The impact of photometric redshifts and CC contamination on parameter uncertainties is a moderate increase (below 20$\%$) for JLA-B diagrams as compared to purely spectroscopic SN Ia HDs. For JLA-P diagrams, the increase is higher, 15 to 30$\%$ on  $\Omega_M$ and a factor of 1.5 to 3 depending on the nuisance parameter and fitting method, the sampling one giving smaller increases. Although both types of fits show close performance, in particular in terms of cosmological biases, the sampling method should be preferred since it propagates a model of photometric redshift uncertainties.

In the following, we discuss the limitations of our procedure as a possible reason for the presence of a residual cosmological bias in fits to contaminated diagrams and then proceed with fits to the actual JLA-P and JLA-B diagrams from data.

\subsection{Discussion}
\label{sec:ccdiscussion}

\begin{figure}[tbp]
\centering
\begin{tabular}{cc}
\includegraphics[width=.55\textwidth]{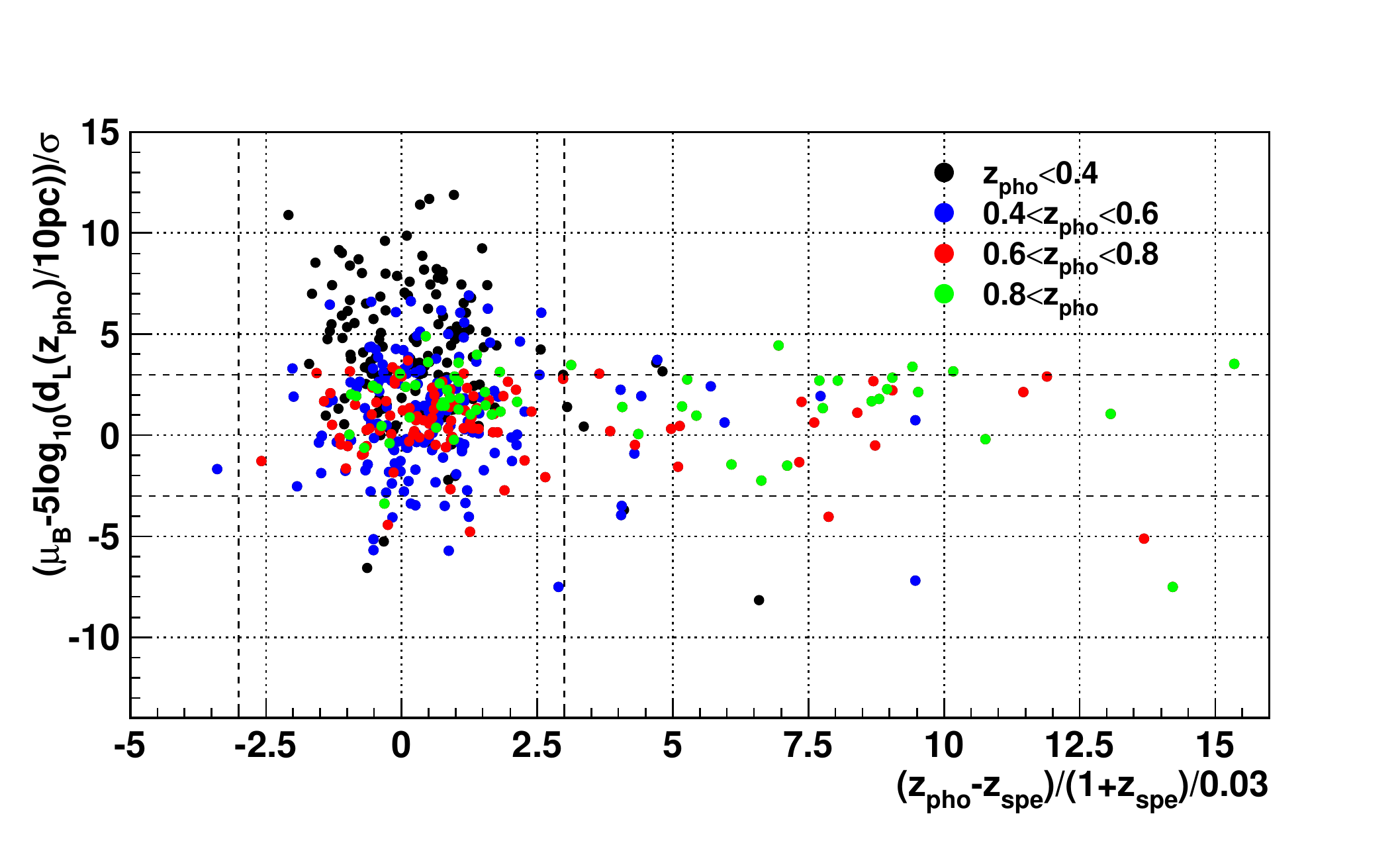} & 
\hspace{-1.4cm} \includegraphics[width=.55\textwidth]{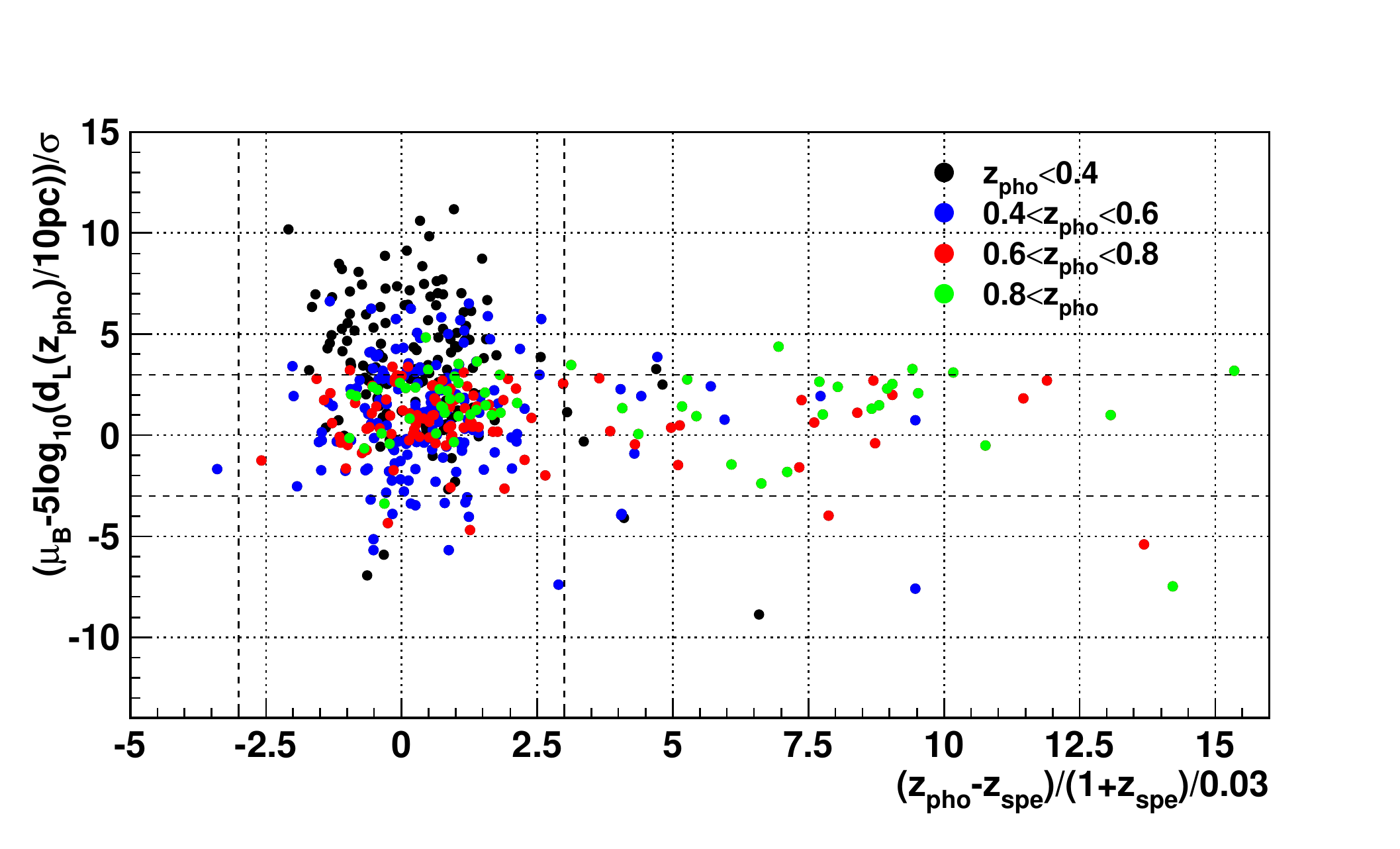} \\ 
\multicolumn{2}{c}{\includegraphics[width=.55\textwidth]{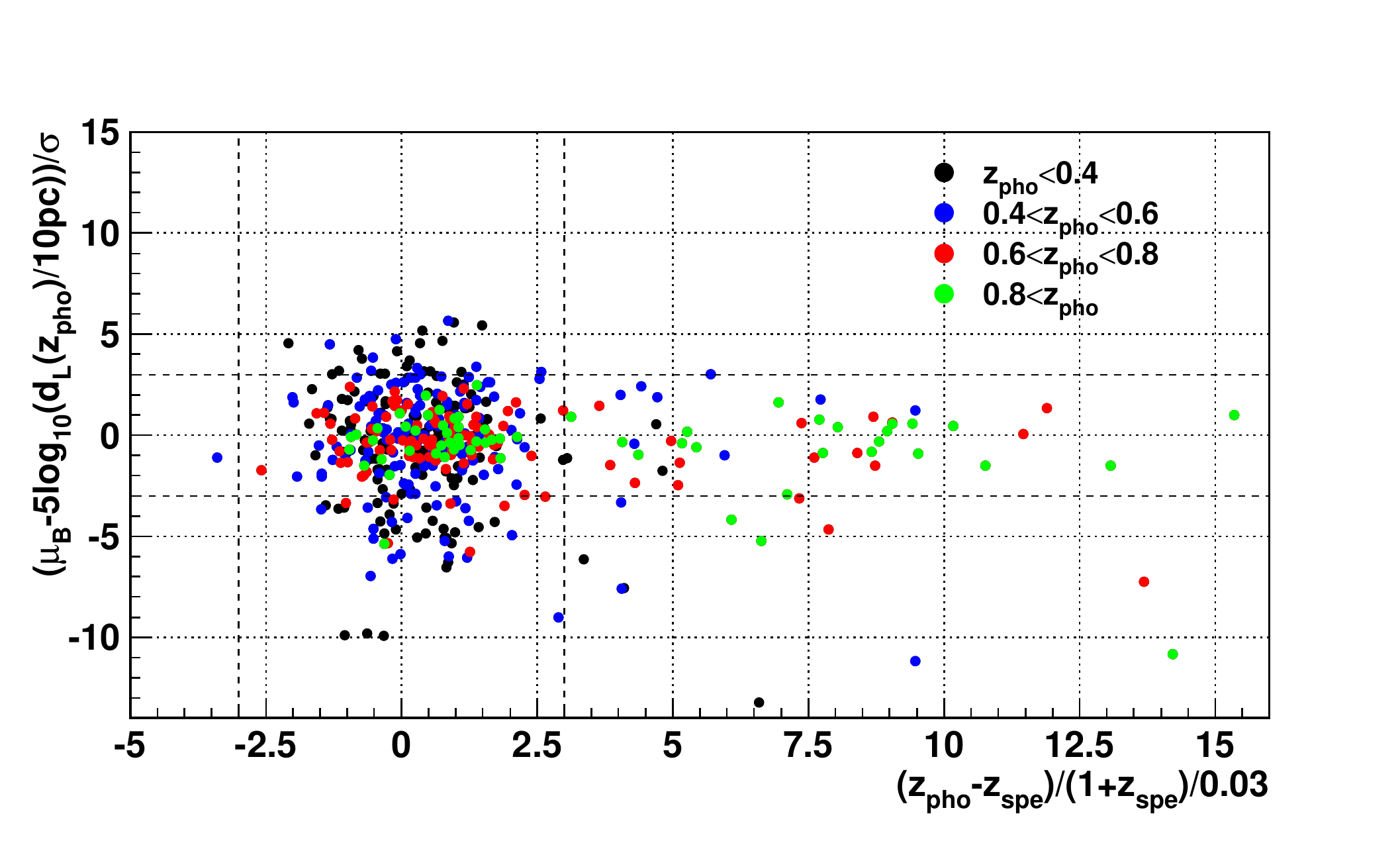}}
\end{tabular}
\caption{\label{fig:outliersCC} Normalised Hubble diagram residual
as a function of host photometric redshift bias normalised by central redshift resolution. We used a sample of around 400 simulated core-collapse light curves passing the SN~Ia photometric selections and adjusted by SALT2 at their host photometric redshift. Different colours refer to different host photometric redshift ranges. The SN magnitude bias correction used in the Hubble diagram residual computation is the SN~Ia intrinsic one in the lefthand plot, the full SNIa-CC combined one in the righthand plot and the full CC one in the bottom plot. Vertical and horizontal dashed lines define $3\sigma$ outliers. Core-collapse contaminants in our photometric selection produce Hubble diagram outliers at low redshift, which cannot be corrected by the full SNIa-CC combined magnitude correction used as baseline in this work. }
\end{figure}

Figure~\ref{fig:outliersCC} shows HD residuals as a function of redshift residuals for simulated CC SNe. As in the case of SNe Ia with photometric redshifts (figure~\ref{fig:outliers}) most cosmological outliers are not redshift outliers, whatever the SN magnitude bias correction applied. CC contamination creates HD outliers mostly at low redshift, $z_{pho}$$<$0.5. With the SN Ia intrinsic bias correction, these are mostly on the fainter side and a negative bias in $\Omega_M$ is to be expected.  This bias is slightly reduced if the full combined bias correction is applied but would vanish if the full CC bias correction was applied. This explains why, in the fits of table~\ref{tab:toyCC}, the full combined bias correction reduces the bias in $\Omega_M$ but in most cases does not correct it as efficiently as in the uncontaminated case. This shows the limit of our attempt to correct the SN magnitude bias with a combined SNIa-CC bias correction rather than trying to apply one or the other correction on each event depending on its probability to be of either type. Methods to include typing probabilities in cosmological fits for contaminated SN~Ia samples have been developed in the case of samples with host spectroscopic redshifts~\cite{Kunz07,Kessler17}. It could be interesting to test whether they can be adapted when part of the sample have host photometric redshifts.

Beside these limitations, our procedure relies on several simplifying but unavoidable assumptions.
 For JLA-B diagrams, whatever the fitting method, the SN Ia intrinsic bias correction is applied to any photometric SN that is assigned a spectroscopic redshift, whatever its type.
Second, the full combined bias correction is applied to SNe which remain with a photometric redshift, although it was computed assuming that all selected SNe (Ia and CC) have a photometric redshift. We thus assume that CC SNe can be treated as SNe Ia once migrated to their spectroscopic redshift and that the mean full magnitude bias of either SN class in each redshift bin is not changed by the spectroscopic redshift assignment. This can be the case if the latter does not correlate with SN properties but operates as a random sampling of the SNLS photometric population in each redshift bin of the diagram, which is the case in simulated diagrams but
may not be the case for data.

Finally, the negative interplay between clipping and the full combined bias correction noticed for the JLA-P diagrams in table~\ref{tab:toyCC} is larger for sampling fits, which may indicate that the clipping from standard fits, also used for sampling ones, may be inappropriate for them.

\section{Results on data}
\label{sec:resudata}
The two JLA mixed diagrams from data (section~\ref{sec:mixed}) were fit with the standard and sampling methods, 
 in the purely diagonal case as was done for simulated HDs, 
 but also in combination with covariance matrices describing the other systematic effects. 
In the following, we first outline how light curve parameters and covariance matrices are obtained, specify how the latter are coupled with the sampling method, before reporting the fit results.

\subsection{Light curve parameters}
\label{sec:LCdata}
The two mixed diagrams contain both JLA spectroscopic SNe and SNLS photometric SNe. We took the published parameters for the JLA SNe and built the photometric SN parameters and their errors as described in section~\ref{sec:method}. Compared to what was done for the simulated HDs, the SN magnitudes are corrected for peculiar velocity effect and the magnitude diagonal errors contain the peculiar velocity and the lensing contributions, but does not include the statistical magnitude bias uncertainty. The latter is accounted for through the bias covariance matrix described in the next section.
According to our simulations, the preferred option for the SN magnitude bias correction for SNe with a host galaxy photometric redshift is  the full combined bias correction (section~\ref{sec:biascc}) since it is expected to provide unbiased cosmological results, but we also provide results for the intrinsic SN~Ia bias correction (section~\ref{sec:bias}) as a comparison.
We recall that the intrinsic SN~Ia bias correction is used for SNLS SNe which can be assigned a spectroscopic redshift in the mixed JLA-B sample.

Values of the magnitude intrinsic dispersion $\sigma_{int}$ are those published in~\cite{Betoule14} for the three surveys other than SNLS. For the latter, we take values that give a $\chi^2$ close to the total number of SNe in the diagram for standard fits performed with diagonal errors. We found that the same values can be taken for both SN magnitude bias corrections but that they become smaller with clipping. The values adopted for the fits are  0.22 (resp. 0.18) for the JLA-B diagrams without (resp. with) clipping and 0.235 (resp. 0.22) for the JLA-P diagrams without (resp. with) clipping.
They are significantly higher than the SNLS value of the JLA sample, 0.08, reflecting the larger dispersion of the mixed diagrams due to their larger sample size, photometric nature, contamination and, to a much lesser extent, to the use of a less precise photometry to reconstruct their light curves~\cite{Bazin11}.

 For the sampling method, the photometric SNLS SN light curves were fit as described in section~\ref{sec:shiftparam} to obtain models of the redshift evolution of their SALT2 parameter shifts w.r.t. values at their initial photometric redshift. We observed that data have similar shift evolutions as simulated SNe, which are as correctly described by degree six polynomial models.
 
\subsection{Covariance matrices for mixed samples}
Systematic uncertainties other than that related to redshift are included via covariance matrices as explained in section~\ref{sec:method}. 
For each mixed sample, the computation was done as follows.

$C_{stat}$ was built with the method of~\cite{Betoule14} applied to all SNe in the diagram. For $C_{cal}$ and $C_{LC mod}$, we followed~\cite{Betoule14} and computed shifts of the photometric SN light curve parameters due to, respectively, calibration and light curve model uncertainties and combined them with those of the non-SNLS SNe of the JLA sample as used in~\cite{Betoule14}.  
$C_{dust}$ and $C_{pecvel}$ were deduced from the corresponding JLA matrices setting all terms where an SNLS photometric SN enters to 0, since these effects do not contribute much for SNLS events, as was checked on the JLA sample.

For $C_{bias}$  we extracted the non-SNLS part of the official JLA matrices, set the covariances between SNLS and other surveys to zero and replaced the SNLS block matrix  with a new one.  
This neglects cross-terms between SNLS and other surveys, which was checked, in the JLA case, to have no significant impact on the cosmological results. To form the SNLS block matrix, bias uncertainties were propagated to the uncertainty on $m_B^*$ and terms from each uncertainty source were added together, as in~\cite{Betoule14}. Popagated uncertainties were statistical and systematic ones, the latter being described in appendix~\ref{app:A} for the SN~Ia bias corrections and in section~\ref{sec:biascc} for the full CC bias correction.
For each mixed diagram, there are thus two $C_{bias}$ matrices, depending on the SN magnitude bias correction used in the fits. Note that since CC contamination is accounted for through the full combined bias correction, no specific contamination covariance matrix ($C_{non Ia}$ in \eqref{eq:eta}) is introduced.

Finally, as for simulated HDs, fits are performed on the initial samples without clipping or with the clipping defined from standard fits. Versions of the covariance matrices for clipped diagrams were thus also produced.

\subsection{Coupling covariance matrices and sampling}
Systematic uncertainties related to photometric redshift resolution could also be accounted for with a covariance matrix that propagates $1\sigma$ redshift uncertainties to both SN distance moduli and model luminosity distances (see e.g.~\cite{Pantheon22}). This requires one to define a reference cosmological model and point in the model parameter space to compute luminosity distance changes with the redshift resolution. The sampling method proposed in this paper allows changes in the luminosity distances and SN parameters to be computed at every point of the parameter space for any model. In this respect, this is a more exact method to test the impact of redshift uncertainties, but it neglects the cross-terms with the other systematic uncertainties and leads to a significant increase in computation time (a factor of 5 for JLA-B and 15 for JLA-P). 

To use the covariance matrices defined in the previous section within the sampling method, the $\chi^2$ computed at each point of the parameter space is:
 \begin{equation}
\label{eq:chi2systz}
 \begin{split}
 \chi^2  = & \big[ \vec{\mu}(\vec{z})  -  \vec{\mu}_{model}(\vec{z}) \big] ^\dag C_{\rm cov}^{-1}(\vec{z}) \big[ \vec{\mu}(\vec{z})  -  \vec{\mu}_{model}(\vec{z}) \big] \\
+ & \frac{1}{N}\sum
\big[ \vec{\mu}(\vec{z}+\vec{\delta z})  -  \vec{\mu}_{model} (\vec{z}+\vec{\delta z})\big]^\dag C_{\rm diag}^{-1}(\vec{z}+\vec{\delta z}) \big[ \vec{\mu}(\vec{z}+\vec{\delta z})  -  \vec{\mu}_{model}(\vec{z}+\vec{\delta z}) \big]  \\
-&  \big[ \vec{\mu}(\vec{z})  -  \vec{\mu}_{model}(\vec{z}) \big]^\dag C_{\rm diag}^{-1}( \vec{z})  \big[ \vec{\mu}(\vec{z})  -  \vec{\mu}_{model}(\vec{z}) \big]  
\end{split}  
 \end{equation}
where the sum extends over photomeric redshift variations $\vec{\delta z}$, and:
\begin{equation}
\label{eq:etasyst}
\begin{split}
C_{\rm cov} &=  A C_{\eta} A^{\dag} + {\rm diag}\left(\frac{5\sigma_z}{z{\rm log10}}\right)^2 + {\rm diag} \left(\sigma^2_{lens}\right) + {\rm diag}  \left(\sigma^2_{int}\right) \\
C_{\eta} &= C_{stat} + C_{cal} + C_{LC mod} + C_{bias}  + C_{dust} + C_{pecvel} \\
C_{\rm diag} &= C_{LC}  + {\rm diag}\left(\frac{5\sigma_z}{z{\rm log10}}\right)^2 + {\rm diag} \left(\sigma^2_{lens}\right) + {\rm diag}  \left(\sigma^2_{int}\right)
\end{split}
\end{equation}
with $ C_{LC}$ given by~\eqref{eq:cstat}. In the next section, we perform fits with diagonal errors only, in which case the $\chi^2$ in~\eqref{eq:chi2systz} reduces to its central term and is identical to that used for simulated HDs, and fits with covariance matrices, first with statistical errors only, $C_{\rm cov}=C_{stat}$ and then with all covariance matrices to include both statistical and systematic uncertainties.

\subsection{Fit results}
\label{sec:fitdata}

The complete set of results of the fits to the two mixed diagrams is presented in table~\ref{tab:data} in appendix~\ref{app:B}.
We performed standard and sampling fits, in each case with or without clipping, and varying the SN magnitude bias corrections as in section~\ref{sec:fitcc}. We recall that the clipping applied in sampling fits is that given by the standard fits performed under the same conditions. We first compare results obtained in different conditions (clipping,  magnitude bias correction, fitting method) before comparing the results on the two diagrams. We summarise our findings at the end of the section.

We note that clipping has a marginal effect on the fit results for all parameters. This was expected from simulated HDs for $M_B'$ and $\alpha$, but not for $\Omega_M$ and $\beta$ (see table~\ref{tab:toyCC}), where a significant effect was observed especially on $\beta$ in the JLA-P case.
We also note that the HD dispersion is significantly higher in data than in simulation, by 24 and 18 $\%$ for  the JLA-B and JLA-P diagrams, respectively. This is in line with the higher intrinsic dispersion in data (see section~\ref{sec:LCdata}) w.r.t. the input value of 0.08 used in the simulation (see table~\ref{tab:simul}). The number of HD outliers is also lower in data, especially in the JLA-P diagram, and their effect is reduced for both diagrams w.r.t. expectations.
 
Comparing the two magnitude bias corrections, the full combined one gives upward shifts in $\Omega_M$, three times more pronounced in the JLA-P diagram than in the JLA-B one. This shift is slightly reduced when statistical and systematic uncertainties are combined. Note that the shifts predicted by the simulation for diagonal errors, around $+0.008$ for JLA-B and $+0.024$ for JLA-P, are in remarkable agreement with what is observed in data for both diagonal and statistical only errors. The simulation predicted similar results for the two corrections on $M_B', \alpha$ and $\beta$, a trend we also see in data. 

The two fitting methods give compatible results for marginalised parameter values and their errors, a trend already observed in simulation. The largest difference between the two methods is observed for the JLA-P diagram and the $\beta$ parameter (see also figure~\ref{fig:postData} in appendix~\ref{app:B}), its marginalised value being 0.4 to 0.6$\sigma$ lower for the sampling method, depending on the covariance matrices included, all other things being equal. Differences for all other parameters do not exceed 0.2$\sigma$. Note that the standard deviation that we used to quantify differences is always that from one of the fits.

Errors on the fit parameters are similar whatever the fitting method and magnitude bias correction, a trend that was also present in the simulation results (table~\ref{tab:toyCC2}). However, errors are larger in data than expected from the 16-84 percentile ranges derived from simulation. 
As an example, restricting to fits with diagonal errors to match what was done in simulation, the error in $\Omega_M$ is $30\%$ higher in data, which we interpret as another effect of the larger dispersion of the actual diagrams. Adding other uncertainties,
errors in $\Omega_M$ are increased by about 20$\%$ in fits with statistical errors only, and almost double in fits with all errors combined. Errors in $\alpha, \beta$ also increase by about 20$\%$ in fits with statistical errors only and remain at a similar level in fits with all errors combined. This is in line with what was published for the JLA sample~\cite{Betoule14} (see also table~\ref{tab:final}).

We finally compare the results obtained for the two diagrams, restricting to the conditions advocated by our simulation study, namely sampling fits with the full combined bias correction, with clipping for JLA-B and with no clipping for JLA-P. Whatever the covariances included, the marginalised value of $\Omega_M$ is higher for the JLA-P diagram than for the JLA-B one. Comparing the difference between the two values to the standard deviation of one of the fits, the two values are within 0.3$\sigma$ when all errors are combined but are different by 1.4$\sigma$ with statistical errors only and by 1.9$\sigma$ with diagonal errors only. In the same conditions, there are 3 out of 50 fits to simulated HDs that show a higher difference. If no clipping is applied to the JLA-B diagram as well, this difference decreases and a higher value is found in 11 simulated HDs.
As for the nuisance parameters, the marginalised values are consistent between the two diagrams, within less than 1$\sigma$  for $\alpha$ and $\beta$, and within less than $2\,$mmag for $M_B'$, as expected for two samples which are composed mostly of the same SNe. 

\begin{table}[tbp]
\centering
\renewcommand{\arraystretch}{1.3} 
\begin{tabular}{|c|c|c|c|c|c|}
\hline
                             & $\Omega_M$ & $M_B'$ & $ \alpha$ & $\beta$ & rms \\ \hline 
JLA-B (stat+syst) &  $0.306^{+0.038}_{-0.036}$ &  24.08 & $0.134\pm0.007$ & $3.091^{+0.082}_{-0.080}$ & 0.265    \\ 
JLA-B (stat)         &  $0.294^{+0.020}_{-0.019}$ &  24.08 & $0.133\pm0.007$ & $3.114^{+0.081}_{-0.079}$ & 0.266 \\  \hline
JLA-P (stat+syst) &  $0.317^{+0.039}_{-0.037}$ &  24.09 & $0.134\pm0.008$ & $3.030^{+0.086}_{-0.083}$ & 0.280 \\
JLA-P (stat)         & $0.322^{+0.022}_{-0.021}$  &  24.08  & $0.132\pm0.008$ &  $3.041^{+0.084}_{-0.082}$ &  0.280 \\ \hline
JLA  (stat+syst)   & $0.303^{+0.036}_{-0.035}$  & 24.08 & $0.142\pm0.007$ & $3.187^{+0.090}_{-0.088}$ & 0.174\\
JLA (stat)             & $0.283^{+0.019}_{-0.019}$  & 24.07 & $0.138\pm0.006$ & $3.221^{+0.087}_{-0.085}$ & 0.174 \\ \hline
\end{tabular}
\caption{\label{tab:final}  
Final cosmological fit results on the JLA-B and JLA-P diagrams, compared to those obtained from standard fits to the JLA sample. Conditions for the mixed samples are those derived from our simulation study, namely sampling fits, using the full SNIa-CC combined magnitude bias correction, with (resp. without) clipping for JLA-B (resp. JLA-P). We report the marginalised constraints on $\Omega_M, M_B', \alpha$ and $\beta$ and the HD dispersion.
The uncertainties on the marginalised values are computed as described in section~\ref{sec:fitter}. $M_B'$ being analytically marginalised over, no uncertainty is provided.
}
\end{table}

\begin{figure}[tpp]
\centering
\begin{tabular}{ccc}
\hspace{-0.5cm}\includegraphics[width=0.38\textwidth]{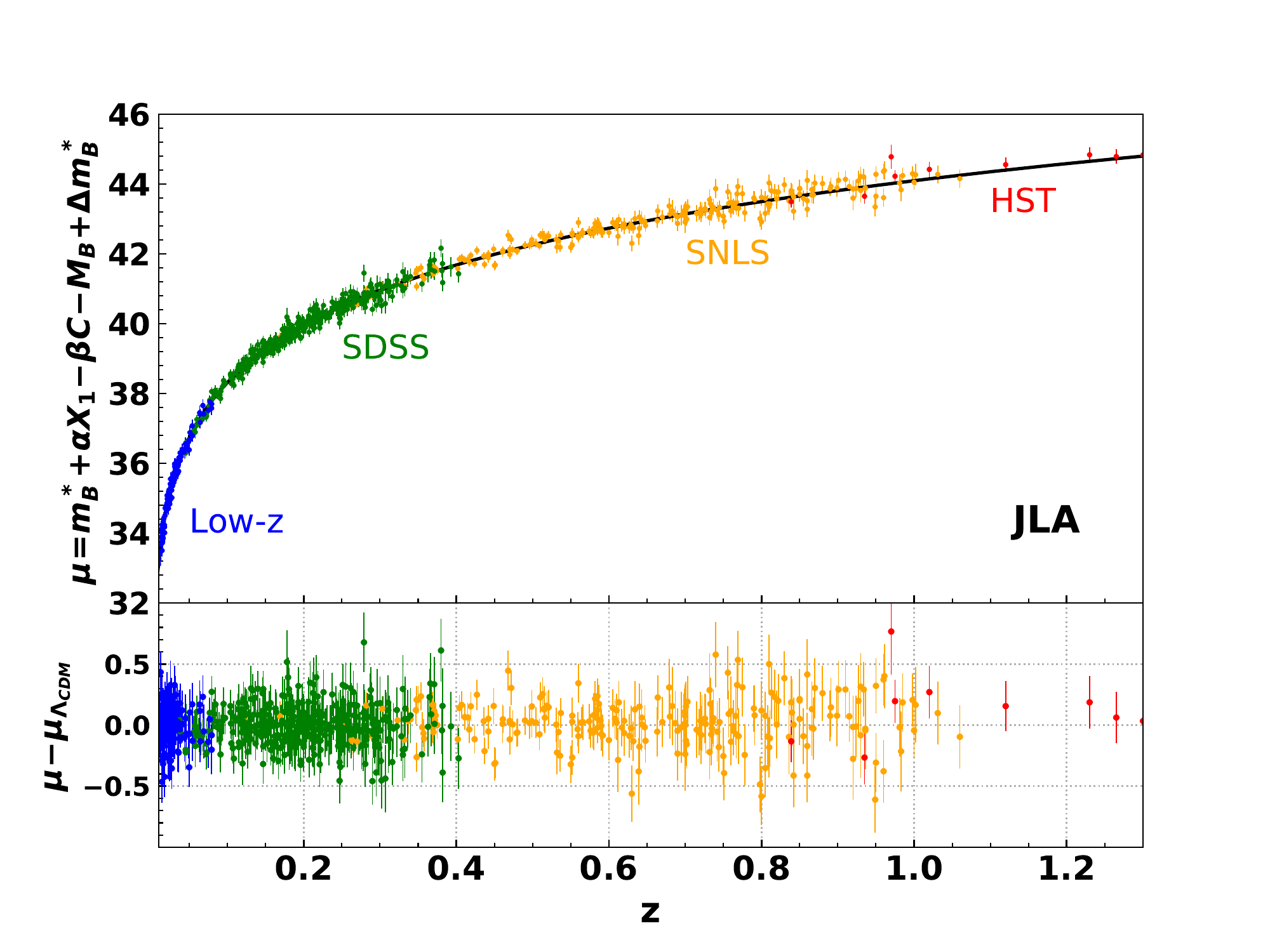} &
\hspace{-0.99cm}\includegraphics[width=0.38\textwidth]{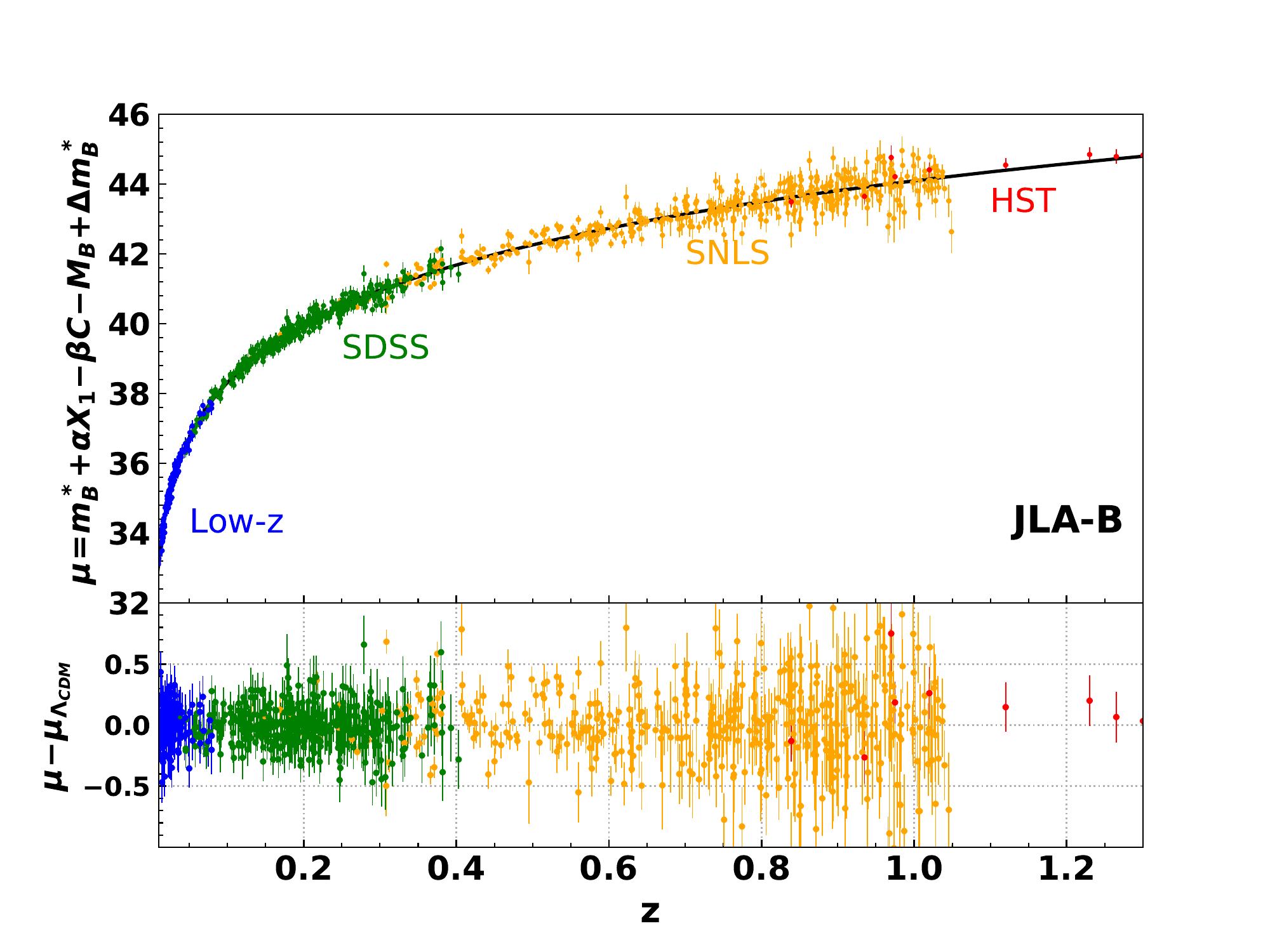} &
\hspace{-0.99cm}\includegraphics[width=0.38\textwidth]{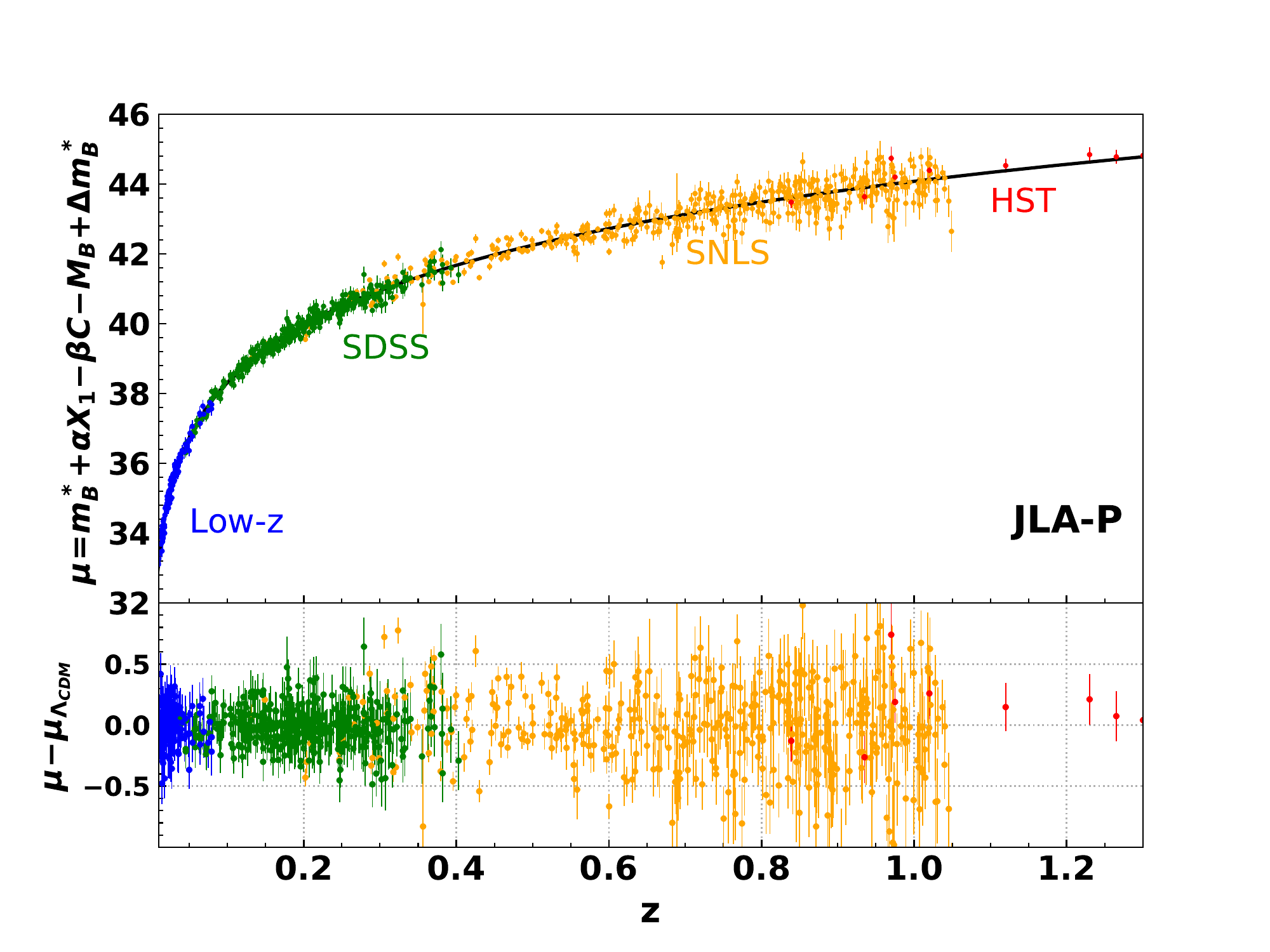} 
\end{tabular}
\caption{\label{fig:HDdata} Hubble diagrams of the JLA, JLA-B and JLA-P samples compared to their flat $\Lambda_{CDM}$ best fit (statistical and systematic uncertainties included). All diagrams share the same low-z, SDSS and HST components. The number of SNLS SNe in these diagrams is 239 for JLA and 437 for both JLA-B and JLA-P, as an effect of clipping being applied for JLA-B only.
}
\end{figure}

\begin{figure}[tpp]
\centering
\begin{tabular}{ccc}
\hspace{-0.5cm}\includegraphics[width=0.38\textwidth]{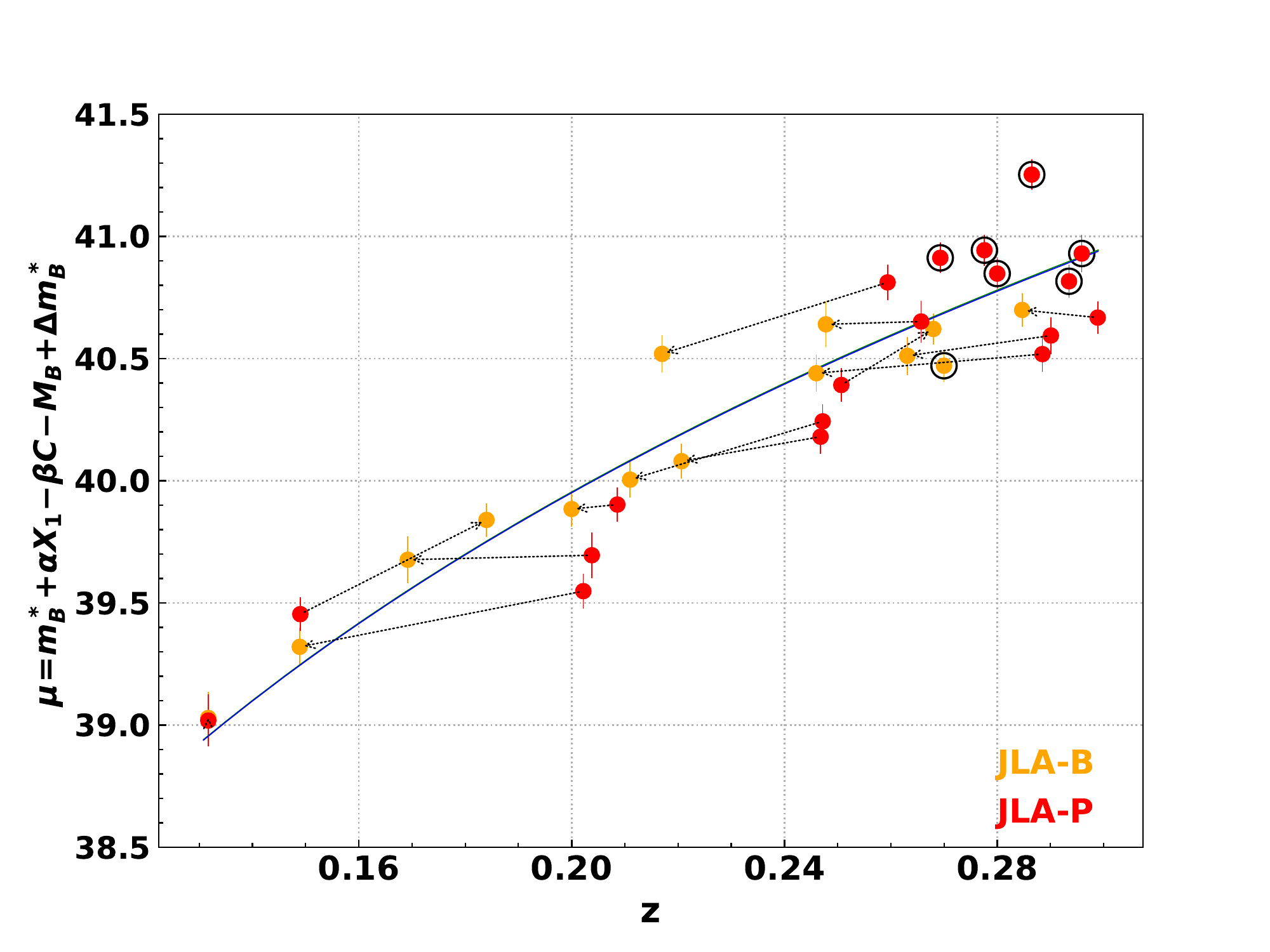} &
\hspace{-0.99cm}\includegraphics[width=0.38\textwidth]{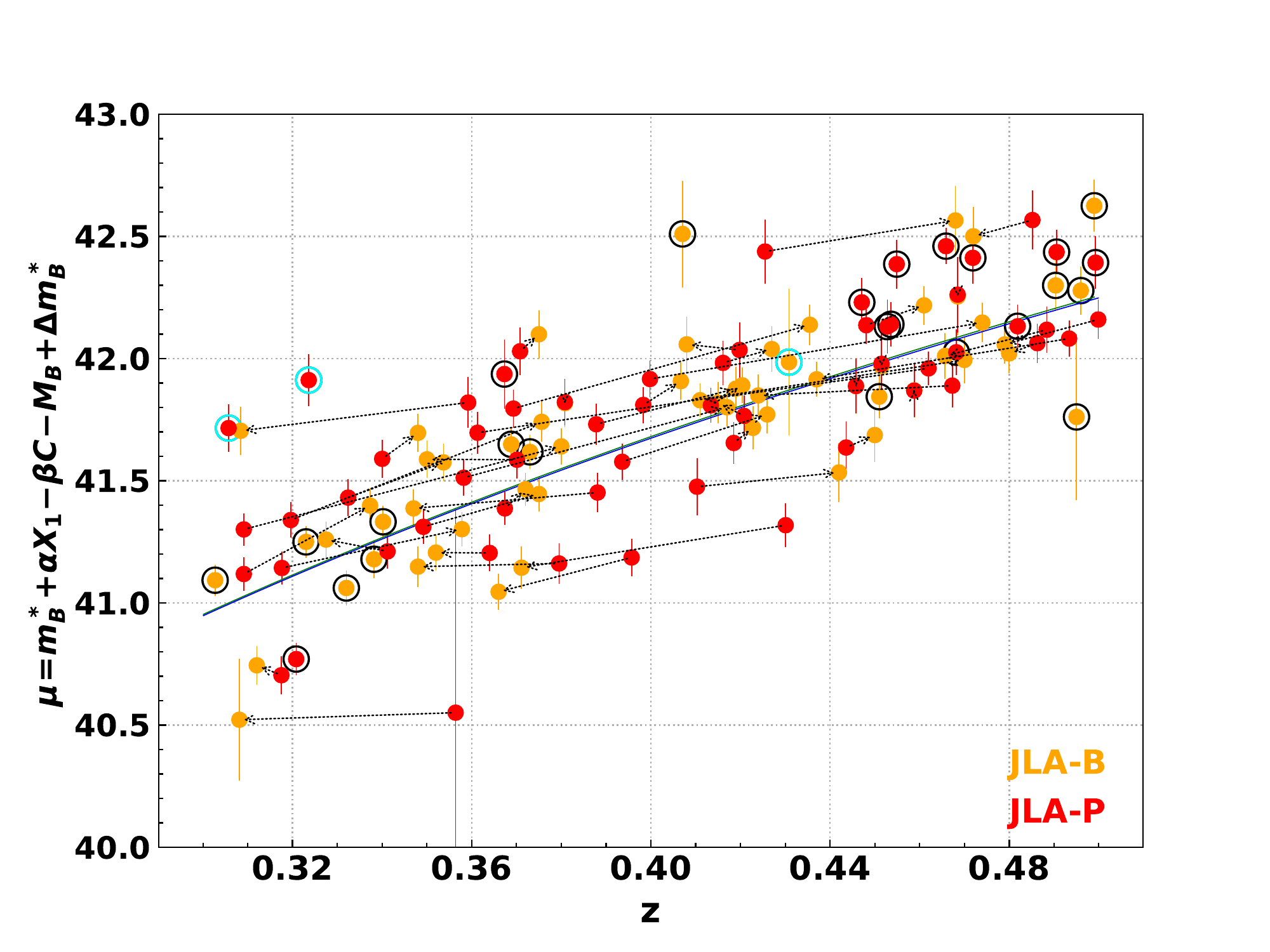} &
\hspace{-0.99cm}\includegraphics[width=0.38\textwidth]{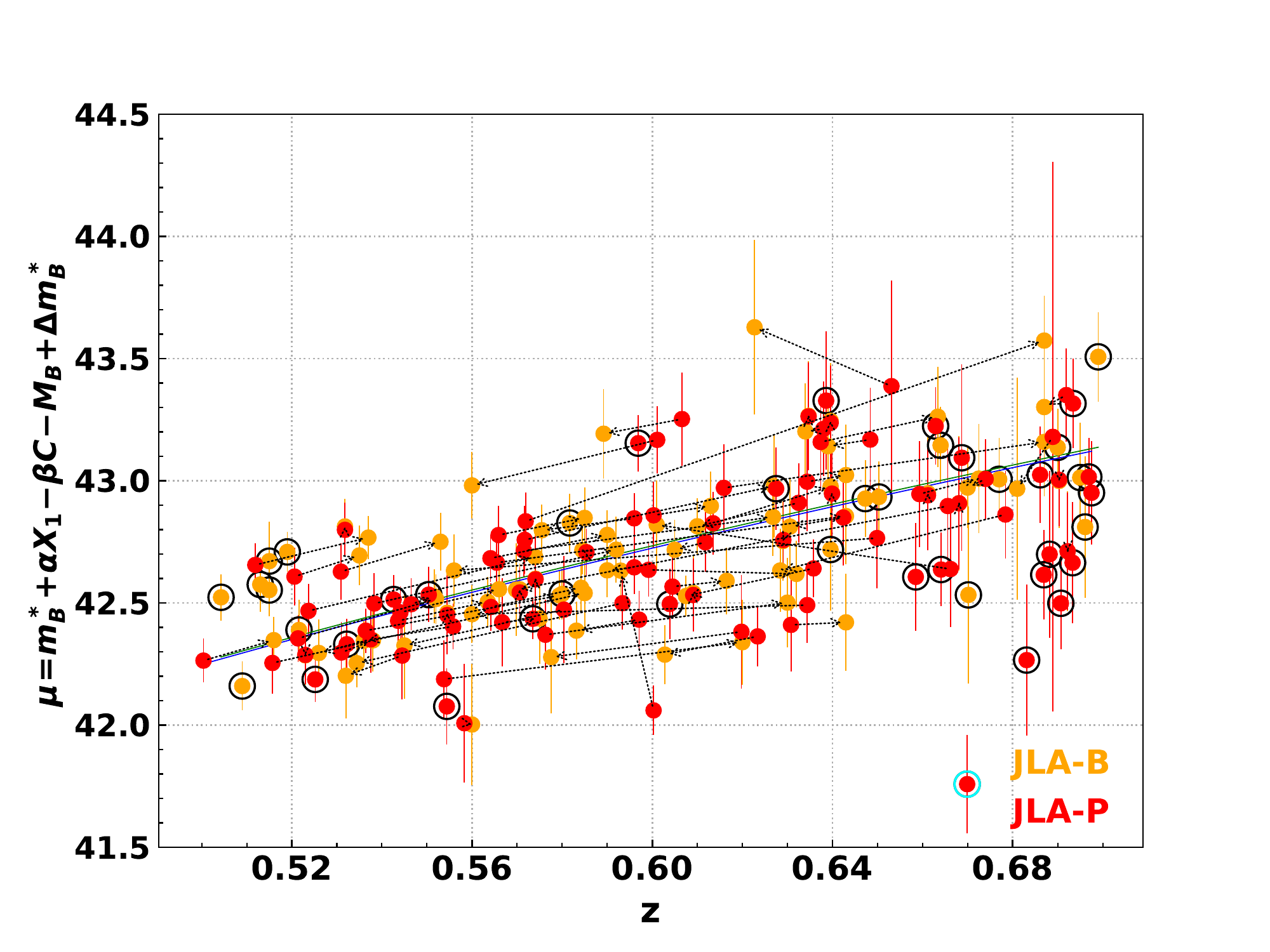} \\
\end{tabular}
\caption{\label{fig:HDcompar} Comparison between the SNLS sub-samples of the JLA-P (red) and JLA-B (yellow) diagrams in three redshift bins. 
In each diagram, the corresponding $\Lambda_{CDM}$ best fit (all uncertainties included) was used to compute distance moduli and luminosity distance expectations. The latter are shown in blue (JLA-P) and green (JLA-B) but are hardly distinguishable in these plots. Arrows indicate how each event migrates when host photometric redshifts are replaced by SN spectroscopic ones. Circled dots are events whose counterparts in the other diagram are in another redshift bin (black) or above 1.05 in redshift (cyan). Note that the variation of luminosity distances with redshift is significantly reduced beyond 0.5 and that redshift migration does not reduce to a shift in redshift but also affects distance moduli.
}
\end{figure}

Our final results on both diagrams, using sampling fits with the full SNIa-CC combined bias correction as advocated by our simulation study, are summarised in table~\ref{tab:final}, which also gives results obtained with standard fits 
on the JLA sample. Although marginalised values of $\Omega_M$ are consistent between the three diagrams (within the error of one of the fits) when all errors are combined, the agreement is preserved only between the JLA and JLA-B diagrams when considering statistical errors only. The marginalised values of $\alpha, \beta$ are slightly different between the JLA and the two mixed diagrams, which reflects the fact that the mixed samples have a significant component (around 330 SNe) not present in the JLA one. However, the agreement on the absolute SN~Ia magnitude $M_B'$ is good. For the sake of completeness, figure~\ref{fig:HDdata} presents the Hubble diagrams for the three samples and figure~\ref{fig:HDcompar}
illustrates how the change in SNLS event redshifts between the JLA-P and JLA-B samples propagates to the Hubble diagram.

\section{Conclusion} 
In this paper, we discussed the case of flat $\Lambda_{CDM}$ fits to Hubble diagrams of photometrically selected SNe Ia, using both host photometric redshifts and spectroscopic redshifts.
The impact of host galaxy photometric redshift uncertainties and contamination by core-collapse SNe was studied. 
As a test case, we used the mixed sample made of the non SNLS part of the spectroscopic JLA sample to which we added the 3-year SNLS photometric sample, which constitutes almost half of the diagram and is its main component at redshift above 0.4. To pass the photometric selection, objects were required to have SN~Ia compatible light curves when fitted by SALT2 at their host photometric redshift. They can enter the Hubble diagram with their SALT2 parameters determined either at these redshifts or at their (SN or host) spectroscopic redshifts  if these exist. Both options were considered, which defines two possible mixed Hubble diagrams, one using only host photometric redshifts for SNLS (JLA-P) and one with only 25$\%$ of the SNLS SNe with photometric redshifts, mostly at redshifts above 0.5 (JLA-B).

The photometric redshift catalogue used in the selection provides a central redshift resolution $\sigma_{\Delta z/(1+z)}$ $ \sim 3\%$ and  a 3$\sigma$ outlier rate of  5.5$\%$ at the end of the selection procedure, as measured from data.
Simulations of SN~Ia and core-collapse SN light curves including a redshift model reproducing these numbers were submitted to the exact same photometric selection algorithm as that applied on data. These simulations were used to estimate the SN magnitude bias due to the photometric selection. We found that the magnitude bias is essentially negligible for a pure sample of SNe~Ia with noiseless light curves treated at their true redshifts, provided one uses $\alpha, \beta$ values that fully account for the cosmological fitting procedure. 
Using light curves with instrumental noise and host galaxy photometric redshifts instead of true SN redshifts makes the SN~Ia magnitude bias increase significantly, especially at low redshifts. This increase is reinforced by core-collapse contamination, but only slightly thanks to the moderate contamination - around 5$\%$ - of the photometric sample.

The expected cosmological performance of both mixed samples was then studied with simulated Hubble diagrams that reproduce the contamination and redshift profile of the two actual diagrams. Treating photometric SNe as pure SNe Ia at their true redshifts in the magnitude bias correction produces a bias in $\Omega_M$ in standard fits to both types of diagrams, at the level of 1.6$\,\sigma$, where $\sigma$ is the expected statistical fit uncertainty. Contamination is responsible for two thirds of this bias, which can be reduced to $0.5\,\sigma$ when applying clipping and a magnitude bias correction that accounts for both redshift migration and contamination. 

To reduce the bias further, we tested two methods that propagate photometric redshift resolution to the cosmological likelihood function, either by refitting the photometric redshifts along with cosmology or by sampling the redshift resolution function when computing the cosmological fit $\chi^2$. Redshift refitting, tested on simulated photometric diagrams without contamination, was found to produce distorted posteriors of the fit parameters due to multiple local maxima in the likelihood surface. Moreover, this method did not reduce the bias in $\Omega_M$. On the other hand, sampling the redshift resolution function produces Gaussian posteriors and thus provides a reliable way to propagate the error redshift uncertainty to the cosmological likelihood (at a significant cost in CPU).  When tested on contaminated simulated diagrams, this method reduces the $\Omega_M$ bias to $0.3\,\sigma$ if the magnitude bias correction accounts for both redshift migration and contamination. Under the same conditions, uncertainties on $\Omega_M$ are increased by 10$\%$ for both diagram types when compared to photometrically selected samples with full spectroscopic redshift coverage. Standard fits show a similar increase.

With the help of the simulation, we investigated why the impact of the sampling method 
appears to be modest. The method acts on photometric redshifts to reduce the Hubble diagram dispersion and the number of outliers. But the latter are not necessarily induced by strong redshift outliers but rather by strong luminosity distance shifts due to small redshift uncertainties in low redshift objects, below 0.5, whether SNe~Ia or contaminants. In simulated diagrams, we observe that acting on photometric redshifts according to redshift resolution does reduce the number of SN~Ia HD outliers w.r.t. standard fitting but not that due to contaminants, probably because changing the redshift of contaminants makes their  SALT2 parameters less compatible with being SNe~Ia. We also observe that the diagram dispersion is not improved w.r.t. standard fitting, which is likely due to the fact that the bulk of the mixed diagrams is at high redshift, a region where acting on the photometric redshifts has no effect on the Hubble residuals. 

The standard and sampling fitting techniques were applied to both mixed samples in data, using the two SN magnitude bias corrections determined from simulations, and extending the sampling method to allow combination with the other systematic uncertainties described by covariance matrices. The change in  $\Omega_M$ we observe between fits using either the pure SN~Ia magnitude bias correction or the correction thats accounts for both redshift migration and contamination, is as expected from the simulations. Both fitting methods provide compatible results, the largest difference being obtained for the $\beta$ parameter and the JLA-P sample. Finally, the JLA-B
diagram provides values of $\Omega_M$ in agreement with those from the JLA sample, whether statistical only or statistical and systematic uncertainties are included in the fit, while the JLA-P diagram gives higher values, only compatible with those from the other two diagrams when all uncertainties are combined. 

For future SN~Ia projects aiming at fitting photometrically selected Hubble diagrams using host photometric redshifts, several conditions appear essential to avoid biasing cosmological parameters: reduce the contamination below the few $\%$ level, obtain spectroscopic host redshifts to cover the photometric sample up to redshifts around 0.5 and account for both contamination and redshift resolution in the SN magnitude bias computation. Propagating the redshift resolution function up to the cosmological likelihood function to correct residual biases and include the redshift resolution in the parameter errors seems a second priority for our sample
but may become necessary for more precise Hubble diagrams.

\acknowledgments
The authors warmly thank Marc Betoule for his help to set up the tools to compute covariance matrices for the JLA sample and for fruitful discussions in the early stage of this project.

\appendix

\section{Systematic uncertainties on the SN~Ia selection magnitude bias}
\label{app:A}

Systematic uncertainties in the SN~Ia photometric selection magnitude bias estimates (both intrinsic and full biases as defined in section~\ref{sec:biasres}) were evaluated with additional simulations. Each simulation comprises 20,000 light curves and corresponds to varying one parameter at a time with respect to values in the main simulation, within ranges given in table~\ref{tab:simul}. Ranges on the cosmological and SN~Ia nuisance parameters are one standard deviations reported in~\cite{Betoule14}. We chose to vary $\sigma_{int}$ by 25$\%$ since $\sigma_{int}$ values reported in~\cite{Betoule14}  for high redshift surveys show such a variation.  $W_{out}$ was varied by $\pm6\%$, a range that allow good modelling of the redshift outlier rates in data.

\begin{figure}[tbp]
\centering
\includegraphics[width=0.9\textwidth]{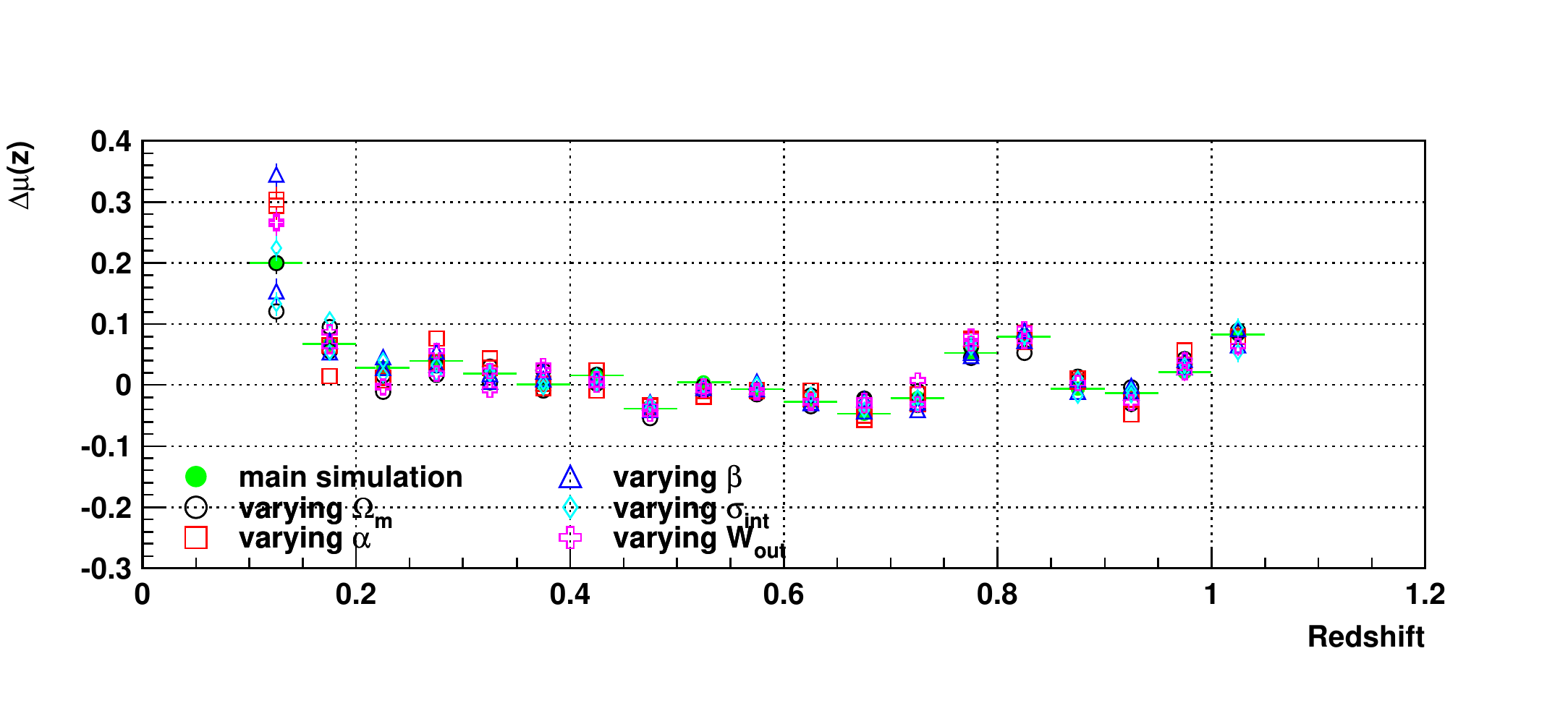}
\caption{\label{fig:biassys} 
SN~Ia photometric selection magnitude bias as a function of the host photometric redshift, from simulated SN~Ia light curves with instrumental noise, adjusted by SALT2 at their host photometric redshift. Alternative simulations were used with one parameter changed at a time (see symbol legend) w.r.t. the main simulation (green dots). In each redshift bin, the symbol for each varied parameter appears twice, as positive and negative changes were simulated. All error bars are statistical.}
\end{figure}

In each case, the bias in the alternative simulation was computed as in the main simulation, with appropriate nuisance parameters. As an example, figure~\ref{fig:biassys} presents the full bias estimate in the ten additional simulations.  Compared to the bias from the main simulation, the changes do not exceed $\pm 0.05\,$mag, except in the first redshift bin between 0.1 and 0.15, where the change is within $\pm 0.15\,$mag.

\section{Additional figure for the redshift refitting method}
\label{app:B}
 
\begin{figure}[tbp]
\centering
\includegraphics[width=0.9\textwidth]{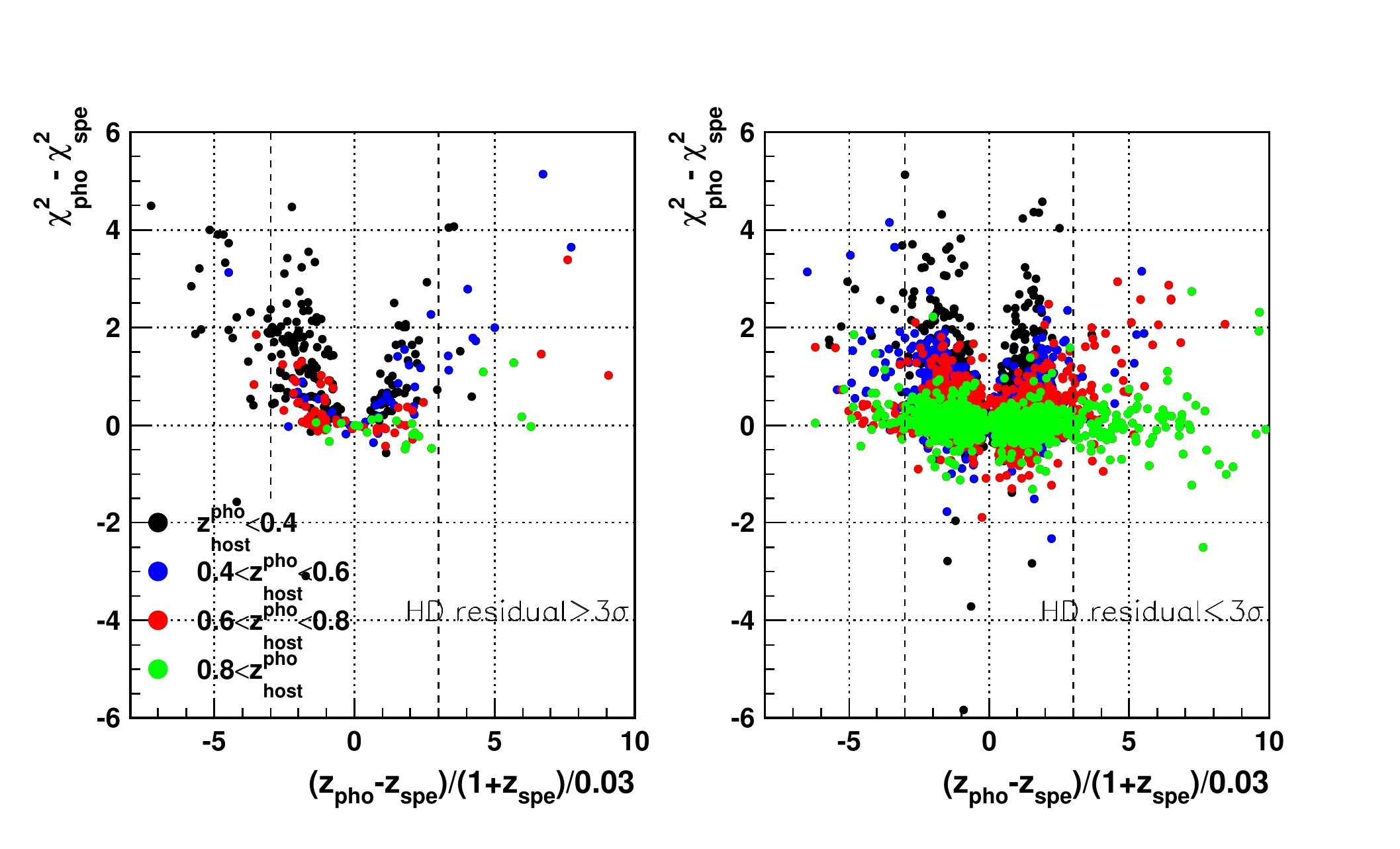}
\caption{\label{fig:chi2} Difference in SALT2 $\chi^2$ between light curve fits at host photometric vs spectroscopic redshifts as a function of host photometric redshift bias normalised by central redshift resolution (from simulation). Results are shown separately for Hubble diagram 3$\sigma$ outliers (left) and the rest of the sample (right). Different colours refer to different host photometric redshift ranges. The intrinsic SN~Ia magnitude bias correction was used in the Hubble diagram residual computation. Vertical dashed lines define $3\sigma$ redshift outliers. }
\end{figure}

 Figure~\ref{fig:chi2} shows the SALT2 $\chi^2$ difference between SN~Ia light curve fits run at host photometric redshifts and spectroscopic redshifts, as a function of the bias between these two redshift choices, normalised by the host photometric redshift central resolution.  This figure is based on a  sample of around 12,000 simulated SN~Ia light curves fulfilling our SN~Ia photometric selections.
 
 This figure shows that the proportion of events with a better SALT2 $\chi^2$ when light curves are fit at their photometric redshift instead of the true one drops for Hubble diagram outliers. 
 This justifies that a SALT2 $\chi^2$ difference term was added to the $\chi^2$ of the redshift refitting method, see eq.~\eqref{eq:chi2refit},  in order to disfavour redshift variations that would improve the cosmological fit at the cost of a degradation of the light curve fit. 

\section{Additional figure and table for results from data}
\label{app:C}

\begin{figure}[thbp]
\centering
\begin{tabular}{c}
\includegraphics[width=0.9\textwidth]{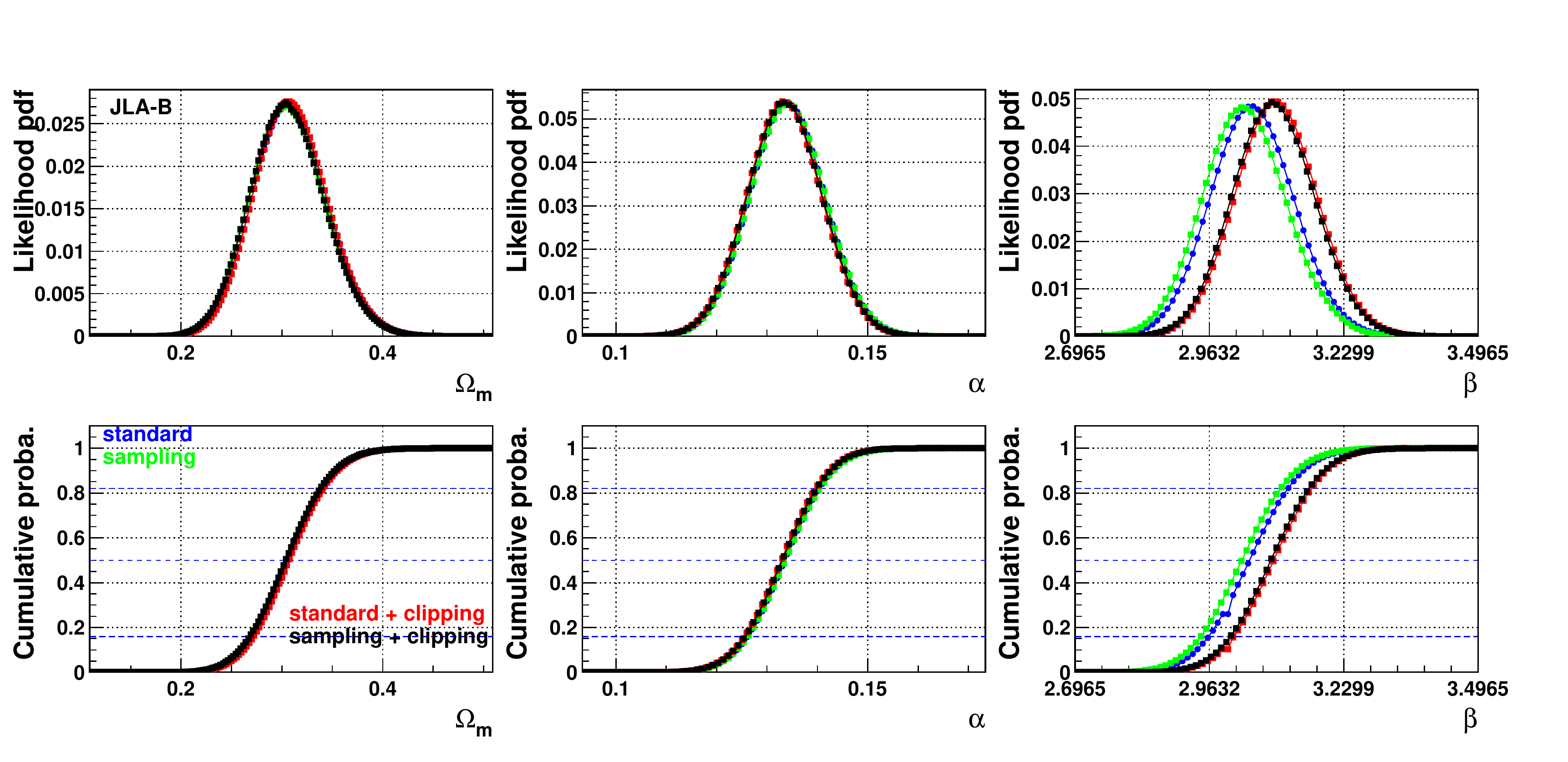} \\
\includegraphics[width=0.9\textwidth]{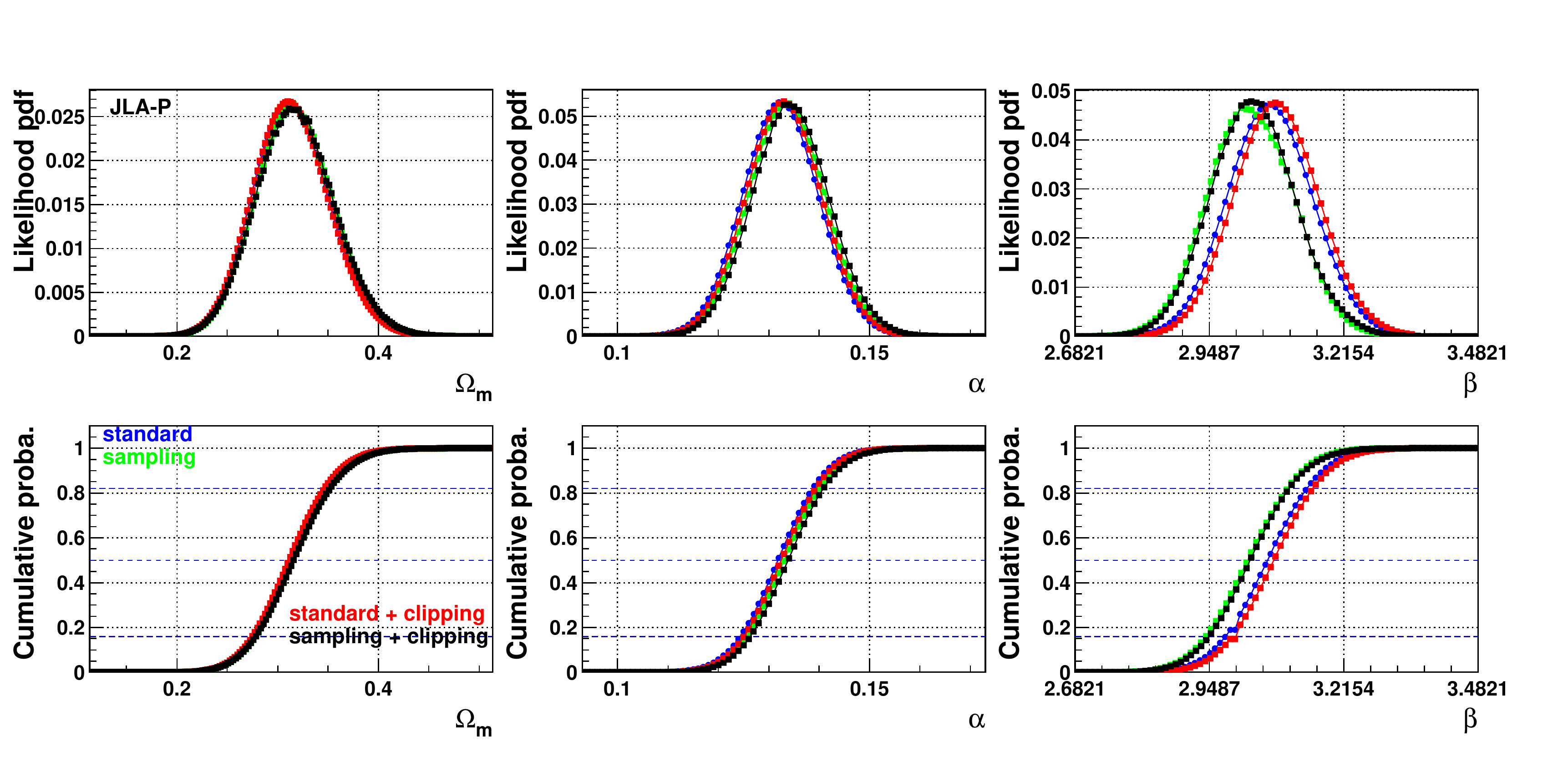} \\
\end{tabular}
\caption{\label{fig:postData} $\Omega_m, \alpha$ and $\beta$ posterior distributions and their cumulative distribution functions obtained for the JLA-B and JLA-P diagrams from data. The results of sampling and standard fits, with or without clipping are shown. All fits include statistical and systematic errors and make use of the full combined magnitude bias correction. Blue dashed lines indicate the median value and error interval deduced from the cumulative probability function. }
\end{figure}

Figure~\ref{fig:postData}  presents posterior distributions from fits to the JLA-P and JLA-B diagrams built from data, 
showing that the posterior distributions for both data diagrams have no distortions and do not depend on the fitting method or conditions, except for the $\beta$ posterior. The latter is modified by clipping but does not change with the fitting method for the JLA-B diagram, while it differs between the two fitting methods but hardly depends on clipping for the JLA-P diagram.

Table~\ref{tab:data} presents the complete set of results from fits to the two mixed Hubble diagrams deduced from data.

\begin{table}[thbp]
\centering
{\scriptsize
\begin{tabular}{|l|c|c|c|c|c|c|}
\hline
 Fit & $\Omega_M$ & $M_B'$ & $ \alpha$ & $\beta$ & $n_{out}$ & rms \\ \hline
\multicolumn{7}{|c|} {{\bfseries JLA-B sample, diagonal errors}}   \\ \hline 
standard              & $0.276+0.018-0.018$  & 24.073  & $0.1284+0.0063-0.0063$  &  $2.966+0.072-0.071$   & 7  & 0.258 \\\hline  
standard, clip       & $0.272+0.017-0.017$  & 24.071  & $0.1267+0.0062-0.0061$  &  $2.957+0.070-0.069$   & 1  & 0.246   \\\hline 
standard, full        & $0.284+0.018-0.018$  & 24.075  & $0.1278+0.0063-0.0063$  &  $ 2.960+0.072-0.071$  & 7 &  0.259   \\\hline 
standard, full, clip & $0.279+0.017-0.017$  & 24.073  & $0.1260+0.0062-0.0062$  &  $2.955+0.070-0.069$   & 1 & 0.247  \\\hline 

sampling                & $0.276+0.018-0.018$ & 24.073  & $0.1283+0.0063-0.0063$  &  $2.953+0.072-0.071$  & 7 & 0.264  \\\hline 
sampling, clip        & $0.273+0.017-0.017$  & 24.071  & $0.1268+0.0062-0.0062$  &  $2.949+0.070-0.069$ & 1 & 0.253 \\\hline 
sampling, full         & $0.283+0.018-0.018$  & 24.074  & $0.1279+0.0063-0.0063$  &  $2.947+0.072-0.071$  & 5 & 0.264 \\\hline 
sampling, full, clip  & $0.277+0.017-0.017$  & 24.072  & $0.1261+0.0062-0.0062$  &  $2.949+0.070-0.069$  & 1 &0.254  \\\hline 

\multicolumn{7}{|c|} {{\bfseries JLA-B sample, statistical errors}}   \\ \hline 
standard              & $0.295+0.021-0.020$ &  24.078  & $0.1339+0.0074-0.0074$ & $3.075+0.082-0.080$ & 5 &  0.265  \\\hline  
standard, clip       & $0.288+0.019-0.019$ &  24.077  & $0.1332+0.0074-0.0074$ & $3.120+0.080-0.079$ & 2 &  0.258   \\\hline 
standard, full        & $0.302+0.021-0.020$ &  24.079  & $0.1333+0.0074-0.0074$  & $3.069+0.082-0.080$ & 5 &  0.265  \\\hline 
standard, full, clip & $0.295+0.020-0.019$ &  24.078  & $0.1325+0.0074-0.0073$ & $3.119+0.080-0.079$ & 2 &  0.259   \\\hline 

sampling                & $0.294+0.021-0.020$ &  24.077 & $0.1337+0.0074-0.0074$ & $3.059+0.082-0.080$ & 5 &  0.271   \\\hline 
sampling, clip        & $0.290+0.019-0.019$ &  24.077 & $0.1333+0.0074-0.0074$ & $3.115+0.081-0.080$  & 2 &  0.266   \\\hline 
sampling, full         & $0.301+0.021-0.020$ &  24.079 & $0.1334+0.0074-0.0074$ & $3.054+0.084-0.081$ & 3 &  0.271   \\\hline 
sampling, full, clip  & $0.294+0.020-0.019$ &  24.078 & $0.1327+0.0074-0.0074$ & $3.114+0.081-0.079$ & 2 &  0.266   \\\hline 

\multicolumn{7}{|c|} {{\bfseries JLA-B sample, statistical + systematic errors}}   \\ \hline 
standard              & $0.303+0.037-0.035$ &  24.084 & $0.1347+0.0075-0.0074$ & $3.055+0.083-0.082$ & 5 &   0.264    \\\hline  
standard, clip       & $0.304+0.037-0.035$ &  24.084 & $0.1340+0.0074-0.0074$ & $3.098+0.082-0.080$ & 2 &  0.257    \\\hline 
standard, full        & $0.309+0.037-0.036$ & 24.085 &  $ 0.1340+0.0075-0.0074$ & $3.049+0.083-0.082$ & 5 &  0.264   \\\hline 
standard, full, clip & $0.309+0.037-0.035$ &  24.085 & $0.1333+0.0074-0.0074$  &  $3.097+0.081-0.080$ & 2 &  0.258   \\\hline 

sampling                & $0.304+0.038-0.036$ &  24.084 & $0.1345+0.0075-0.0074$ & $3.038+0.083-0.081$ & 5 &  0.270   \\\hline 
sampling, clip        & $0.308+0.037-0.035$ &  24.085 & $0.1341+0.0074-0.0074$ & $3.090+0.083-0.081$ & 2 &  0.264    \\\hline 
sampling, full         & $0.306+0.038-0.036$ &  24.084 & $0.1341+0.0075-0.0074$ & $3.034+0.084-0.082$ & 4 &  0.270   \\\hline 
sampling, full, clip  & $0.306+0.038-0.036$ &  24.084 & $0.1335+0.0074-0.0074$ & $3.091+0.082-0.080$ & 2 &  0.265    \\\hline 

\multicolumn{7}{|c|} {{\bfseries JLA-P sample, diagonal errors}}   \\ \hline 
standard                & $0.284+0.019-0.018$ &  24.076  & $0.1265+0.0064-0.0064$ & $2.993+0.074-0.073$ & 3 &  0.266  \\\hline 
standard, clip        & $0.282+0.019-0.018$ &  24.075  & $0.1269+0.0063-0.0063$ &  $2.975+0.073-0.072$ & 0 &  0.259  \\\hline 
standard, full         & $0.308+0.020-0.019$ &  24.079  & $0.1258+0.0064-0.0063$ & $2.978+0.074-0.073$ & 3 &  0.263    \\\hline 
standard, full, clip & $0.307+0.019-0.019$ &  24.078  & $0.1261+0.0063-0.0063$ & $2.963+0.073-0.072$ & 0 &  0.257   \\\hline

sampling               & $0.286+0.019-0.018$ &  24.076  & $0.1271+0.0064-0.0064$ & $2.953+0.073-0.072$ & 3 &  0.277  \\\hline 
sampling, clip        & $0.284+0.019-0.018$ &  24.075  & $0.1276+0.0065-0.0064$ & $2.936+0.073-0.072$ & 0 &  0.271  \\\hline 
sampling, full          & $0.309+0.020-0.019$ &  24.080  & $0.1268+0.0064-0.0064$ & $2.948+0.072-0.071$ & 1 &  0.274  \\\hline 
sampling, full, clip   & $0.308+0.019-0.019$ &  24.079  & $0.1271+0.0064-0.0063$ & $2.930+0.073-0.072$ & 0 &  0.268   \\\hline 

\multicolumn{7}{|c|} {{\bfseries JLA-P sample, statistical errors}}   \\ \hline 
standard                & $0.298+0.021-0.020$ &  24.079  & $0.1319+0.0075-0.0075$ & $3.095+0.084-0.083$ & 3 &  0.271   \\\hline 
standard, clip        & $0.296+0.021-0.020$ &  24.078  & $0.1323+0.0075-0.0074$ & $3.077+0.083-0.082$ & 0  & 0.265    \\\hline 
standard, full         & $0.322+0.022-0.021$ &  24.083  & $0.1310+0.0075-0.0074$ &  $3.079+0.084-0.082$ & 2 &  0.269   \\\hline 
standard, full, clip & $0.319+0.021-0.021$ &  24.082  & $0.1318+0.0075-0.0074$ & $3.093+0.083-0.082$ & 1 &  0.265   \\\hline

sampling               & $0.300+0.021-0.020$ &  24.080  & $0.1328+0.0075-0.0075$ & $3.044+0.083-0.081$ & 2 &  0.283   \\\hline 
sampling, clip        & $0.298+0.021-0.020$ &  24.079  & $0.1336+0.0079-0.0076$ & $3.028+0.083-0.081$ & 0 &  0.277  \\\hline 
sampling, full           & $0.322+0.022-0.021$ &  24.083  & $0.1324+0.0075-0.0075$ &  $3.041+0.084-0.082$ & 1 &  0.280  \\\hline 
sampling, full, clip   & $0.320+0.022-0.021$ &  24.083  & $0.1332+0.0075-0.0075$ & $3.049+0.083-0.081$ & 0 &   0.277  \\\hline 

\multicolumn{7}{|c|} {{\bfseries JLA-P sample, statistical + systematic errors}}   \\ \hline 
standard                & $0.294+0.037-0.035$ &  24.083  & $0.1327+0.0075-0.0075$ & $3.081+0.086-0.084$ & 3 &  0.271  \\\hline  
standard, clip        & $0.297+0.037-0.035$ &  24.083  & $0.1330+0.0075-0.0074$ & $3.062+0.085-0.083$ &  0 &  0.265  \\\hline 
standard, full         & $0.314+0.039-0.037$ &  24.083  & $0.1323+0.0075-0.0075$ & $3.070+0.085-0.084$ & 2 &  0.269    \\\hline 
standard, full, clip & $0.313+0.038-0.036$ &  24.083  & $0.1330+0.0075-0.0075$ & $3.082+0.085-0.083$ & 1 &  0.265    \\\hline

sampling               & $0.301+0.039-0.036$ &  24.087  & $0.1334+0.0076-0.0075$ & $3.029+0.085-0.083$ & 2 &  0.282  \\\hline 
sampling, clip        & $0.305+0.041-0.038$ &  24.087  & $0.1342+0.0077-0.0076$ & $3.010+0.084-0.082$ & 0 &  0.276   \\\hline 
sampling, full           & $0.317+0.039-0.037$ &  24.086  & $0.1336+0.0076-0.0075$ & $3.030+0.086-0.083$ & 1 &  0.280   \\\hline 
sampling, full, clip   & $0.318+0.040-0.038$ &  24.086  & $0.1344+0.0077-0.0075$ & $3.036+0.084-0.083$ & 0 &  0.277   \\\hline 

\end{tabular}
}
\caption{\label{tab:data}  
Cosmological fit results on the mixed JLA samples. We report the marginalised constraints on $\Omega_M, M_B', \alpha$ and $\beta$,
the number of  $3\sigma$ outliers and the HD dispersion.
Fits with $3\sigma$ clipping are marked clip, those using the full combined magnitude bias correction instead of the SN~Ia intrinsic one are labelled full. 
The uncertainties on the marginalised values are computed as described in section~\ref{sec:fitter}. $M_B'$ being analytically marginalised over, no uncertainty is provided.}
\end{table}



\begin{thebibliography}{99}


\bibitem{Planck18}
Planck Collaboration: N. Aghanim, Y. Akrami, M. Ashdown et al.,  \emph{Planck 2018 results. VI. Cosmological parameters}, \emph{A\&A} {\bf 641} (2020) A6, [arXiv:1807.06209].

\bibitem{eBOSS20}
eBOSS Collaboration: S. Alam, M. Aubert, S. Avila et al.,  \emph{The completed SDSS-IV extended Baryon Oscillation Spectroscopic Survey: Cosmological Implications from two Decades of Spectroscopic Surveys at the Apache Point observatory}, \emph{Phys. Rev.} {\bf D 103} (2021) 083533, [arXiv:2007.08991]

\bibitem{Pantheon22}
D. Brout, D. Scolnic, B. Popovic et al.,  \emph{The Pantheon+ Analysis: Cosmological Constraints}, submitted to  \emph{ApJ}, [arXiv:2202.04077]

\bibitem{Perrett12}
K. Perrett, M. Sullivan, A. Conley et al., \emph{Evolution in the Volumetric Type Ia Supernova Rate from the Supernova Legacy Survey}, \emph{AJ} {\bf 144} (2012) 59P, [arXiv:1206.0665].

\bibitem{Bazin11}
G. Bazin, V. Ruhlmann-Kleider, N. Palanque-Delabrouille et al., \emph{Photometric selection of Type Ia supernovae in the Supernova Legacy Survey}, \emph{A\&A} {\bf 534} (2011) A43, [arXiv:1109.0948].

\bibitem{Lidman12}
C. Lidman, V. Ruhlmann-Kleider, M. Sullivan et al., \emph{An Efficient Approach to Obtaining Large Numbers of Distant Supernova Host Galaxy Redshifts},   
\emph{Pub. Astr. Soc. Austr.} {\bf 30} (2013) 1L, [arXiv:1205.1306]. 

\bibitem{Lidman20}
C. Lidman, B.E. Tucker, T.M. Davis et al.,  \emph{OzDES multi-object fibre spectroscopy for the Dark Energy Survey: Results and second data release}, \emph{MNRAS} {\bf 496} (2020) 19L, [arXiv:2006.00449].

\bibitem{Jones16}
D.O. Jones, D.M. Scolnic, A.G. Riess et al.,  \emph{Measuring the Properties of Dark Energy with Photometrically Classified Pan-STARRS Supernovae. I. Systematic Uncertainty from Core-Collapse Supernova Contamination}, \emph{ApJ} {\bf 843} (2017) 6J, [arXiv:1611.0742].

\bibitem{Moller22}
A. Möller, M. Smith, M. Sullivan et al.,  \emph{The Dark Energy Survey 5-year photometrically identified Type Ia Supernovae}, \emph{MNRAS} {\bf 514} (2022) 5159, [arXiv:2201.11142].

\bibitem{Hlozek11}
R. Hlozek, M. Kunz, B. Bassett et al.,  \emph{Photometric Supernova Cosmology with BEAMS and SDSS-II}, \emph{ApJ} {\bf 752} (2012), [arXiv:1111.5328]

\bibitem{Sako11}
M. Sako, B. Bassett, B. Connolly et al.,  \emph{Photometric SN~Ia Candidates from the Three-Year SDSS-II SN Survey Data}, \emph{ApJ} {\bf 738} (2011) 162S, [arXiv:1107.5106].

\bibitem{Campbell12}
H. Campbell, C.B. D'Andrea, R.C. Nichol et al.,  \emph{Cosmology with Photometrically-Classified Type Ia Supernovae from the SDSS-II Supernova Survey}, \emph{ApJ} {\bf 763} (2013) 88C, [arXiv:1211.4480]

\bibitem{Jones17}
D.O. Jones, D.M. Scolnic, A.G. Riess et al.,  \emph{Measuring the Properties of Dark Energy with Photometrically Classified Pan-STARRS Supernovae. II. Cosmological parameters}, \emph{ApJ} {\bf 857} (2018) 51J, [arXiv:1710.00846].

\bibitem{Vincenzi21}
M. Vincenzi, M. Sullivan, A. Möller et al., \emph{The Dark Energy Survey Supernova Program: Cosmological biases from supernova photometric classification},  to appear in \emph{MNRAS},  [arXiv:2111.10382]

\bibitem{Kunz07}
 M. Kunz, B.A. Bassett and R.A. Hlozek,  \emph{Bayesian Estimation Applied to Multiple Species: Towards cosmology with a million supernovae}, \emph{Phys. Rev.} {\bf D 75} (2007) 103508, [arXiv:astro-ph/0611004]
 
\bibitem{Kessler17}
 R. Kessler and D. Scolnic,  \emph{Correcting Type Ia Supernova Distances for Selection Biases and Contamination in Photometrically Identified Samples}, \emph{ApJ} {\bf 836} (2017) 56K, [arXiv:1610.04677]

\bibitem{Chen22}
R. Chen, S. Scolnic, E. Rozo et al., \emph{Measuring Cosmological Parameters with Type Ia Supernovae in redMaGiC galaxies}, \emph{ApJ} {\bf 938}(2022) 62C,  [arXiv:2202.10480]


\bibitem{Mitra20}
A. Mitra and E. Linder,  \emph{Cosmology Requirements on Supernova Photometric Redshift Systematics for Rubin LSST and Roman Space Telescope}, \emph{Phys. Rev.} {\bf D 103} (2021) 103b3524M, [arXiv:2011.08206]

\bibitem{Linder19}
E. Linder and A. Mitra,  \emph{Photometric Supernovae Redshift Systematics Requirements}, \emph{Phys. Rev.} {\bf D 100} (2019) 043542, [arXiv:1907.00985]

\bibitem{Dai17}
M. Dai, S. Kuhlmann, Y. Wang, E. Kovacs,,  \emph{Photometric classification and redshift estimation of LSST Supernovae}, \emph{MNRAS} {\bf 477} (2018) 4142D, [arXiv:1701.05689]

\bibitem{Betoule14}
M. Betoule, R. Kessler, J. Guy et al.,  \emph{Improved cosmological constraints from a joint analysis of the SDSS-II and SNLS supernova samples},
\emph{A\&A} {\bf 568A} (2014) 22B,  [arxiv:1401.4064].  

\bibitem{Conley11}
A. Conley, J. Guy, M. Sullivan et al., \emph{Supernova Constraints and Systematic Uncertainties from the First Three Years of the Supernova Legacy Survey}, \emph{ApJS} {\bf 192} (2011) 1C, [arXiv:1104.1443].

\bibitem{Sullivan11}
M. Sullivan, J. Guy, A. Conley et al., \emph{SNLS3: Constraints on Dark Energy Combining the Supernova Legacy Survey Three Year Data with Other Probes}, \emph{ApJ} {\bf 737} (2011) 102S, [arXiv:1104.1444].

\bibitem{Ilbert06}
O. Ilbert, S. Arnouts, H.J. McCracken et al., \emph{Accurate photometric redshifts for the CFHT Legacy Survey calibrated using the VIMOS VLT Deep Survey}, \emph{A\&A} {\bf 457} (2006)  841I, [arXiv:astro-ph/0603217].


\bibitem{Guy10}
J. Guy, M. Sullivan, A. Conley et al.,  \emph{The Supernova Legacy Survey 3-year sample: Type Ia Supernovae photometric distances and cosmological constraints}, \emph{A\&A} {\bf 523} (2010) A7, [arXiv:1010.4743].

\bibitem{Regnault09}
N. Regnault, A. Conley, J. Guy et al., \emph{Photometric Calibration of the Supernova Legacy Survey Fields}, \emph{A\&A} {\bf 506} (2009) 999, [arXiv:0908.3808].

\bibitem{Betoule13} 
M. Betoule, R. Kessler, J. Guy et al.,  \emph{Improved Photometric Calibration of the SNLS and the SDSS Supernova Surveys}, \emph{A\&A} {\bf 568A} (2014) 22B,  [arXiv:1212.4864].  

\bibitem{Mosher}
J. Mosher, J. Guy, R. Kessler et al., \emph{Cosmological parameter uncertainties from SALT-II Type Ia supernova light curve models}, \emph{ApJ} {\bf 793} (2014) 16M, [arxiv:1401.4065].

\bibitem{Bazin09}
G. Bazin, N. Palanque-Delabrouille, J. Rich et al., \emph{The Core-collapse rate from the Supernova Legacy Survey},  \emph{A\&A} {\bf 499} (2009) 653, [arXiv:0904.1066] 





\end{thebibliography}
\end{document}